\documentclass{ws-ijmpa}
\usepackage{url}
\include{babarsym}

\def\Heff  {\ensuremath{{\cal H}}\xspace}

\def\state#1{\ensuremath{\vert{#1}\rangle}\xspace}
\def\GenAmp#1#2#3{\ensuremath{\langle{#1}\vert{#2}\vert{#3}\rangle}\xspace}
\def\Amp#1#2{\GenAmp{#1}{\Heff}{#2}\xspace}
\def\Ampf       {\Amp{f}{\Dz}}
\def\Ampfb      {\Amp{\bar f}{\Dz}}
\def\Ampbf      {\Amp{f}{\Dzb}}
\def\Ampbfb     {\Amp{\bar f}{\Dzb}}

\def\pislow    {\ensuremath{\pi_s}\xspace}
\def\pislowp   {\ensuremath{\pip_s}\xspace}

\def\deltaKpi {\ensuremath{{\delta_{K\pi}}}\xspace}
\def\phikpi   {\ensuremath{\phi_{K\pi}}\xspace}
\def\phikk    {\ensuremath{\phi_{KK}}\xspace}
\def\phifCP   {\ensuremath{\phi_{\fCP}}\xspace}
\def\Akppim   {\ensuremath{A_{\Kp\pim}}\xspace} 
\def\Abkppim  {\ensuremath{{\bar A}_{\Kp\pim}}\xspace} 
\def\Akmpip   {\ensuremath{A_{\Km\pip}}\xspace} 
\def\Abkmpip  {\ensuremath{{\bar A}_{\Km\pip}}\xspace}

\def\CPV      {\ensuremath{CPV}\xspace}
\def\AM       {\ensuremath{A_{M}}\xspace}
\def\AD       {\ensuremath{A_{D}}\xspace}

\def\ADkk     {\ensuremath{A_{D}^{KK}}\xspace}
\def\ACP      {\ensuremath{{A}_{CP}}\xspace}
\def\Real       {\ensuremath{{\rm Re}}\xspace}
\def\Imaginary  {\ensuremath{{\rm Im}}\xspace}
\def\Rdcs       {\ensuremath{R_{D}}\xspace} % BaBar doesn't like 
\let\Re=\Real\relax
\let\Im=\Imaginary\relax

\def\RdcsM      {\ensuremath{R_{D}^{-}}\xspace}
\def\RdcsP      {\ensuremath{R_{D}^{+}}\xspace}
\def\RdcsPm     {\ensuremath{R_{D}^{\pm}}\xspace}

\def\Rws        {\ensuremath{R_{WS}}\xspace}
\def\RwsP       {\ensuremath{R^{+}_{WS}}\xspace}
\def\RwsM       {\ensuremath{R^{-}_{WS}}\xspace}
\def\RwsPm      {\ensuremath{R^{\pm}_{WS}}\xspace}

\def\Rm         {\ensuremath{R_{M}}\xspace}

\def\xPrimeSq   {\ensuremath{{x^{\prime}}^2}\xspace}
\def\yPrimeSq   {\ensuremath{{y^{\prime}}^2}\xspace}
\def\xPrime     {\ensuremath{x^{\prime}}\xspace}
\def\xPrimeP    {\ensuremath{x^{\prime+}}\xspace}
\def\xPrimeM    {\ensuremath{x^{\prime-}}\xspace}
\def\yPrime     {\ensuremath{y^{\prime}}\xspace}
\def\yPrimeP    {\ensuremath{y^{\prime+}}\xspace}
\def\yPrimeM    {\ensuremath{y^{\prime-}}\xspace}
\def\xPrimePm   {\ensuremath{x^{\prime\pm}}\xspace}
\def\yPrimePm   {\ensuremath{y^{\prime\pm}}\xspace}
\def\xPrimePmSq {\ensuremath{{x^{\prime\pm}}^2}\xspace}
\def\yPrimePmSq {\ensuremath{{y^{\prime\pm}}^2}\xspace}

\def\xPrimePSq  {\ensuremath{{x^{\prime+}}^{+^2}}\xspace}
\def\xPrimeMSq  {\ensuremath{{x^{\prime-}}^{-^2}}\xspace}
\def\A       {\ensuremath{A}\xspace}
\def\Abar    {\ensuremath{\bar A}\xspace}
\def\Abarcc    {\ensuremath{{\bar A}^*}\xspace}
\def\Af      {\ensuremath{A_{\f}}\xspace}
\def\Afb     {\ensuremath{A_{\fbar}}\xspace}
\def\Abfb    {\ensuremath{\Abar_{\fbar}}\xspace}
\def\Abf     {\ensuremath{\Abar_{\f}}\xspace}
\def\Afcc      {\ensuremath{A^*_{\f}}\xspace}

\def\Abfbcc    {\ensuremath{\Abarcc_{\fbar}}\xspace}

\def\AfCP    {\ensuremath{A_{\fCP}}\xspace}
\def\AbfCP   {\ensuremath{{\bar A}_{\fCP}}\xspace}
\def\f       {\ensuremath{f}\xspace}
\def\fbar    {\ensuremath{\bar f}\xspace}
\def\fCP     {\ensuremath{f_{\CP}}\xspace}

\def\absx    {\ensuremath{\left|x\right|}\xspace}
\def\absy    {\ensuremath{\left|y\right|}\xspace}
\def\lf      {\ensuremath{\lambda_{\f}}\xspace}
\def\lfinv   {\ensuremath{\lambda_{\f}^{-1}}\xspace}
\def\lfCP      {\ensuremath{\lambda_{\fCP}}\xspace}
\def\lfCPinv      {\ensuremath{\lambda^{-1}_{\fCP}}\xspace}

\def\lfb     {\ensuremath{\lambda_{\fbar}}\xspace}
\def\linvkppim {\ensuremath{\lambda^{-1}_{\Kp\pim}}\xspace}
\def\lkmpip  {\ensuremath{\lambda_{\Km\pip}}\xspace}
\def\lkpkm   {\ensuremath{\lambda_{\Kp\Km}}\xspace}

\def\Di      {\ensuremath{D_1}\xspace}
\def\Dii     {\ensuremath{D_2}\xspace}

\def\slowpi      {\ensuremath{\pi_s}\xspace}
\def\slowpip     {\ensuremath{\slowpi^+}}

\def\KmKp       {\ensuremath{K^{-}K^{+}}\xspace}
\def\KpKm       {\ensuremath{K^{+}K^{-}}\xspace}

\def\pimp       {\ensuremath{\pi^{\mp}}\xspace}
\def\pipm       {\ensuremath{\pi^{\pm}}\xspace}

\def\mDz        {\ensuremath{m_{\Dz}}\xspace}

\def\m          {\ensuremath{M^{0}}\xspace}

\def\pstar      {\ensuremath{p^{*}}\xspace}
\newcommand{\kevcc}{\ensuremath{{\mathrm{\,Ke\kern -0.1em V\!/}c^2}}\xspace}

\def\hp         {\ensuremath{h^{+}}\xspace}
\def\hm         {\ensuremath{h^{-}}\xspace}

\def\qoverp     {\ensuremath{\frac{q}{p}}\xspace}
\def\poverq     {\ensuremath{\frac{p}{q}}\xspace}

\def\decayt {\ensuremath{t}\xspace}
\def\terr   {\ensuremath{\sigma_{\decayt}}\xspace}
\def\mKpi   {\ensuremath{m_{K\pi}}\xspace}
\def\DeltaM {\ensuremath{\Delta m}\xspace}
\def\Like#1{\ensuremath{{\cal L}{#1}}\xspace}
\def\changeL{\ensuremath{-2\Delta\log\Like{}}\xspace}
\def\SUthree {\ensuremath{SU(3)_F}\xspace}

\begin{document}

\markboth{Chavez, Cowan, and Lockman}{Review of $D$ Mixing}

%%%%%%%%%%%%%%%%%%%%% Publisher's Area please ignore %%%%%%%%%%%%%%%
%
\catchline{}{}{}{}{}
%
%%%%%%%%%%%%%%%%%%%%%%%%%%%%%%%%%%%%%%%%%%%%%%%%%%%%%%%%%%%%%%%%%%%%

\title{Charm Meson Mixing: An Experimental Review}
%\footnote{For the title, try not to use more than 
%3 lines. Typeset the title in 10 pt roman, uppercase and 
%boldface.}

\author{Carlos A. Chavez}
\address{PH Department, CERN\\ 
1211 Geneva 23\\
Switzerland\\
carlos.chavez.barajas@cern.ch
}

\author{Ray F. Cowan}

\address{Lepton Quark Studies Group, Laboratory for Nuclear Science, M.I.T.,
77 Massachusetts Avenue\\
Cambridge, Massachusetts 02139, USA\\
rcowan@mit.edu
%Group, Laboratory, Address\\
%City, State ZIP/Zone, Country\\
%second\_author@domain\_name}
}

\author{W.S. Lockman}
\address{University of California at Santa Cruz, Institute for Particle Physics\\
Santa Cruz, California 95064, USA\\
lockman@slac.stanford.edu
}
\maketitle

\begin{history}
\received{Day Month Year}
\revised{Day Month Year}
\end{history}

\begin{abstract}
We review current experimental results on charm mixing 
and \CP violation.  We survey experimental techniques,
including time-dependent, time-independent, and quantum-correlated
measurements.  We review techniques that use a slow pion tag from 
$\Dstarp\ra\pip\Dz$ + c.c. decays and those that do not, and cover
two-body and multi-body \Dz decay modes.  We provide a summary of 
$D$-mixing results to date and comment on future experimental prospects
at the LHC and other new or planned facilities.

\end{abstract}

\ccode{PACS numbers: 13.25Ft, 11.30Er, 12.15Ff, 14.40Lb}

%%------- INTRODUCTION --------

\section{Introduction}
\label{sec:Introduction}
\par
Quantum-mechanical mixing between neutral meson particle and anti-particle flavor 
eigenstates provides important information about electroweak interactions and the 
Cabibbo-Kobayashi-Maskawa (CKM) matrix, as well as the virtual particles that are exchanged 
in the mixing process itself. 
The two parameters characterizing
\DzdashDzb mixing are
\begin{eqnarray}
{x\equiv\frac{\Delta M}{\Gamma}},&&\Delta M \equiv M_1 - M_2\label{eq:xformula}\\
{y\equiv\frac{\Delta \Gamma}{2\Gamma}},&&\Delta\Gamma \equiv \Gamma_1-\Gamma_2
\label{eq:yformula}
\end{eqnarray}
where $M_{1,2}$ are the masses of \Donetwo, $\Gamma_{1,2}$ are the decay widths,
and $\Gamma\equiv(\Gamma_1+\Gamma_2)/2$ is the mean decay width.
\par
\begin{figure}[h!]
\begin{center}
\hbox to\hsize{%
  \includegraphics[scale=0.4]{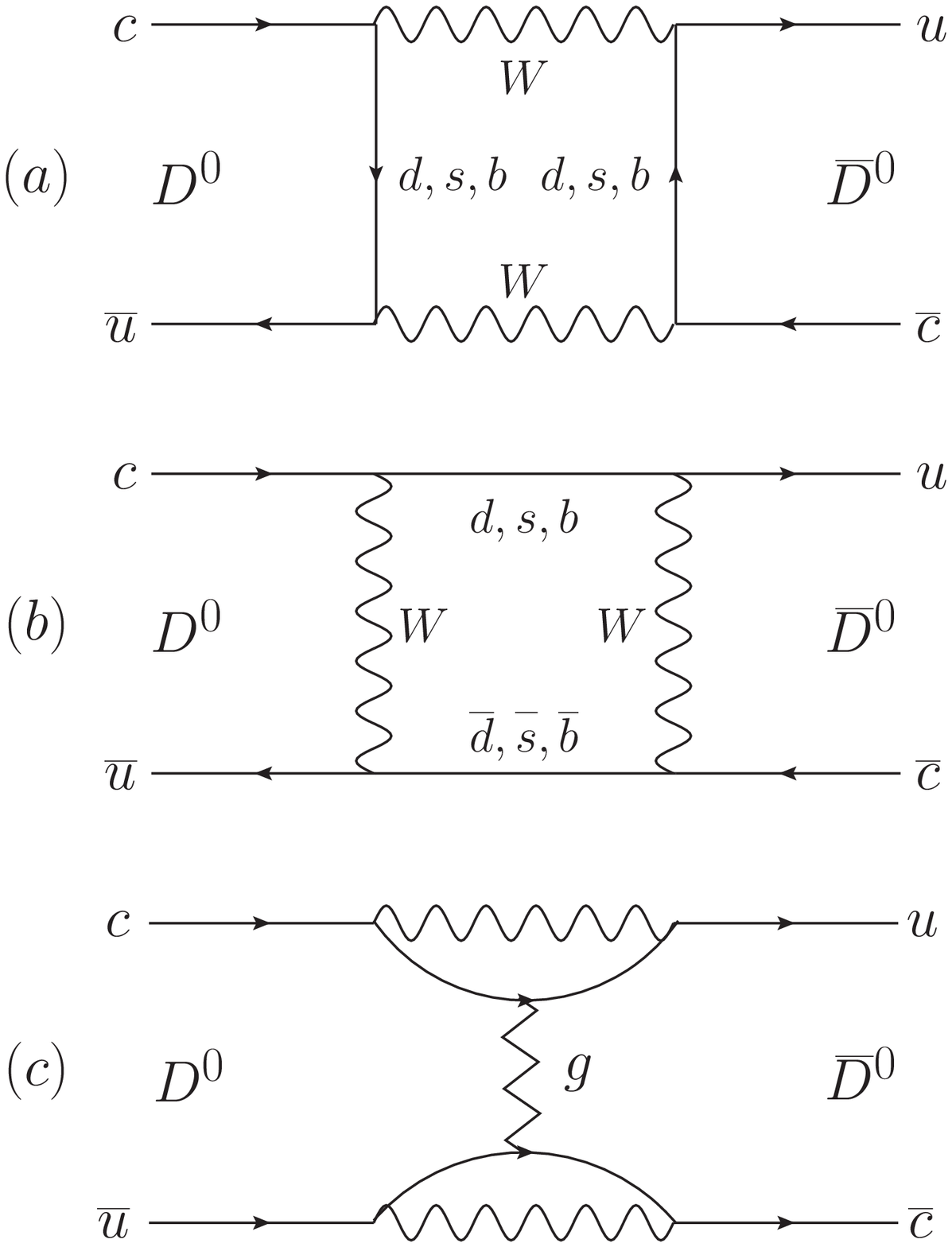}%
  \hfil
  \includegraphics[scale=0.4]{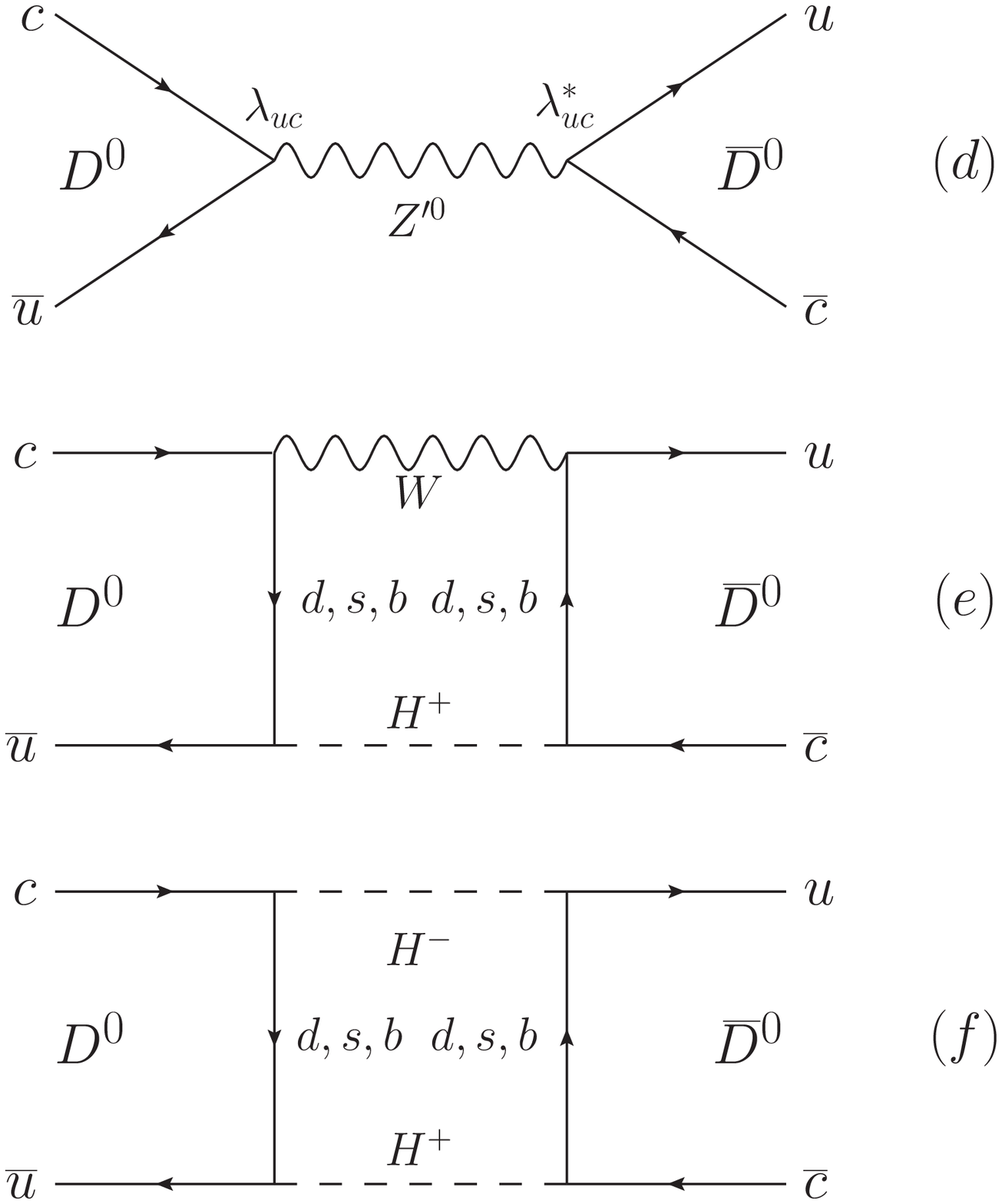}%
}
\caption{Possible short-distance amplitudes contributing to \DzdashDzb mixing. (a--b) SM boxes; (c) SM di-penguin; (d): new physics flavor-changing neutral current process mediated by a heavy $Z^{\prime0}$; (e--f): charged Higgs in the mixing loop.}
\label{fig:short_distance_diagrams}
\end{center}
\end{figure}

Mixing between the states \Kz and \Kzb, \Bz and \Bzb, and \Bs and \Bsb is 
well established. Mixing in these systems is well described  by standard model (SM) box 
diagrams containing up-type ($u$, $c$, $t$) quarks. 
In contrast, the \DzdashDzb SM 
mixing amplitude at short distances involves loops containing down-type $(d,\,s,\,b)$
quarks. 
The $s$ and $d$ box amplitudes\cite{Datta:1984jx} 
together are suppressed by by a factor $(m_s^2-m_d^2)^2/(m_W^2 m_c^2)$ due to the Glashow-Iliopoulos-Maiani (GIM) mechanism, while the contribution from loops involving $b$ 
quarks is further suppressed by the 
Cabibbo-Kobayashi-Maskawa (CKM) factor $|V_{ub}V_{cb}^*|^2/|V_{us}V_{cs}^*|^2=\order(10^{-6})$.
The contribution of the box diagrams shown in 
Fig.~\ref{fig:short_distance_diagrams}(a--b) to $x$ is $x^{box}\approx5\times 10^{-6}\left[m_s/0.2\gevcc\right]^4$.\cite{Petrov:1997ch} The di-penguin diagram shown in Fig.~\ref{fig:short_distance_diagrams}(c) 
contributes at a similar level,
but with opposite sign.\cite{Petrov:1997ch}. 
Such diagrams contribute only to $x$.
A perturbative QCD next-to-leading order (NLO) analysis of \DzdashDzb mixing\cite{Golowich:2005pt}
using an operator product expansion\cite{Georgi:1992as,Ohl:1992sr,Bigi:2000wn} 
to evaluate $\Delta\Gamma$ in terms of
local $|\Delta C|=1$ operators, followed by a 
dispersion relation to evaluate $\Delta M$,\cite{Falk:2004wg} 
obtains: $x,y\simeq 6\times10^{-7}$.
Taken together, the short-distance SM predictions 
are $x\sim\order(10^{-5})$, $y\sim\order(10^{-7})$, 
far below the current
measurements, $x,y\sim\order(10^{-2})$. 
\par
\begin{figure}[h]
\begin{center}
\includegraphics[scale=0.5]{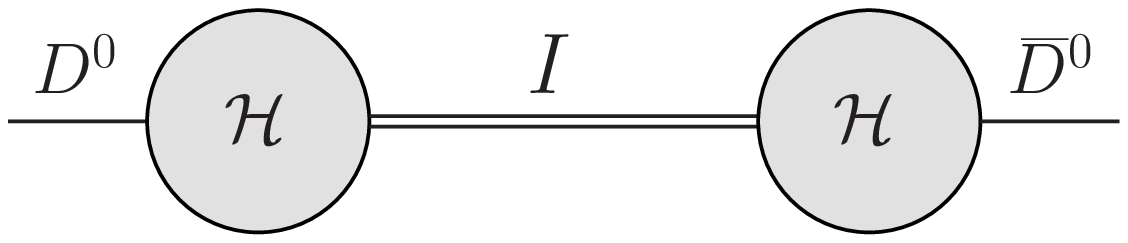}
\caption{Long-distance contribution from an intermediate state $I$ to \DzdashDzb mixing. ${\cal H}$ is the Hamiltonian governing weak decays. From Ref.~\protect\refcite{Donoghue:1985hh}.} 

\label{fig:long_distance_diagram}
\end{center}
\end{figure}

The long-distance contributions to \DzdashDzb mixing are inherently 
nonperturbative and thus difficult to estimate. 
There are two approaches to estimating \DzdashDzb mixing in the SM:
An inclusive approach
uses the Operator Product Expansion and quark-hadron duality to expand $x$ and $y$ in terms of local operators.\cite{Georgi:1992as,Ohl:1992sr,Bigi:2000wn}
If the charm quark mass $m_c$ is large compared to the scale of strong interactions, the series can be truncated after a few terms. 
Such calculations typically yield $x,y\leq10^{-3}$.
The exclusive approach
sums over intermediate hadronic states to which both \Dz and \Dzb can decay, as 
shown schematically in Fig.~\ref{fig:long_distance_diagram}. 
Here, non-vanishing $y$ arises from 
\SUthree breaking in decay rates when summing over intermediate states within an \SUthree multiplet. 
Ref.~\refcite{Falk:2001hx} found that \SUthree violation in the final state phase space could provide enough
\SUthree breaking to generate $y\sim 10^{-2}$. Ref.~\refcite{Falk:2004wg} used a dispersion relation to relate $x$
to $y$ and found $x$ to be in the range $(-1.0, -0.1)\times y$. 
\par
New physics (NP) processes, some examples of which are shown in 
Figs.~\ref{fig:short_distance_diagrams}(d--f), could enhance the \DzdashDzb mixing rate to the 
level of experimental detection, but the predictions for these rates also span many orders of magnitude.\cite{Nelson:1999fg,Petrov:2006nc,Golowich:2007ka}. 
Given the uncertainties in both 
the SM and NP calculations, observation of \DzdashDzb mixing at $\order(10^{-2)}$ does not unambiguously indicate
the presence of new physics.
See Refs.~\refcite{Nelson:1999fg} and~\refcite{Petrov:2006nc} for a summary of \DzdashDzb 
mixing parameter predictions. 
\par 
Evidence for \DzdashDzb mixing was reported in 2007 using
high-luminosity data sets acquired at the $B$~factories\cite{Aubert:2007wf,Staric:2007dt} and 
Tevatron collider.\cite{Aaltonen:2007uc} 
While 
the significance of the current world average for \DzdashDzb mixing is greater than ten standard
deviations ($10\sigma$),\cite{Asner:2010qj} to date
no one single \DzdashDzb mixing measurement exceeds $5\sigma$, 
the commonly accepted criterion for observation.
\subsection{\DzdashDzb Mixing Formalism}
\label{sec:Neutral_Meson_Mixing}
The \Dz and \Dzb mesons  are produced as flavor eigenstate with charm quantum numbers $C=+1$ and $-1$, respectively.
They propagate and decay
according to the Schr\"odinger equation:
\begin{equation}
i\frac{\partial}{\partial t}\left(\begin{array}{c}
\Dz(t)\\\Dzb(t)
\end{array}\right)=\left({\bf M}-\frac{i}{2}
{\bf\Gamma}\right)\left(\begin{array}{c}
\Dz(t)\\\Dzb(t)
\end{array}\right)\label{eq:schroedinger}.
\end{equation}
Mixing between \Dz and \Dzb occurs because these flavor states are not the eigenstates \Done and \Dtwo of the 
\DzdashDzb mass matrix ${\bf M}-i{\bf\Gamma}/2$, but linear combinations of them.
Assuming that the product of charge conjugation, parity and time reversal (CPT) is conserved,\cite{Bigi:2009zz} 
the eigenstates of Eq.~\ref{eq:schroedinger},
$|D_{1,2}\rangle$
are given by:\cite{Bigi:2009zz,Kabir:1968zz}
\begin{eqnarray}
\begin{array}{rcc}
|D_1\rangle\ &=&\ p|\Dz\rangle+q|\Dzb\rangle, \\
|D_2\rangle\ &=&\ p|\Dz\rangle-q|\Dzb\rangle,
\end{array}\label{eq:mass_to_quark_eigen}
\end{eqnarray}
and inversely
\begin{eqnarray}
\begin{array}{rcc}
|\Dz\rangle\ &=&\  \frac{1}{2p}\left(|\Done\rangle+|\Dtwo\rangle\right), \\
|\Dzb\rangle\ &=&\ \frac{1}{2q}\left(|\Done\rangle-|\Dtwo\rangle\right),
\end{array}\label{eq:quark_to_mass_eigen}
\end{eqnarray}
where the complex quantities $p$ and $q$ satisfy
\begin{equation}
\left(\qoverp\right)^2=\frac{M^*_{12}-\frac{i}{2}\Gamma^*_{12}}
{M_{12}-\frac{i}{2}\Gamma_{12}},\hskip 10pt\left|p\right|^2+\left|q\right|^2=1\label{eq:pandq},
\end{equation}
where $M_{12}$ and $\Gamma_{12}$ are the complex off-diagonal elements of the $2\times2$ matrices
${\bf M}$ and ${\bf\Gamma}$, respectively. 
In the limit of \CP conservation, \Done is \CP-even and \Dtwo is \CP-odd.\footnote{%
We use the \CP phase convention: $\CP\state\Dz=+\state\Dzb$ and $\CP\state\Dzb=+\state\Dz$.}  
The eigenvalues of Eq.~\ref{eq:schroedinger} are:
\begin{equation}
\gamma_{1,2}\equiv M_{1,2}-\frac{i}{2}\Gamma_{1,2}=M_{11}-\frac{i}{2}\Gamma_{11}\pm\qoverp\left(M_{12}-\frac{i}{2}\Gamma_{12}\right)\label{eq:eigenvalues}
\end{equation}
The eigenstates of Eq.~\ref{eq:schroedinger} develop in time as follows:\cite{Bigi:2009zz,Kabir:1968zz}
\begin{equation}
|\Donetwo(t)\rangle = e_{1,2}(t)|\Donetwo(0)\rangle,\hskip20pt 
e_{1,2}(t)\equiv\exp\left[-i\left(M_{1,2}-\frac{i}{2}\Gamma_{1,2}\right)t\right],
\label{eq:mass_eigen_time_evol}
\end{equation}
Using
Eq.~\ref{eq:mass_to_quark_eigen}, Eq.~\ref{eq:quark_to_mass_eigen} and Eq.~\ref{eq:mass_eigen_time_evol},
the proper time evolution of a state which is initially a pure \Dz (\Dzb) is given by:
\begin{eqnarray}
\state{\Dz(t)}&=&g_+(t)\state{\Dz}+\qoverp g_-(t)\state{\Dzb}\label{eq:DzEvol},\\
\state{\Dzb(t)}&=&g_+(t)\state{\Dzb}+\poverq g_-(t)\state{\Dz}\label{eq:DzbEvol}.
\end{eqnarray}
where:
\begin{equation}
g_\pm(t)=\left[e_1(t)\pm e_2(t)\right]/2.\label{eq:goft}
\end{equation}
The probabilities for obtaining a \Dz or \Dzb at proper time $t$, starting from an initially pure
\Dz or \Dzb are:
\begin{eqnarray}
|\langle\Dz|\Dz(t)\rangle|^2&=&|\langle\Dzb|\Dzb(t)\rangle|^2\nonumber\\
&=&\left|g_+(t)\right|^2=\frac12e^{-\Gamma t}\left[\cosh(y\Gamma t)+\cos(x\Gamma t)\right],\label{eq:probnomix}\\
|\langle\Dzb|\Dz(t)\rangle|^2&=&\left|\qoverp\right|^2|g_-(t)|^2=\frac12\left|\qoverp\right|^2e^{-\Gamma t}\left[\cosh(y\Gamma t)-\cos(x\Gamma t)\right]\label{eq:probmixDz},\\
|\langle\Dz|\Dzb(t)\rangle|^2&=&\left|\poverq\right|^2|g_-(t)|^2=\frac12\left|\poverq\right|^2e^{-\Gamma t}\left[\cosh(y\Gamma t)-\cos(x\Gamma t)\right]\label{eq:probmixDzb},
\end{eqnarray} 
If both $x$ and $y$ are zero, then the probability for a \Dz to mix to a \Dzb or for a \Dzb to
mix to a \Dz
will be identically 
zero for all proper times. If either $x$ or $y$ is non-zero, then \DzdashDzb mixing will occur. 
\par
\begin{figure}[h!]
\begin{center}
  \includegraphics[width=13cm]{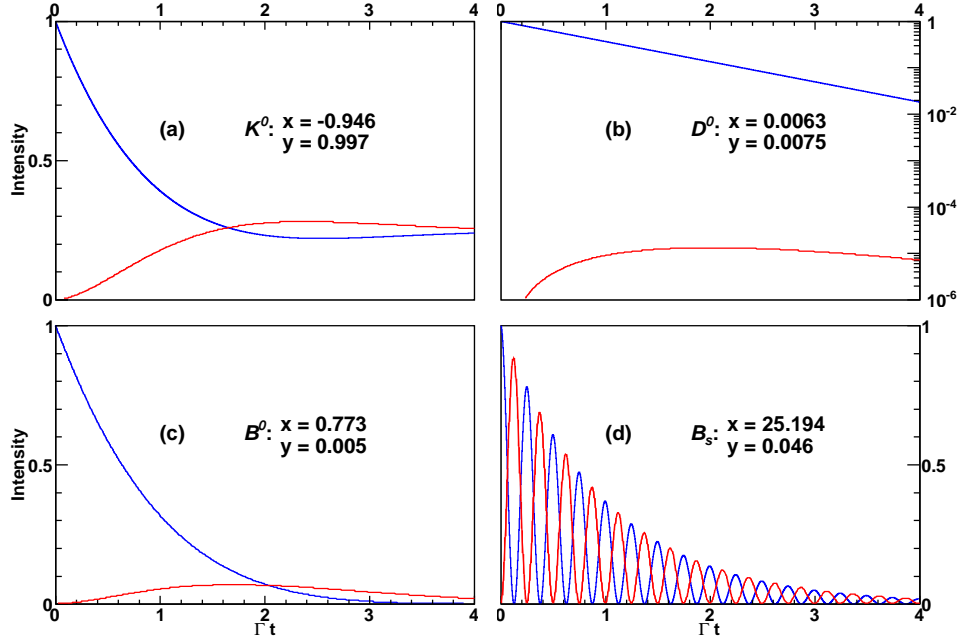}
\caption{The unmixed (blue) and mixed (red) intensities for an initially pure  
(a) $K^0$; (b) \Dz; 
(c) $B^0$; (d) $B_s$ state. The vertical scale in (b) is logarithmic, the others linear.
The values of the mixing parameters as defined in 
Eqs.~\ref{eq:xformula} and \ref{eq:yformula} are obtained using data from
Ref.~{\protect\refcite{Nakamura:2010zzi}}, assuming $\left||q/p\right|=1$.}
\label{fig:mixing_intensities}
\end{center}
\end{figure}

$M_{12}$ and $\Gamma_{12}$ 
determine the mass and width splittings $\Delta M$ and $\Delta\Gamma$, respectively:
\begin{eqnarray}
\Delta M &\equiv& M_1 - M_2 = 2\Real\left[\qoverp(M_{12} - \frac{i}{2}\Gamma_{12})\right]\label{eq:DM}\\
\Delta \Gamma &\equiv& \Gamma_1-\Gamma_2 = -4\Imaginary\left[\qoverp(M_{12} - \frac{i}{2}\Gamma_{12})\right]\label{eq:DGamma},
\end{eqnarray}
and therefore the characteristics of \DzdashDzb mixing.  
We show the
unmixed and mixed intensities as a function of the dimensionless variable, $\Gamma t$, for 
initially pure states of  $K^0$, \Dz, \Bz and $B_s$, in Figs.~\ref{fig:mixing_intensities}(a--d), 
respectively. 
Of the four lowest-lying neutral pseudoscalar meson systems, the \DzdashDzb system shows the
smallest mixing, as noted earlier. 
In the $K^0$ system, both \absx and \absy are both of order 1; in the \Dz system, \absx and \absy are both of order 1\%; in the \Bz and $B_s$ systems, $\absx\gg\absy$.    
\par
\begin{figure}[h!]
\begin{center}
  \includegraphics[width=5cm]{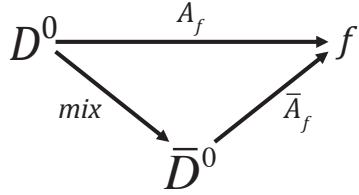}
\caption{Diagram illustrating the interference between the amplitude for direct
\Dz decay to \f and that for $\Dz\to\Dzb$ mixing followed by \Dzb decay to \f.}  
\label{fig:twopath}
\end{center}
\end{figure}

From Eq.~\ref{eq:DzEvol} (Eq.~\ref{eq:DzbEvol}),
the amplitude that a \Dz (\Dzb) produced at $t=0$ will develop into a linear combination of \Dz and \Dzb 
and decay into $f$ (\fbar) at time $t$ is:
\begin{eqnarray}
\Amp{f}{\Dz(t)}&=&\Af g_+(t)+\Abf \qoverp g_-(t)
%=\Abf\qoverp\left[\lfinv g_+(t)+g_-(t)\right]
\label{eq:AmpEvol},\\
\Amp{\fbar}{\Dzb(t)}&=&\Abfb g_+(t)+ \Afb \poverq g_-(t)
%=\Afb\poverq\left[\lfbphan g_+(t)+g_-(t)\right]
\label{eq:AmpEvolb},
\end{eqnarray}
where \Af and \Abf are the \Dz and \Dzb decay amplitudes to a final state \f; \Afb and \Abfb are the \Dz and \Dzb decay amplitudes to a final 
state \fbar:
\begin{eqnarray}
\Af&\equiv&\Ampf,\label{eq:Af}\\
\Abf&\equiv&\Ampbf,\label{eq:Abf}\\
\Afb&\equiv&\Ampfb,\label{eq:Afb}\\
\Abfb&\equiv&\Ampbfb\label{eq:Abfb}.
\end{eqnarray}
where ${\cal H}$ is the Hamiltonian governing weak decays.
Written in terms of the decay amplitudes, the general expressions for the time-dependent decay rates $\Gamma(\Dz(t)\to\f)$ and $\Gamma(\Dzb(t)\to\fbar)$ are:
\begin{eqnarray}
\lefteqn{\Gamma(\Dz(t)\to f)=}\nonumber\\ 
&\quad&\frac{e^{-\Gamma t}}{2}\left[|\Af|^2[\cosh(y\Gamma t)+\cos(x\Gamma t)]
+\left|\qoverp\right|^2|\Abf|^2[\cosh(y\Gamma t) - \cos(x\Gamma t)]\right.\nonumber\\
&\quad&\qquad\left.-2\Real\left(\Afcc\Abf\qoverp\right)\sinh(y\Gamma t) + 2\Imaginary\left(\Afcc\Abf\qoverp\right)\sin(x\Gamma t)\right],\label{eq:Dztof}\\
\lefteqn{\Gamma(\Dzb(t)\to\fbar)=}\nonumber\\ 
&\quad&\frac{e^{-\Gamma t}}{2}\left[|\Abfb|^2[\cosh(y\Gamma t)+\cos(x\Gamma t)]
+\left|\poverq\right|^2|\Afb|^2[\cosh(y\Gamma t) - \cos(x\Gamma t)]\right.\nonumber\\
&\quad&\qquad\left.-2\Real\left(\Abfbcc\Afb\poverq\right)\sinh(y\Gamma t) + 2\Imaginary\left(\Abfbcc\Afb\poverq\right)\sin(x\Gamma t)\right].\label{eq:Dzbtofb}
\end{eqnarray}
To describe the time dependence for the ``wrong-sign'' (WS) decay 
such as $\Dz\to\Kp\pim$ ($\Dzb\to\Km\pip$), we rewrite Eq.~\ref{eq:Dztof} (Eq.~\ref{eq:Dzbtofb}) in terms of of the Cabibbo-favored (CF) amplitude \Abf (\Afb) and the parameter \lfinv (\lfb), where
\begin{equation}
\lf\equiv\qoverp\frac{\Abf}{\Af}.\label{eq:lf}
\end{equation}
For decay times $t$ satisfying $|x\Gamma t|,|y\Gamma t|\ll 1$, the decay rates are given by:
\begin{eqnarray}
\lefteqn{\Gamma(\Dz(t)\to f)=}\nonumber\\
&\quad&e^{-\Gamma t}\left|\Abf\right|^2\left|\qoverp\right|^2
\left[\left|\lfinv\right|^2-\Real\left(\lfinv\right)y\Gamma t-\Imaginary\left(\lfinv\right)x\Gamma t + \frac{x^2+y^2}{4}(\Gamma t)^2\right],\qquad
\label{eq:Dztofsmall}\\
\lefteqn{\Gamma(\Dzb(t)\to\fbar)=}\nonumber\\
&\quad&e^{-\Gamma t}\left|\Afb\right|^2\left|\poverq\right|^2
\left[\left|\lfb\right|^2-\Real\left(\lfb\right)y\Gamma t-\Imaginary\left(\lfb\right)x\Gamma t+\frac{x^2+y^2}{4}(\Gamma t)^2\right].\qquad\qquad
\label{eq:Dzbtofbsmall}
\end{eqnarray}
Under similar conditions, the time-dependent rates for \Dz and \Dzb decaying to a \CP-eigenstate $\f=\fbar=\fCP$ 
can be written as:
\begin{eqnarray}
\lefteqn{\Gamma(\Dz(t)\to\fCP)=}\nonumber\\
&&e^{-\Gamma t}\left|\AfCP\right|^2\left[1-\Real\left(\lfCP\right)y\Gamma t +\Imaginary\left(\lfCP\right)x\Gamma t
+ \left|\lfCP\right|^2\frac{x^2+y^2}{4}(\Gamma t)^2
\right]\label{eq:Dztofcpsmall},\qquad\\
\lefteqn{\Gamma(\Dzb(t)\to\fCP)=}\nonumber\\
&&e^{-\Gamma t}\left|\AbfCP\right|^2\left[1-\Real\left(\lfCPinv\right)y\Gamma t +\Imaginary\left(\lfCPinv\right)x\Gamma t
 + \left|\lfCPinv\right|^2\frac{x^2+y^2}{4}(\Gamma t)^2
\right]\label{eq:Dzbtofcpsmall},\qquad
\end{eqnarray}
where the terms proportional to $e^{-\Gamma t}(\Gamma t)^2$ are due to mixing, those proportional to $e^{-\Gamma t}\Gamma t$ are
due to the interference between mixing and decay, while those proportional to $e^{-\Gamma t}$ are due to direct decay. 
Fig.~\ref{fig:twopath} illustrates the two interfering decay paths from an initial \Dz to to a final state \f. 
\subsection{\CP violation}
\label{sec:CP_Violation}
There are three different types of \CP-violating effects in meson 
decays:\footnote{For a complete review, see 
Refs.~\refcite{Bigi:2009zz,Nir:1999mg,Nir:2005js}.}
\begin{enumerate}
\item \CP violation (\CPV) in mixing;
\item \CPV in decay, also known as direct \CPV;
\item \CPV in the interference between a direct decay, $\Dz\to f$, and a decay involving mixing, $\Dz\to\Dzb\to\f$. 
\end{enumerate}
\CPV in mixing and in the interference between mixing and decay is referred to as indirect \CPV.
\par
\CPV in mixing occurs when the mixing probability of \Dz to \Dzb is different than the mixing probability of \Dzb to \Dz.
As can be seen from Eqs.~\ref{eq:probmixDz} and \ref{eq:probmixDzb}, this happens if and only if
$\left|q/p\right|\ne1$. This type of $CP$-violating effect depends only on the mixing parameters
and not the final state of the decay. 
\par
\begin{figure}[h!]
\begin{center}
\includegraphics[width=7.1cm]{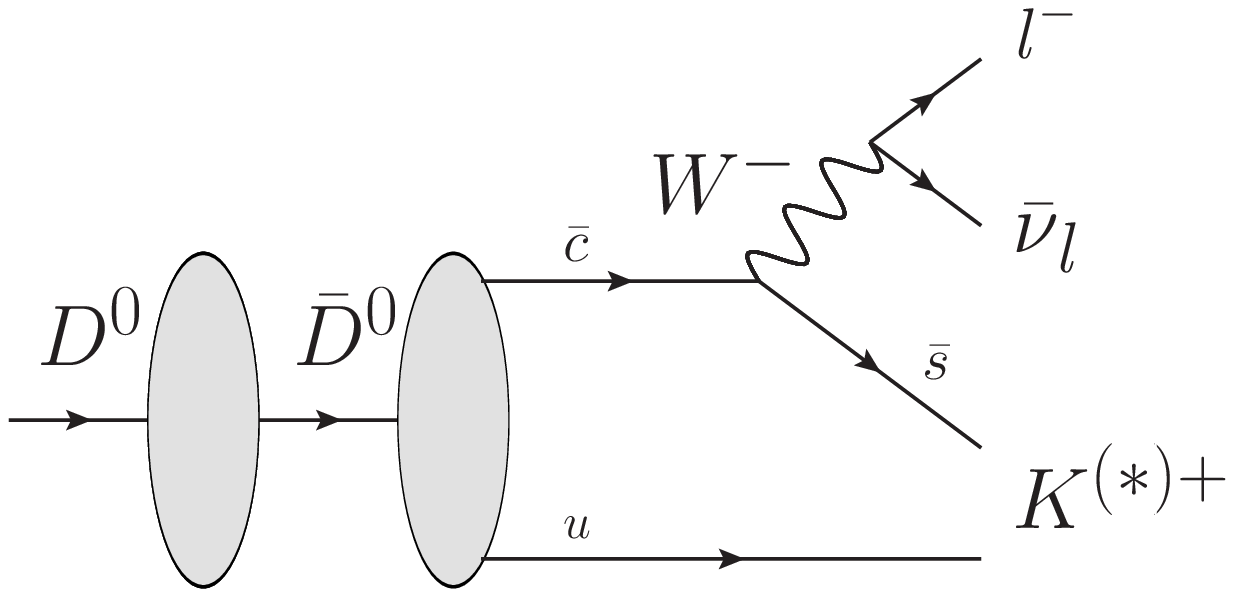}
\caption{Diagram for the decay
$\Dz\to K^{(*)+}l^-{\bar\nu}_l$. }
\label{fig:semilepfeyn}
\end{center}
\end{figure}

As an example, consider the decay
$\Dz\to K^{(*)+}l^-{\bar\nu}_l$. The diagram for this decay is illustrated in 
Fig.~\ref{fig:semilepfeyn}. Within the SM, the \Dz must first mix to \Dzb, followed by the
direct decay $\Dzb\to K^{(*)+}l^-{\bar\nu}_l$. There is no direct decay of \Dz to the final state 
$K^{(*)+}l^-{\bar\nu}_l$ in the SM. 
Therefore, the time-dependent \CP asymmetry:
\begin{eqnarray}
{\cal A}_{SL}&=&\frac{d\Gamma/dt(\Dz\to K^{(*)+}l^-{\bar\nu}_l)-d\Gamma/dt(\Dzb\to K^{(*)-}l^+{\nu}_l)}
{d\Gamma/dt(\Dz\to K^{(*)+}l^-{\bar\nu}_l)+d\Gamma/dt(\Dzb\to K^{(*)-}l^+{\nu}_l)}\label{eq:ASL}
\end{eqnarray}
is equal to the mode-independent quantity: 
\begin{eqnarray}
{\AM}&=&\frac{|q/p|^2-|p/q|^2}{|q/p|^2+|p/q|^2}\label{eq:A_M}
\end{eqnarray}
which is used to characterize \CPV in mixing.
The techniques used to analyze the decay $\Dz\to K^{(*)+}l^-{\bar\nu}_l$ for \DzdashDzb mixing
are discussed in Sec.~\ref{Analysis_Techniques_Semileptonic_Decays}; the results are presented 
in Sec.~\ref{Semileptonic_Decays_results}. 
\par
\CPV in decay occurs when the amplitude for a decay and its $CP$ conjugate process have
different magnitudes: $\left|\Abfb/\Af\right| \ne 1$. 
In charged $D$ modes where no mixing can occur, \CPV in decay is characterized 
by non-zero values for the time-integrated asymmetry:
\begin{equation}
{\cal A}_{f^\pm}\equiv\frac{\Gamma(\Dm\to f^-)-\Gamma(\Dp\to f^+)}{\Gamma(\Dm\to f^-)+\Gamma(\Dp\to f^+)}
= \frac{\left|\Abar_{f^-}/\A_{f^+}\right|^2-1}{\left|\Abar_{f^-}/\A_{f^+}\right|^2+1}
\label{eq:AsyDch}
\end{equation}
Consider the case where two amplitudes:
\begin{equation}
\begin{array}{rcl}
\Af &=& |a_1|e^{i(\delta_1+\phi_1)}+|a_2|e^{i(\delta_2+\phi_2)}\\
\Abfb &=& |a_1|e^{i(\delta_1-\phi_1)}+|a_2|e^{i(\delta_2-\phi_2)}
\end{array}\label{eq:twoamp}
\end{equation}
mediate the decay,
where $\delta_{1,2}$ and $\phi_{1,2}$ are the strong and weak phases, respectively, 
of amplitude $A_{1,2}$. The weak
phase changes sign under \CP, whereas the strong phase does not. The \CP asymmetry 
${\cal A}_{f^\pm}$ can then be written as:
\begin{equation}
{\cal A}_{f^\pm}=-\frac{2|a_1||a_2|\sin(\delta_2-\delta_1)\sin(\phi_2-\phi_1)}
{|a_1|^2+|a_2|^2+2|a_1||a_2|\cos(\delta_2-\delta_1)\cos(\phi_2-\phi_1)}.\label{eq:twoamp_asy}
\end{equation}
Thus, direct \CPV will occur (${\cal A}_{f^\pm}\ne0$) only if the differences between the \CP-conserving strong phases and the 
differences between the \CP-violating weak phases of the two contributing amplitudes are not zero or multiples of $\pi$.
In neutral $D$ decays, direct \CPV is characterized by the mode-dependent parameter $A_D$:\cite{Nakamura:2010zzi}
\begin{eqnarray}
%\sqrt{\frac{1+A_D}{1-A_D}}=\frac{\left|\Af/\Abf\right|}{\left|\Abfb/\Afb\right|}.\label{eq:A_D}
\AD&\equiv&\frac{\left|\Af/\Abf\right|^2-\left|\Abfb/\Afb\right|^2}{\left|\Af/\Abf\right|^2+\left|\Abfb/\Afb\right|^2}\label{eq:A_D}.
\end{eqnarray}
\CPV in the interference between a decay without mixing, $\Dz\to\f$, and a decay with mixing, $\Dz\to\Dzb\to\f$, 
where
\f can be reached from both \Dz and \Dzb decays, can also occur.
Consider the time-dependent rate asymmetry of neutral meson decays to a final \CP eigenstate:
\begin{equation}
\ACP(t)=\frac{\Gamma(\Dz(t)\to\fCP)-\Gamma(\Dzb(t)\to\fCP)}
{\Gamma(\Dz(t)\to\fCP)+\Gamma(\Dzb(t)\to\fCP)}\label{eq:ACP}
\end{equation}
If both \CPV in mixing and decay are absent, then $|q/p|=1$,  $|\AbfCP/\AfCP|=1$ and therefore, 
$|\lfCP| = 1$. 
As can be seen by comparing Eqs.~\ref{eq:Dztofcpsmall} and \ref{eq:Dzbtofcpsmall},
$\ACP(t)$ will be nonzero when ${\lfCPinv}\ne\lfCP$, 
Therefore, \CPV in the interference between mixing and decay will be present 
if $\Imaginary\lfCP\ne0$, which implies $\sin\phifCP\ne0$, where
$\lfCP\equiv|\lfCP|\exp({i\phifCP})$. The phase $\phifCP$ is the sum of the phase difference between $q$ and $p$,
\begin{equation}
\varphi\equiv\arg\left(q/p\right)\label{eq:varphi}
\end{equation}
and
the (weak) phase difference between \Abf and \Af. 
In general, the weak phase component of $\arg(\lf)$ is said to characterize \CPV in 
the interference between mixing and decay.
\par
The quantities \lfinv and \lfb can
be evaluated for the modes $f=\Kp\pim$ and $\f=\KpKm$ 
in terms of \AM and 
the mode-dependent quantities \Rdcs, \AD, \ADkk, \phikpi and \phikk:\cite{Nakamura:2010zzi}
\begin{eqnarray}
\linvkppim&=&-\sqrt{\Rdcs}\sqrt[4]{\frac{(1+\AD)(1-\AM)}{(1-\AD)(1+\AM)}}e^{-i(\deltaKpi+\phikpi)},\label{eq:linvkppim}\\
\lkmpip&=&-\sqrt{\Rdcs}\sqrt[4]{\frac{(1-\AD)(1+\AM)}{(1+\AD)(1-\AM)}}e^{-i(\deltaKpi-\phikpi)},\label{eq:lkmpip}\\
\lkpkm&=&\sqrt[4]{\frac{(1-\ADkk)(1+\AM)}{(1+\ADkk)(1-\AM)}}e^{i\phikk}\label{eq:lkpkm}.
\end{eqnarray}
For $f=\Kp\pim$ the quantity \Rdcs is defined to be
\begin{equation}
\begin{array}{ccc}
\Rdcs\equiv\sqrt{\RdcsP\RdcsM},\hskip10pt & \RdcsP\equiv|\Akppim/\Abkppim|^2, \hskip10pt &\RdcsM\equiv|\Abkmpip/\Akmpip|^2.
\end{array}\label{eq:R_D}
\end{equation}
In the absence of direct \CPV, 
\begin{equation}
\begin{array}{cc}
\RdcsP=|\Akppim/\Akmpip|^2, \hskip10pt &\RdcsM\equiv|\Abkmpip/\Abkppim|^2,
\end{array}
\end{equation}
are the ratios of the doubly-Cabibbo-suppressed (DCS) to Cabibbo-favored (CF) decay widths for \Dz and \Dzb decays,
respectively. These ratios are ${\cal O}(\tan^4(\theta_c))\approx0.3\%$, where $\theta_c$ is the Cabibbo angle. 
Additionally, \deltaKpi is the relative strong phase between \Akppim and \Abkppim. The quantities \AD and 
$\phi$ are in general mode-dependent. However, assuming that SM tree-level amplitudes dominate the decays, 
$\phi$ appearing in Eqs.~\ref{eq:linvkppim}, \ref{eq:lkmpip} and \ref{eq:lkpkm} will be the same for the 
$K\pi$ and $KK$ modes.\cite{Nir:2005js,Kagan:2009gb}
\par
Traditional SM estimates for \CP asymmetries in $D$-meson decays
are small, less than ${\cal O}(0.01\%)$.\cite{Buccella:1994nf,Bianco:2003vb,Petrov:2004gs}
This is because, to a very good approximation, only two generations of quarks are involved in
charm mixing and decay, while the CKM mechanism\cite{Kobayashi:1973fv} requires three 
quark generations to produce 
\CPV.\cite{Harrison:1998yr} Present experimental uncertainties on time-integrated \CP asymmetries in $D$ decays
are ${\cal O}(0.1\%)$.\cite{Asner:2010qj} Through 2010, all measured \CP asymmetries in $D$ decays were consistent with 
zero within experimental errors.\cite{Asner:2010qj,Nakamura:2010zzi} 
\par
In 2011, the LHCb Collaboration presented evidence for direct \CPV by
measuring the difference in time-integrated \CP asymmetries between two singly Cabibbo suppressed
$D$ decay modes: $\Delta \ACP\equiv\ACP(D\to\Kp\Km)-\ACP(D\to\pip\pim)=(-0.82\pm0.21({\rm stat})\pm0.11({\rm syst}))\%$.\cite{Charles:2011nx} In this 
difference, the mode-independent indirect contribution cancels. 
A new standard 
model calculation\cite{Brod:2011re} of this difference, while uncertain to a factor of a few, may
accommodate this intriguingly large experimental result. 
New physics, such as supersymmetric
gluino-squark loops, could also yield direct \CP asymmetries as large as ${\cal O}(1\%)$
.\cite{Grossman:2006jg} 

\subsection{Outline}
\label{sec:Outline}
This paper discusses the experimental status of \DzdashDzb mixing and \CPV as
of the end of the 2011 calendar year.  
We review the current results from recent
colliding-beam and fixed-target experiments and discuss in some detail the 
techniques involved.  
We survey the primary analysis methods used to study two-body and
multi-body hadronic and semileptonic $\Dz$ decays.  Then we 
present results from experimental measurements of mixing and searches 
for \CPV from time-independent analyses (those that do not use the 
proper decay time of the $\Dz$ to search for mixing) and time-dependent
analyses (which do use the proper $\Dz$ decay time) as well as 
quantum-correlated decays.  Finally we review
future prospects for measurements in the near- and longer-term future and
summarize the overall status of $\Dz$-$\Dzb$ mixing experiments.

%% Various sections and subsections below can be split out into
%% separate files as desired.

%%------- ANALYSIS TECHNIQUES --------

\section{Analysis Techniques for Measuring Charm Mixing and CP Violation}
\label{sec:Analysis_Techniques}

\subsection{Time-independent Methods}
\label{sec:Analysis_Techniques_Time-independent_Methods}
    
% Time-independent analysis methods

\par
Time-independent methods provide an important technique
for measuring \Dz-\Dzb mixing and searching for 
\CP violation in charm decays.  They also yield
information on relative strong phases between
mixed and direct decays for several different hadronic modes of 
interest to mixing studies.  Knowledge of the strong phase $\deltaKpi$ between
\Dz\to\Kp\pim and \Dzb\to\Kp\pim 
allows conversion of the observable \yPrime to mixing parameter $y$ (see
Sec.~\ref{sec:Analysis_Techniques_Hadronic_Two-body_Decay_Modes}).

\par
As the name implies, these methods do not make use of decay-time information.
Instead, they count the numbers of \Dz and \Dzb decays to specific modes 
when the $\DzdashDzb$ pair has been produced in a quantum-coherent,
charge conjugation ($C$) eigenstate.  Relative numbers
of decay modes of 
both singly-tagged (ST) events, where one \Dz or \Dzb is fully 
reconstructed, 
and doubly-tagged (DT) events, where both the \Dz and the \Dzb mesons are 
fully reconstructed,
provide information on the mixing parameters $x$ and $y$, the strong phase
differences $\delta_i$ for each decay mode~$i$, and their DCS decay rates.
This method is especially useful when separating the individual \Dz
and \Dzb decay vertices is difficult, as in the case of non-asymmetric
energy \epem colliders.

\par
At present, these methods have
been performed\cite{Rosner:2008fq,Sun:2010zz} only
at the 3.770~\gev resonance, but could also
be done at higher center-of-mass energies through initial-state radiation,
if the number of ISR photons can be determined (so that the \CP state of
the coherent $\DzdashDzb$ pair is known).

\subsubsection{Correlated Decays at 3.770~\gev}
\label{sec:correlated_decays_method}

In \epem collisions at or above 3.770~\gev that produce a $\DzdashDzb$ pair, 
the production of the 
pair may be assumed to proceed through a single virtual photon with
$J^{PC} = 1^{--}$.
At 3.770~\gev, the final state will have $C = -1$.
At higher energies, additional pions and photons may be
produced:\cite{Goldhaber:1976fp}
\begin{equation}
\epem\to\Dz\Dzb + m(\piz) + n (\gamma)
\end{equation}
where $m$, $n \ge 0$.  
Therefore the \Dz-\Dzb pair will be produced with $C(\Dz\Dzb) = -1^{n+1}$.  When
$n=0$, $C(\Dz\Dzb)$ will be~$-1$.  The value of $m$ is not a factor
since $C(\piz) = +1$.

Additionally, if the $\DzdashDzb$ pair has relative angular momentum $l$, then 
$C(\Dz\Dzb) = P(\Dz\Dzb) = -1^l$ where $P$ is the parity operator.
Assuming \CP is conserved, we can write the wavefunction of the \DzdashDzb
state in 
the center of mass system (where the mesons have momentum $\vec p$ and $-\vec p$, respectively) in
terms of either the flavor eigenstates \Dz and \Dzb or the $CP$ eigenstates \Di, \Dii (with $CP = +1$, $-1$ 
respectively).  If $n$ is even (0, 2, $\dots$), the 
produced $\Dz\Dzb$ state is 
\begin{equation}
|\Dz(\vec p)\rangle|\Dzb(-\vec p)\rangle - |\Dzb(\vec p)\rangle|\Dz(-\vec p)\rangle = |\Dii(\vec p)\rangle|\Di(-\vec p)\rangle - |\Di(\vec p)\rangle|\Dii(-\vec p)\rangle
\end{equation}
which has $C = P = -1$.
If $n$ is odd, the produced state is
\begin{equation}
|\Dz(\vec p)\rangle|\Dzb(-\vec p)\rangle + |\Dzb(\vec p)\rangle|\Dz(-\vec p)\rangle = |\Di(\vec p)\rangle|\Di(-\vec p)\rangle - |\Dii(\vec p)\rangle|\Dii(-\vec p)\rangle
\end{equation}
with $C = P = +1$. 
Therefore at 3.770~\gev when both the \Dz and \Dzb decay to \CP-eigenstates,
they will have opposite \CP.  If any same-\CP decays occur, the
number produced will be a measure of the rate of charm mixing.

To connect the number of like-\CP and opposite-\CP events to the mixing
rate and strong phase~$\delta_i$ for a given decay mode $i$, expressions
for time-integrated rates for ST or DT events can be calculated from decay
amplitudes.  Observed rates for one or more modes can be investigated
simultaneously, with mixing parameters and strong phases obtained from a
simultaneous fit to all decay modes under consideration.

As an example, consider the DT decay to (\Km\pip, \Km\pip).  From a
$\CP = -1$ $\Dz\Dzb$ coherent state, this rate should be zero in the absence
of mixing. A short calculation\cite{Gronau:2001nr} yields
\begin{eqnarray}
\Gamma^{(\CP=-1)}(\Km\pip, \Km\pip) & = & \frac{1}{2} |A_{\Km\pip}|^4 
\left| 1 - r_{K\pi}^2 e^{-2i\delta_{K\pi}^{\prime}}\right|^2 (x^2 + y^2)\\
& \approx & \frac{1}{2} |A_{\Km\pip}|^4  (x^2 + y^2),
\end{eqnarray}
where $A_{K\pi} = \langle \Km\pip | \Dz\rangle$, 
$\overline{A}_{K\pi} = \langle \Km\pip | \Dzb\rangle$,
$r_{K\pi} \equiv  |\overline{A}_{K\pi}/A_{K\pi}|$,
and
$\delta_{K\pi}^{\prime}$ is the strong phase difference between
$\overline{A}_{K\pi}$ and $A_{K\pi}$:
\begin{equation}
\overline{A}_{K\pi}/A_{K\pi} \equiv  r_{K\pi} e^{-i\delta_{K\pi}^{\prime}} =
-r_{K\pi} e^{-i(\deltaKpi + \pi)},
\end{equation}
where we have incorporated the phase convention used by CLEO in the
definition of $\delta_{K\pi}^\prime$.
Note that if the mixing rate $\Rm \equiv (x^2 + y^2)/2$ vanishes,
then $\Gamma^{(\CP=-1)}(\Km\pip, \Km\pip)$ will vanish.  A non-zero rate
will be an indication of the presence of mixing.
This rate can be contrasted with the DT $(\Km\pip, \Kp\pim)$ decay rate
\begin{eqnarray}
\Gamma^{(\CP=-1)}(\Km\pip, \Kp\pim) & = &  |A_{\Km\pip}|^4 
\left| 1 - r_{K\pi}^2 e^{-2i\delta_{K\pi}^{\prime}}\right|^2 \left[ 1 - \frac{1}{2}(x^2 - y^2)\right]\\
& \approx & |A_{\Km\pip}|^4\left[ 
  1 - 2 r_{K\pi}^2 \cos 2\delta_{K\pi}^{\prime} - \frac{1}{2}(x^2 - y^2) \right]\,.
\end{eqnarray}
Comparison of these rates yields information on $\Rm \equiv (x^2+y^2)/2$.  Inclusion of other
DT decay mode pairs permits measurement of $x$, $y$, $\delta_i$, and 
\Dz branching fractions.  See Table~\ref{tab:coherentModes}. 

\begin{table}[ht]
\tbl{Correlated and uncorrelated decay rates for ST and DT events
used in analysis of coherent $\Dz\Dzb$ decays by CLEO-c.\protect\cite{Rosner:2008fq,Asner:2008ft}
Rates are normalized to the branching fraction(s) of reconstructed 
mode(s) (note that a normalization is used where $A_j^2$ is the 
\Dz branching fraction to mode $j$ when no mixing is present).
$S_+ (S_-)$ denotes a decay
to a $\CP=+1 (-1)$ eigenstate; $e^{-}$ denotes a semileptonic 
decay containing $e^{-}$. Rates are given to leading order in $x$, $y$, and $\Rws$, the WS-to-RS decay rate ratio.
Effects of \CP violation are negligible. Charge-conjugate modes are implied.}
{\label{tab:coherentModes}
\begin{tabular}{lll}
\toprule
ST mode & Uncorrelated rate & Correlated rate \\
\colrule
$\Km\pip $ & $ 1 + \Rws $ & $ 1 + \Rws $ \\
$S_{\pm} $ & $ 2 $        & $ 2 $        \\
\colrule
DT mode & Uncorrelated rate & Correlated rate \\
\colrule
$ \Km\pip, \Km\pip $ & $\Rws$       & $\Rm$ \\
$ \Km\pip, \Kp\pim $ & $1 + \Rws^2$ & $(1 + \Rws)^2 - 4r\cos\deltaKpi(r\cos\deltaKpi + y) $ \\
$ \Km\pip, S_{\pm} $ & $1 + \Rws$   & $ 1 + \Rws \pm 2r \cos\deltaKpi \pm y $ \\
$ \Km\pip, e^{-}   $ & $1$          & $1 - ry \cos\deltaKpi - rx \sin\deltaKpi $ \\
$ S_{\pm}, S_{\pm} $ & $1$          & $ 0 $ \\
$ S_{+}, S_{-}     $ & $2$          & $ 4 $ \\
$ S_{\pm}, e^{-}   $ & $1$          & $ 1 \pm y $ \\
\botrule
\end{tabular}}
\end{table} 

\par
\Dz final states used by CLEO-c\cite{Rosner:2008fq} are $\Kpm\pimp$, $\Kp\Km$, $\pip\pim$, $\KS\piz\piz$,
$\KL\piz$, $\KS\piz$, $\KS\eta$, $\KS\omega$, and inclusive semileptonic decays $X e^+\nu_e$, $X e^-\overline{\nu}_e$.
Events containing neutral $D$ candidates are selected using two quantities, the beam-constrained mass~$M$:
\begin{equation}
M \equiv \sqrt{ E_0^2 - {\vec p}_D^2/c^2}
\end{equation}
and the energy difference $\Delta E \equiv E_D - E_0$ where $E_0$ is the beam energy, $E_D$ is the sum of
energies of the \Dz candidate decay products, and ${\vec p}_D$ is the \Dz candidate momentum. Well-reconstructed
\Dz candidates will have  distributions that peak at the \Dz mass in $M$ and at zero in $\Delta E$.  After mode-dependent cuts on
$\Delta E$ are imposed, ST yields are obtained by fitting the $M$ distribution and DT yields by counting
events in a signal region in the two-dimensional $M$ distribution.

\par
Semileptonic decays are reconstructed inclusively, with only the electron required to be identified.  
Electron identification is performed by use of multivariate techniques.\cite{Coan:2005iu}  Decays
involving \KL mesons or neutrinos are reconstructed 
using a missing-mass technique only in DT events.\cite{He:2007aj}

Measurements of $x^2$, $y$, $r^2$, $rx\sin\deltaKpi$, and $r\cos\deltaKpi$ are obtained from the observed 
ST and DT yields and external branching fraction measurements using a least-squares fit.\cite{Sun:2005ip}
The DT yields provide information on mixing and strong phase parameters.  Use of ST and DT yields 
simultaneously provides normalization, so that independent measurements 
of the absolute $\DzdashDzb$ production rate and the integrated
luminosity are not required.
This method is described fully in Refs.~\refcite{Asner:2008ft,Asner:2005wf},
and~\refcite{Asner:2005wf-erratum},
including event selection and global fit techniques.
Quantum-correlated results are presented in Section~\ref{sec:correlated_decays_results}.

\subsection{Time-dependent Analyses of Two-body Decays}
\label{sec:Analysis_Techniques_Time-dependent_Two-body_Modes}

\subsubsection{$\Dz\to\Kp\pim$ Wrong-sign Analysis}
\label{sec:Analysis_Techniques_Hadronic_Two-body_Decay_Modes}
\begin{figure}[h!]
\begin{center}
  \includegraphics[width=12cm]{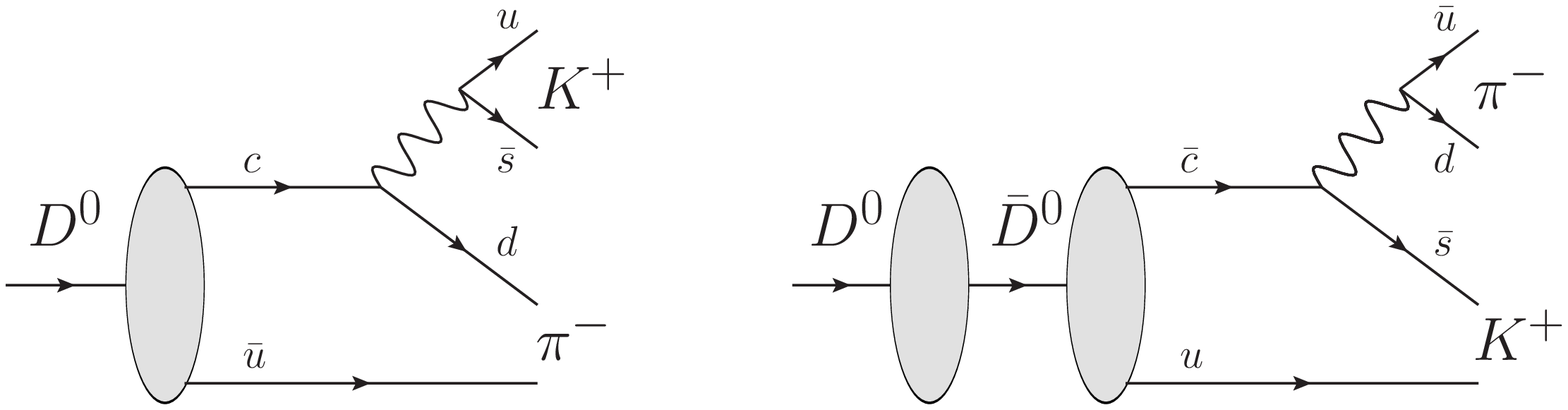}
\caption{Diagrams illustrating two ways to reach the $\Kp\pim$ final state 
from an initial \Dz. Left:
direct DCS decay, $\Dz\to\Kp\pim$. Right: mixing, $\Dz\to\Dzb$, followed by CF decay, 
$\Dzb\to\Kp\pim$.}
\label{fig:twopathKpi}
\end{center}
\end{figure}

In the wrong-sign (WS) \Dz decay, 
$\Dz\to\Kp\pim$, the $\Kp\pim$ 
final state may be reached either through a direct, doubly Cabibbo-suppressed (DCS) decay, or
through mixing, $\Dz\to\Dzb$, followed by the Cabibbo-favored (CF) right-sign (RS)  decay, $\Dzb\to\Kp\pim$. 
Since the two processes involve the same initial- and final states and are therefore 
indistinguishable, interference between the two amplitudes will occur.
For \Dz decays to $\Kp\pim$ and \Dzb decays to $\Km\pip$, we define the 
WS decay rates relative to the RS decay rates as follows:
\begin{equation}
\RwsP(t)\equiv\frac{\Gamma(\Dz(t)\to\Kp\pim)}{e^{-\Gamma t}|\Abkppim|^2},\hskip10pt
\RwsM(t)\equiv\frac{\Gamma(\Dzb(t)\to\Km\pip)}{e^{-\Gamma t}|\Akmpip|^2}.\label{eq:RwsPmDef}
\end{equation} 
From Eqs.~\ref{eq:Dztofsmall}, \ref{eq:Dzbtofbsmall}, \ref{eq:linvkppim} and \ref{eq:lkmpip} these are given by:
\begin{equation}
\RwsPm(t)=\RdcsPm+\yPrimePm\sqrt{\RdcsPm}
(\Gamma t)+\frac{\xPrimePmSq+\yPrimePmSq}{4}(\Gamma t)^2,\label{eq:RwsPm}
\end{equation}
where the mixing parameters \xPrimeP, \yPrimeP (\xPrimeM, \yPrimeM):
\begin{eqnarray}
\xPrimePm&\equiv&\sqrt[4]{\frac{1\pm\AM}{1\mp\AM}}\left[\xPrime\cos\phi\pm\yPrime\sin\phi\right],\\
\yPrimePm&\equiv&\sqrt[4]{\frac{1\pm\AM}{1\mp\AM}}\left[\yPrime\cos\phi\mp\xPrime\sin\phi\right].
\end{eqnarray}
are the mixing parameters
measured in the \Dz (\Dzb) decay modes, $\phi$ is the weak phase characterizing \CPV in the
interference between mixing and decay,
and the parameters \xPrime and \yPrime are related to the mixing parameters $x$ and $y$ through 
a rotation by the strong
phase, \deltaKpi:
\begin{eqnarray}
\xPrime&\equiv&x\cos\deltaKpi+y\sin\deltaKpi,\label{eq:xPrime}\\
\yPrime&\equiv&y\cos\deltaKpi-x\sin\deltaKpi.\label{eq:yPrime}
\end{eqnarray}
The DCS branching fraction for \Dz and \Dzb decays is related to the direct \CPV
asymmetry parameter \AD as follows:
\begin{eqnarray}
\RdcsPm&\equiv&\Rdcs\sqrt{\frac{1\pm\AD}{1\mp\AD}}.
\end{eqnarray}
In the limit of \CP conservation ($\AD=\AM=\phi=0$), Eq.~\ref{eq:RwsPm} reduces
to:
\begin{equation}
\Rws(t) = \Rdcs+\yPrime\sqrt{\Rdcs}(\Gamma t) +\frac{\xPrimeSq+\yPrimeSq}{4}(\Gamma t)^2.
\label{eq:Rws}
\end{equation}
The relative WS decay rate 
allows a determination of 
\xPrimeSq, \yPrime and \Rdcs, but not the strong phase \deltaKpi. 
For small mixing parameter values, the main sensitivity to
mixing comes through the interference term which is linear in $\yPrime\Gamma t$.
Tree diagrams for the two amplitudes mediating the $\Dz\to\Kp\pim$ decay are shown in 
Fig.~\ref{fig:twopathKpi}.
\par
\begin{figure}[h!]
\begin{center}
  \includegraphics[width=6cm]{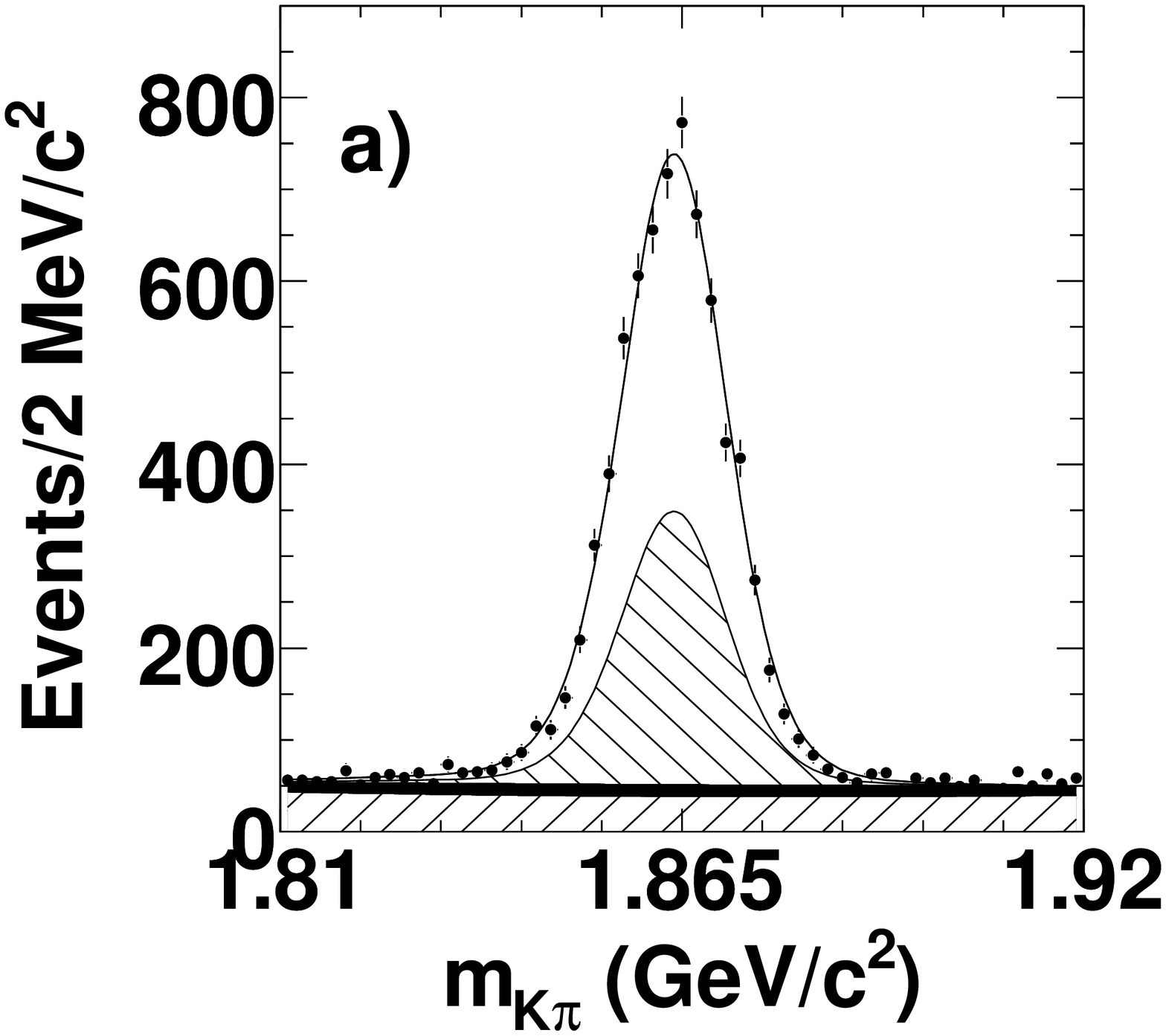}
  \includegraphics[width=6cm]{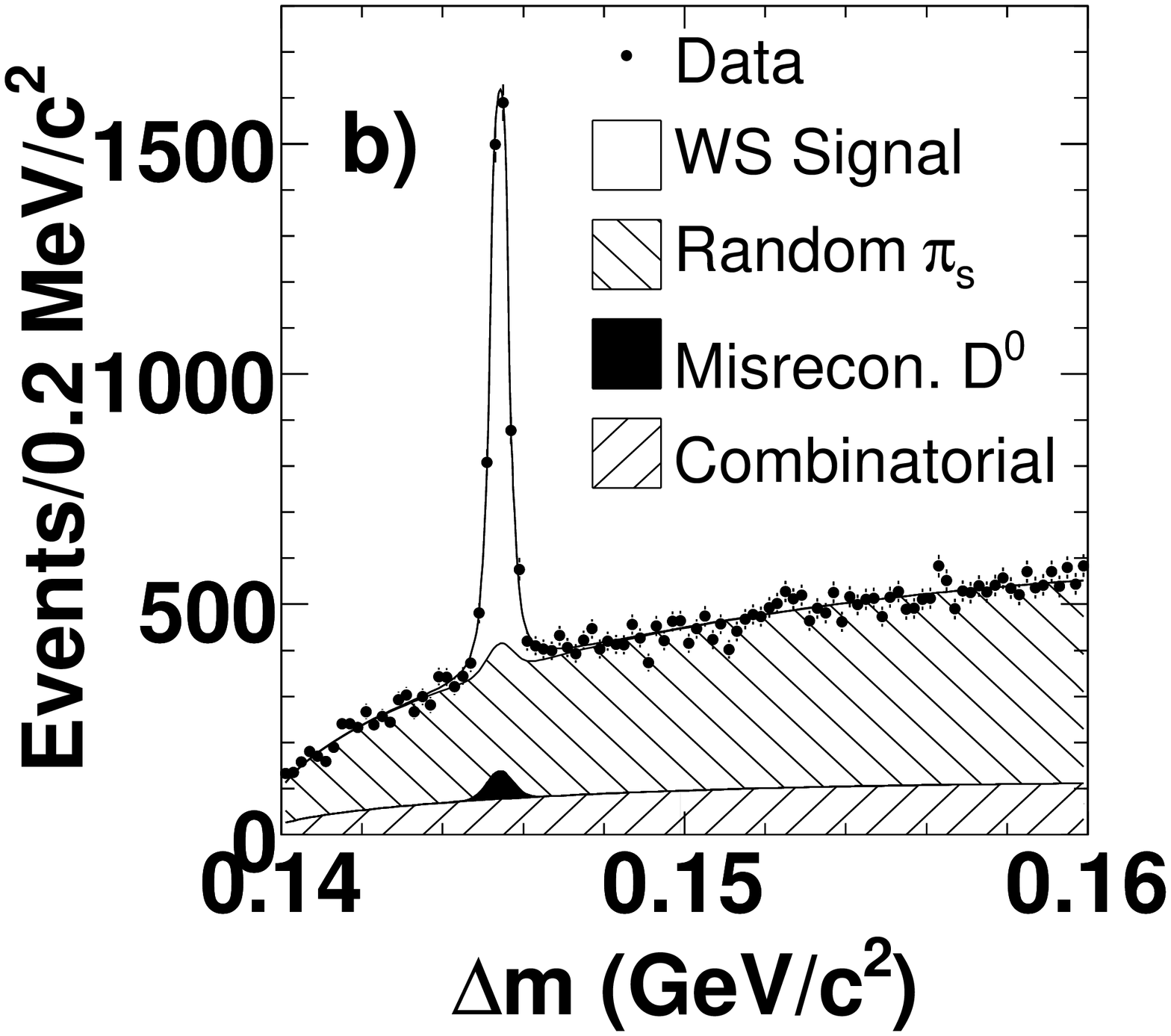}
\caption{\babar\ distributions of (a): \mKpi from WS candidates with $0.1445<\DeltaM<0.1465\gevcc$ and 
(b): \DeltaM for WS candidates with $1.843<\mKpi<1.883\gevcc$. The projections of the signal- and background
fits are overlaid, where the random \pislow background sample peaks in \mKpi but not in \DeltaM;
the misreconstructed \Dz 
paired with a \pislow from a \Dstar 
peaks in \DeltaM but not in \mKpi; the combinatoric background sample 
peaks neither in \mKpi nor in \DeltaM.  
Reprinted figure with permission from 
B. Aubert {\it et al.}, {\it Phys. Rev. Lett.} 98, 211802 (2007). 
Copyright 2007 by the American Physical Society.\protect\cite{Aubert:2007wf}}
\label{fig:KpiWSMdM}
\end{center}
\end{figure}

Experiments use the slow pion \pislowp in the strong decay $\Dstarp\to\pislowp\Dz$ 
to tag the charm flavor of the neutral $D$ at 
production.\footnote{Unless otherwise stated, reference to a given
decay mode implies reference to its \CP-conjugate mode as well.} 
The charge of the \pislow, together with the
charge of the kaon from the decay of the neutral $D$ allows the signal sample to be divided
into four categories: two WS decay samples,
$\Dstarp\to\pislowp\Dz,\thinspace\Dz\to\Kp\pim+c.c.$, and two much
larger right-sign (RS) decay control samples, $\Dstarp\to\pislowp\Dz,\thinspace\Dz\to\Km\pip+c.c.$. 
A simultaneous fit to the RS and WS distributions is performed to determine the 
direct CF and DCS lifetime
and the parameters of the decay-time resolution model (from the RS and WS samples)
and the parameters \Rdcs, \xPrimeSq, \yPrime (from the WS sample).
The independent variables of the fit are \mKpi,
the reconstructed $K\pi$ invariant mass; \DeltaM, the $\Dstarp$-$\Dz$ mass difference; 
\decayt, the reconstructed decay time, and its measured uncertainty, \terr. 
The variables \mKpi and \DeltaM are used to separate signal from background. 
At \babar, the vertical height of the beam spot is $\approx6\mum$. This beam spot information is 
used to constrain the location of the \Dstar vertex, thus substantially 
improving the determination of
\DeltaM and the reconstructed decay time, \decayt.
Fig.~\ref{fig:KpiWSMdM} shows the projections of the \mKpi and \DeltaM data and 
signal and background fit functions
from the 2007 
384 \invfb WS \babar\ data set.\cite{Aubert:2007wf}

\subsubsection{$D^0$ Lifetime Ratio Analysis}
\label{sec:Analysis_Techniques_Lifetime_Ratio}

In this section we describe analysis techniques which measure the decay-time distributions of neutral $D$ mesons decaying to \CP eigenstates and \CP mixed states.
The potential of this method was first described
in Ref.~\refcite{Liu:1994ea}, and the first experimental results
were presented utilizing these techniques in Ref.~\refcite{Aitala:1999dt}.
In the last ten years, several experimental collaborations 
have measured \DzdashDzb mixing and \CP violation observables
with increasing precision by comparing the rate for \Dz mesons 
decaying to flavor-specific final states. 

In particular, the mixing parameter $y$ (Eq.~\ref{eq:yformula}) may be measured  
by comparing the rate of \Dz decays to \CP eigenstates with decays to non-\CP eigenstates.
If decays to \CP eigenstates have a shorter effective lifetime than those decaying to non-\CP eigenstates,
then $y$ is positive.

\begin{figure}[h]
\begin{center}
\hbox to\hsize{%                                                                                                                                                                                                 
  \includegraphics[scale=0.4]{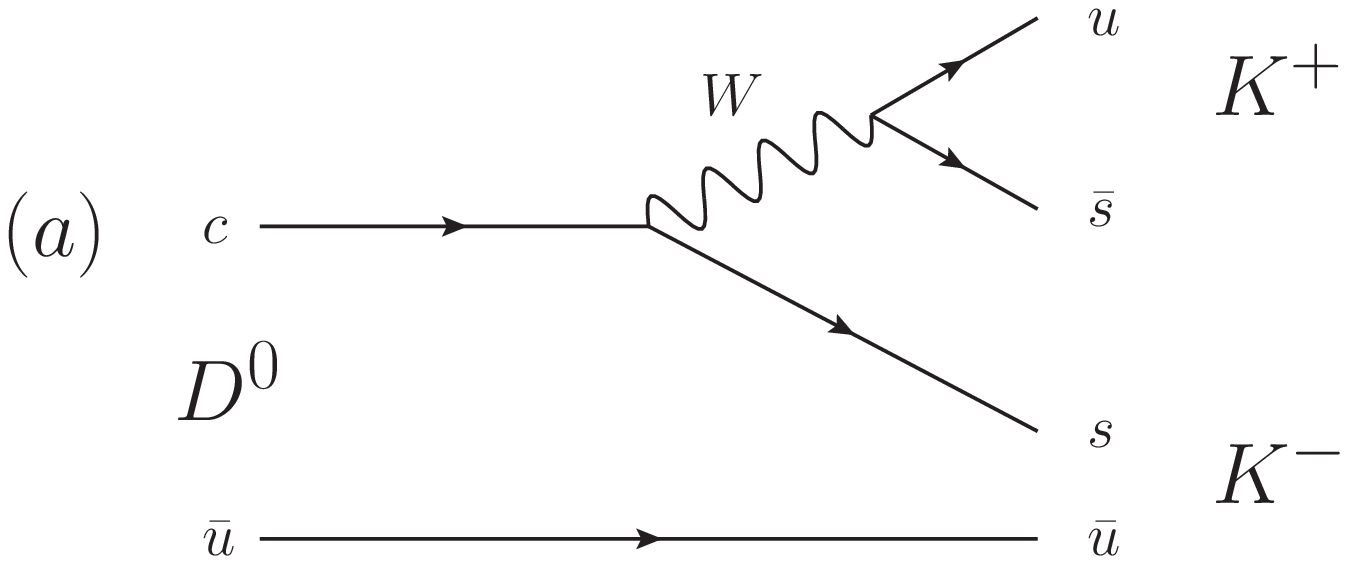}%

  \hfil
  \includegraphics[scale=0.4]{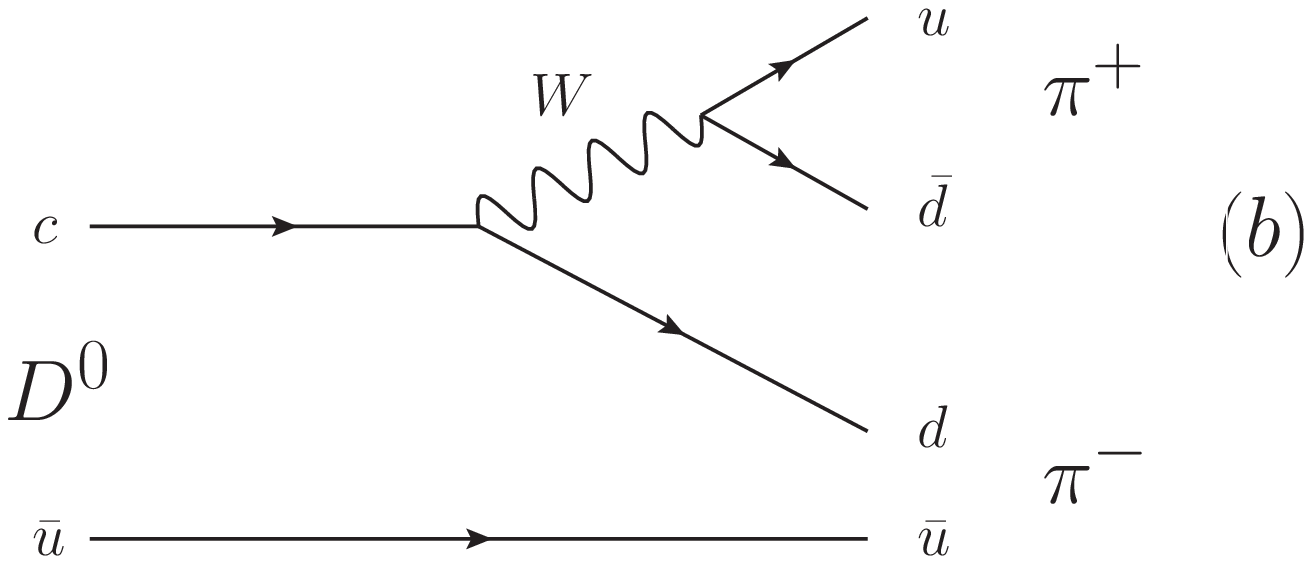}%  
}

\caption{Feynman diagrams for singly-Cabibbo-suppressed decays of \Dz to \CP eigenstates, $D^0~\to~K^+~K^-$~(a), and $D^0 \to \pi^+ \pi^-$~(b).}

\label{fig:DztoKKpipi}
\end{center}
\end{figure}

\begin{figure}[h]
\begin{center}
  \includegraphics[scale=0.4]{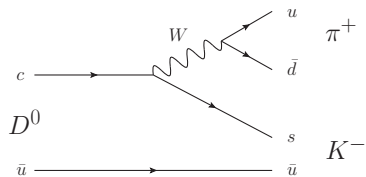}

\caption{Feynman diagram of the Cabibbo-favored decay $D^0 \to K^- \pi^+ $ .} 

\label{fig:DztoKmPip}
\end{center}
\end{figure}

Experimentally, one is interested in collecting high purity samples with large statistics of \Dz decays
to final states of specific \CP content.
The two-body SM processes with these characteristics are 
the singly-Cabibbo-suppressed decays to \CP-even eigenstates, 
$D^0\rightarrow K^+K^-$ and $D^0\rightarrow \pi^+\pi^-$, 
shown in Figs.~\ref{fig:DztoKKpipi}(a)~and~\ref{fig:DztoKKpipi}(b) respectively, 
and the Cabibbo-favored decay to \CP-mixed final state, $D^0\rightarrow K^-\pi^+$, 
shown in Fig.~\ref{fig:DztoKmPip}, and the corresponding \CP-conjugate decay processes.

Similarly to the two-body final states,
the mixing parameter $y$ can also be measured by analyzing the \CP-odd component of $\Dz\to \KS\Kp\Km$ decays,
by means  of comparing the mean decay times for different regions of the three-body phase space distribution of the final state.\footnote{Details of this measurement will be discussed in Sec.~\ref{sec:Analysis_Techniques_Multibody_Modes}.}

Neglecting the quadratic (mixing) terms in Eqs.~\ref{eq:Dztofcpsmall} and \ref{eq:Dzbtofcpsmall}, an approximation 
valid when $|\lfCP|\approx 1$ and $|x\Gamma t|, |y\Gamma t|\ll 1$, we obtain the
the following expressions for the time-dependent decay rates $\Gamma(D^0(t)\rightarrow h^+h^-)$ and 
$\Gamma(\Dzb (t)\rightarrow h^+h^-)$:
\begin{eqnarray}
\begin{array}{lcc}
\Gamma(D^0(t)\rightarrow h^+h^-)\ &=&\ e^{-\Gamma t}|A_{h^+h^-}|^2
\left\{1-[\Re(\lambda_{h^+h^-})y-\Im(\lambda_{h^+h^-})x]\Gamma t\right\},\\
\Gamma(\Dzb(t)\rightarrow h^+h^-)\ &=&\ e^{-\Gamma t}|\bar A_{h^+h^-}|^2
\left\{1-[\Re(\lambda^{-1}_{h^+h^-})y-\Im(\lambda^{-1}_{h^+h^-})x]
\Gamma t\right\},
\end{array}
\end{eqnarray}
\begin{eqnarray}
\begin{array}{lcc}
\Gamma(D^0(t)\rightarrow\Km\pip)\ &=&\ e^{-\Gamma t}|A_{K^-\pi^+}|^2,\\
\Gamma(\Dzb(t)\rightarrow\Kp\pim)\ &=&\ 
e^{-\Gamma t}|\bar A_{K^+\pi^-}|^2,
\end{array}
\end{eqnarray}
where $h=\pi,K$. In the absence of direct \CP violation (as expected in the SM), but allowing for a small indirect \CP violation (with a weak phase $|\phi|\ll 1$), we can write $\lambda_{h^+h^-}=\left|q/p \right|e^{i\phi}$. To a good approximation, these decay-time distributions can be treated as exponentials with effective lifetimes given by Ref.~\refcite{Bergmann:2000id}
\begin{eqnarray}
\begin{array}{lclcl}
\tau_{K\pi} &=& \tau(\Dz\rightarrow K^-\pi^+)\ &=& \tau(\Dzb\rightarrow K^+\pi^-),\\
\tau^+_{hh} &=& \tau(\Dz\rightarrow h^+h^-)\ &=& \tau_{K^-\pi^+}
\left[1+\left|\frac{q}{p}\right|(y\cos{\phi}-x\sin{\phi})\right]^{-1},\\
\tau^-_{hh} &=& \tau(\Dzb\rightarrow h^+h^-)\ &=& \tau_{K^-\pi^+}
\left[1+\left|\frac{p}{q}\right|(y\cos{\phi}+x\sin{\phi})\right]^{-1},
\end{array}
\end{eqnarray}
as before $h=\pi,K$. Combining these quantities we then define the parameters $y_{CP}$, $A_{\Gamma}$ and $\Delta Y$ as: 
\[\begin{array}{l}
y_{CP} = \frac{\displaystyle \tau_{K\pi}}{\displaystyle \langle \tau_{hh} \rangle}-1 , \\
A_{\Gamma} = \frac{\displaystyle
\tau(\Dzb\rightarrow h^- h^+) - \tau(\Dz\rightarrow h^+h^-)}
{\displaystyle
\tau(\Dzb\rightarrow h^- h^+) + \tau(\Dz\rightarrow h^+h^-)} , \\[10pt]
\Delta Y = \frac{\displaystyle \tau_{K\pi}}{
\displaystyle\langle\tau_{hh}\rangle}A_{\Gamma} ,
\end{array}
\]
where $\langle\ldots\rangle$ implies average over flavors,  $\langle {\displaystyle \tau_{hh}} \rangle = ({\displaystyle \tau^+_{hh}} + {\displaystyle \tau^-_{hh} } ) / 2 $ and $h=\pi,K$.
In the limit of \CP conservation, $y_{CP} = y$ and $\Delta Y = 0$. 
In the absence of \DzdashDzb mixing, both $y_{CP}$ and $\Delta Y$ are zero.

Measurements of $y_{CP}$ have been conducted at $e^+e^-$  colliders (\babar, Belle and CLEO) as well as fixed-target experiments (FOCUS and E791).  
Historically, experiments at $e^+e^-$  colliders have relied on the kinematic separation of
charm decays at high center-of-mass momentum (from $e^+e^- \to c\bar{c}$) to reduce backgrounds. 
In addition, excellent particle identification and tracking capabilities for hadrons over a large range of momenta are required
when measuring $y_{CP}$. 

\begin{figure}[!ht]
\hbox to \hsize{
 \includegraphics[width=0.5\linewidth]{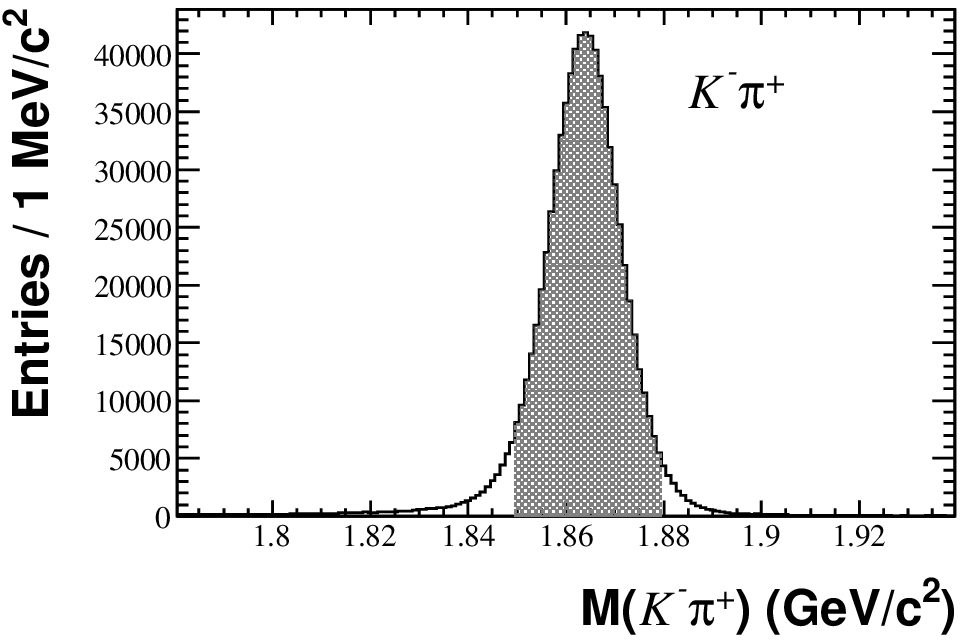}\hfill
 \includegraphics[width=0.5\linewidth]{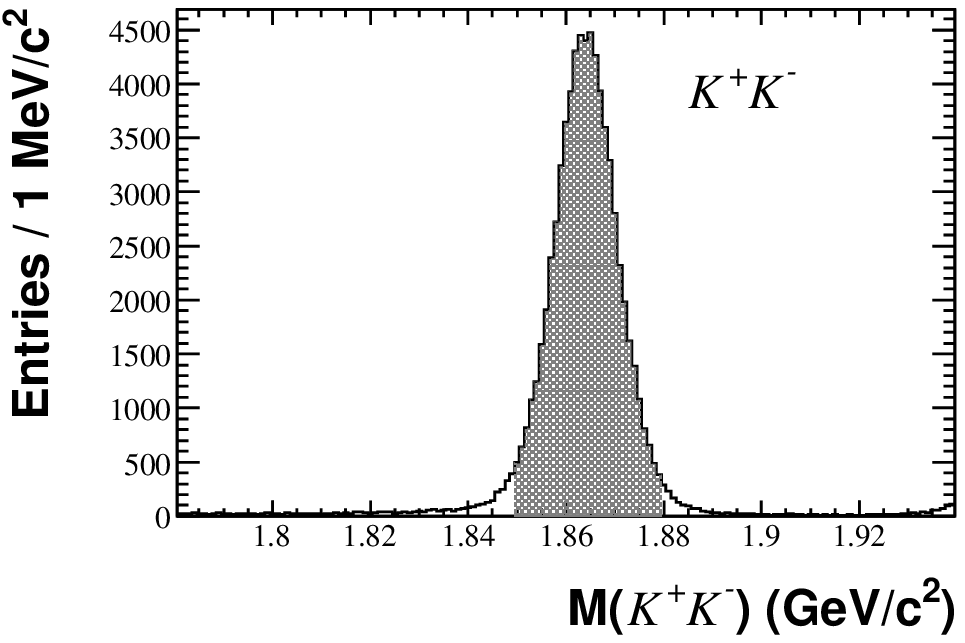}
}
\hbox to \hsize{
 \includegraphics[width=0.5\linewidth]{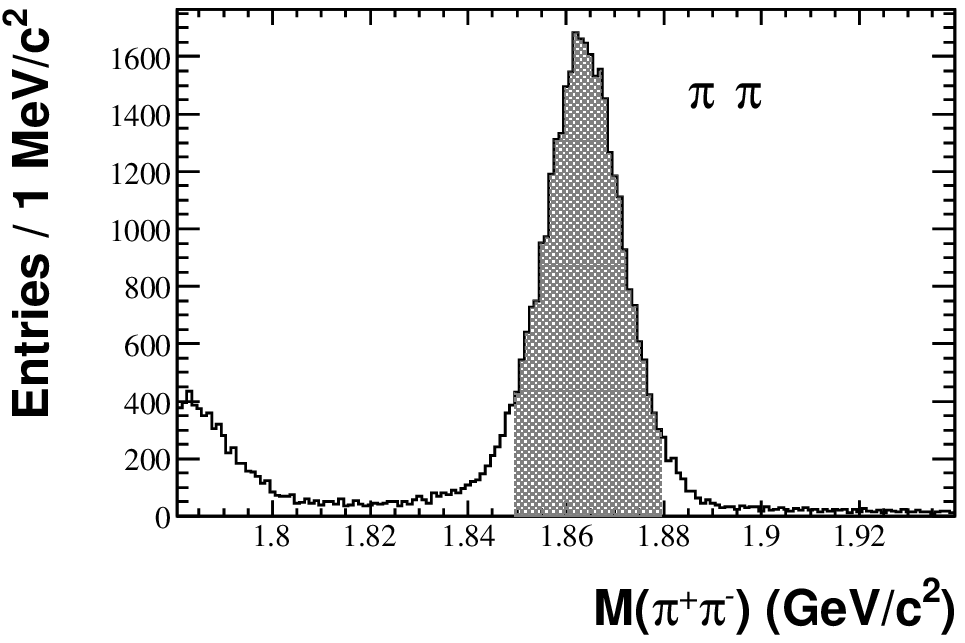}\hfill
\raise2.25cm\hbox{\begin{tabular}{lcr}\hline\hline
Sample &  Size & Purity (\%) \\
\hline
$K^-\pi^+$         & 730,880 & 99.9  \\
$K^-K^+$           &  69,696 & 99.6  \\
$\pi^-\pi^+$       &  30,679 & 98.0  \\
\hline\hline
\end{tabular}}
}
\caption{\label{fig:BabarTwoBodyMassPlotTagged}The \babar\ reconstructed $D^0$ mass distributions
for the three $D^0$ tagged samples, within $\pm$0.8~\mevcc of the peak of $\Delta
m$. The shaded region indicates the events used
in the lifetime fit. (The structures appearing above 1.92 \gevcc in the \KmKp decay mode, and below 1.81 \gevcc in the
\pipi decay mode, are mainly due to candidates
with misidentified kaons or pions.) Also shown are the yield and purity of the three $D^0$ samples
as calculated inside the $\pm 15\mevcc$ mass window used in the lifetime measurements. 
Reprinted figure with permission from 
B. Aubert {\it et al.}, {\it Phys. Rev.} D~78, 011105(R) (2008). 
Copyright 2008 by the American Physical Society.\protect\cite{Aubert:2007en}}
\end{figure}

The \babar\ and Belle experiments have both produced measurements of $y_{CP}$ and $A_{\Gamma}$\cite{Staric:2007dt,Aubert:2007en,Aubert:2009ck} 
by means of selecting highly pure samples with high statistics of \Dz candidates decaying to $K^+K^-,\pip\pim$ and $K^-\pip$ final states, as shown in Fig.~\ref{fig:BabarTwoBodyMassPlotTagged}.
In these experiments, $D$ mesons are produced from $c\bar{c}$ initial states and as secondaries from $B$ decays,
those produced from $c\bar{c}$ events are used in the lifetime ratio measurements by choosing
high momentum $D$ mesons as well as minimizing other backgrounds. 
However, slightly different strategies were followed by different experiments, but 
the description of what follows is in general correct for all $y_{CP}$ measurements.
We focus primarily on the measurements done at the $B$-factories as those are the most precise.

The so-called tagged technique is implemented by reconstructing $D^{*\pm}\to \Dz \slowpi^{\pm}$ decays, 
where the slow $\slowpi^{\pm}$~determines the flavor of the decaying neutral meson as \Dz or \Dzb at creation.
The \Dz candidates are selected by kinematically combining pairs of oppositely-charged $K^{\pm}$ and $\pi^{\pm}$ tracks that have a common vertex. 
and have an invariant mass typically in the range between 1.8~\gev and 1.92~\gev (approximately $\pm 60~\mev$ around the nominal \Dz mass\cite{Nakamura:2010zzi}).
The kinematic fit provides the \Dz decay position and its momentum vector $p_{\Dz}$, 
which is required to point back to the $e^+e^-$ interaction region. 
The \Dz candidate and the slow $\slowpi^{\pm}$~are also required to form a common vertex in the interaction region.
For each \Dz candidate the proper decay time $t$ and its error $\terr$ are  calculated using the decay length $l=\beta \gamma ct = ctp_{\Dz}/m_{\Dz}$.

To further suppress background events, Belle exploits the distribution of the energy released in the $D^{*\pm}$ decay given by $Q=m_{D*} - m_{\Dz} - m_{\pi} $; 
equivalently, \babar\ uses the distribution of the mass difference $\Delta m$ of the reconstructed  $D^{*\pm}$ and the \Dz candidates in the event.
Typically \Dz candidates are required to be within  $\pm0.1~\mev$ of the peak of the $\Delta m$ or $Q$ distributions.

The \babar\ \Dz invariant mass distribution of tagged events of different decay channels are shown in Fig.~\ref{fig:BabarTwoBodyMassPlotTagged}.\cite{Aubert:2007en} 
The shaded area shows the sample of events used in the lifetime measurements.
These events were selected after particle identification, tracking, and vertex probability requirements.

In the tagged sample, the charge of the reconstructed $D^{*\pm}$ allows the determination of the lifetime separately for \Dz or \Dzb decays.
The lifetimes are determined by performing a simultaneous maximum likelihood fit to the reconstructed decay time and its error to all five decay samples 
($\Dz\to K^-\pi^+$ and $\Dzb\to K^+\pi^-$ decay samples are combined into one).
In general, there are  three main PDF components entering the lifetime fit: signal, combinatoric background, and misreconstructed charm decays.

In the so-called untagged technique, there is no reconstructed $D^{*\pm}$, hence it is not possible to identify the initial flavor of the decaying \Dz meson.
In the untagged analysis, the \Dz may be produced directly from a $\ccbar$ state or as a decay
product of a higher mass resonance (other than \Dstarpm).
The \Dz momentum is required to point back to the beam spot in order to reduce backgrounds. 
In order to exclude \Dz mesons coming from \B decays,  \Dz candidates with momentum in the \epem center-of-mass
(CM) frame less than 2.5 GeV/c.

In the untagged \babar\ analysis,\cite{Aubert:2009ck} all events appearing in the tagged data sample are removed from the untagged sample
in order to treat the tagged and untagged results as statistically independent from one another.
The signal yields in the untagged data samples are about 3.5 times larger than those in the respective tagged samples; 
however, their purity is lower and the systematic uncertainties due to the higher backgrounds are more challenging.

The signal PDF is generally described by an exponential convolved with a resolution function, 
which is composed of three Gaussian functions sharing some common parameters between them.
The high statistics of the Cabibbo-favored $\Dz\to K^-\pi^+$ sample drives the determination of the resolution function parameters in the lifetime fit.  
In the tagged analysis, the random combinatoric background in the signal region is determined from a sideband region in \Dz invariant mass (\mDz) and $\DeltaM$. In the untagged analysis, two sideband regions in \mDz are defined, one above the \Dz
mass peak and one below.
A small background component corresponding to misreconstructed charm decays that have long lifetimes and can thus mimic 
the decay time of signal events is included.
The proper time distribution for this background is taken from 
Monte Carlo (MC).
\par
While systematic uncertainties are expected to cancel in the lifetime ratio, 
the sources of backgrounds are different for each final state, 
hence systematics from background sources are not necessarily expected to cancel. 
The main signal model systematic uncertainties are the selection of the \Dz invariant mass signal region window (central position and size), 
opening angle distributions, and variations of the signal resolution model.
The systematic uncertainties associated with backgrounds are the combinatorial PDF model
and its normalization, and the misreconstructed charm PDF model (taken from simulation) and its normalization.

Results from these measurements are discussed in section \ref{sec:KK_pipi_results} below. 

% Hadronic Multi-body Decay Modes
%

\subsection{Time-dependent Analyses of Hadronic Multi-body Decay Modes}
\label{sec:Analysis_Techniques_Multibody_Modes}

Amplitude analyses of 
multi-body \Dz decay modes provide what are potentially the
most definitive measurements of charm mixing parameters.
Advantages include the ability, for some decay modes,
to measure mixing without the ambiguity of an unknown strong phase or insensitivity to the sign of $\xPrime$ that 
limits the measurement to \xPrimeSq and \yPrime rather than $x$ and $y$,
as is the case with the time-dependent analysis of $\Dz\to\Kp\pim$ decays.
Multi-body decays useful in this regard include $\Dz\to\KS\Kp\Km$ or 
$\KS\pip\pim$, which we will generically designate as $\KS\hp\hm$ where
$h$ represents $K$ or $\pi$.  Three-body decays also include $\Dz\to\Kp\pim\piz$.
Four-body decays include $\Dz\to\Kp\pim\pip\pim$.
Three-body decays are amenable to ``Dalitz-plot analysis,'' 
while higher-order decays require other methods.  

\subsubsection{$\Dz\to\KS\hp\hm$ Analysis}
\label{sec:Analysis_Techniques_Hadronic_Multi-body_Kshh}

\babar, Belle, and CLEO have performed studies of \Dz-\Dzb mixing using
the decays $\Dz\to\KS\pip\pim$, $\Dz\to\KS\Kp\Km$, or 
both.\cite{delAmoSanchez:2010xz,Asner:2005sz,Abe:2007rd}  The idea is
to fit the Dalitz-plot distribution of selected \Dz decays using the
time-dependent formalism given in Eq.~\ref{eq:Dztof} (for \Dz) and 
Eq.~\ref{eq:Dzbtofb} (for \Dzb).  The variation of the decay amplitudes \Af, \Afb and their
conjugates across the Dalitz plot must be taken 
into account.  We define $A(s_+,s_-)$ to be the amplitude for $\Dz\to\KS\hp\hm$
and $\overline{A}(s_+,s_-)$ to be the amplitude for  $\Dzb\to\KS\hp\hm$,
where $s_+$ and $s_-$ are the coordinates of a given position in the 
Dalitz plot, e.g., $s_+$, $s_- \equiv m^2_{\KS\pim}, m^2_{\pim\pip}$ (CLEO)
or $s_+$, $s_- \equiv m^2_{\KS\pip}, m^2_{\KS\pim} $ (Belle, \babar). In order
to fit the Dalitz-plot distribution as a function of time, it is necessary
to assume a Dalitz fit model.
These models typically 
include a coherent sum of ten to twelve 
quasi-two-body intermediate
resonances plus a non-resonant component.  $P$- and $D$-wave amplitudes
are modeled by Breit-Wigner or Gounaris-Sakurai functional
forms, including Blatt-Weisskopf centrifugal barrier factors.
In the \babar\ analysis, 
the $S$-wave dynamics are modeled using a $K$-matrix formalism ($\pi\pi$),
a Breit-Wigner plus non-resonant contribution ($K\pi$), and a
coupled-channel Breit-Wigner model describing the $a_0(980)$ and Breit-Wigner
models for the $f_0(1370)$ and $a_0(1450)$ ($KK$).

As far as mixing is concerned, the interesting Dalitz plot regions are
where the CF and DCS amplitudes interfere and regions where \CP eigenstates
predominate.  

\begin{figure}[ht]
\begin{center}
\hbox to\hsize{%
  \includegraphics[scale=0.30]{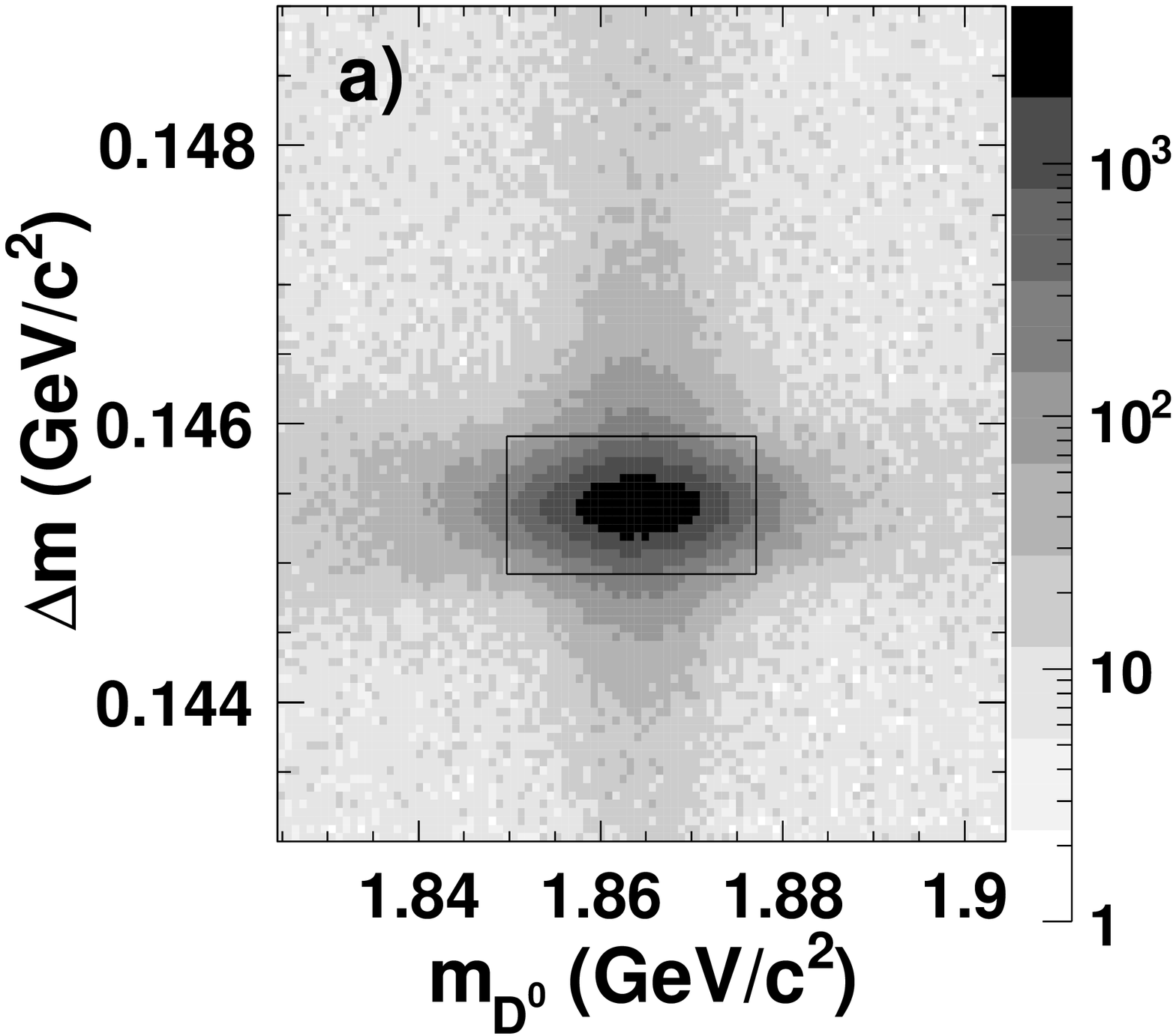}
  \hfil
  \includegraphics[scale=0.30]{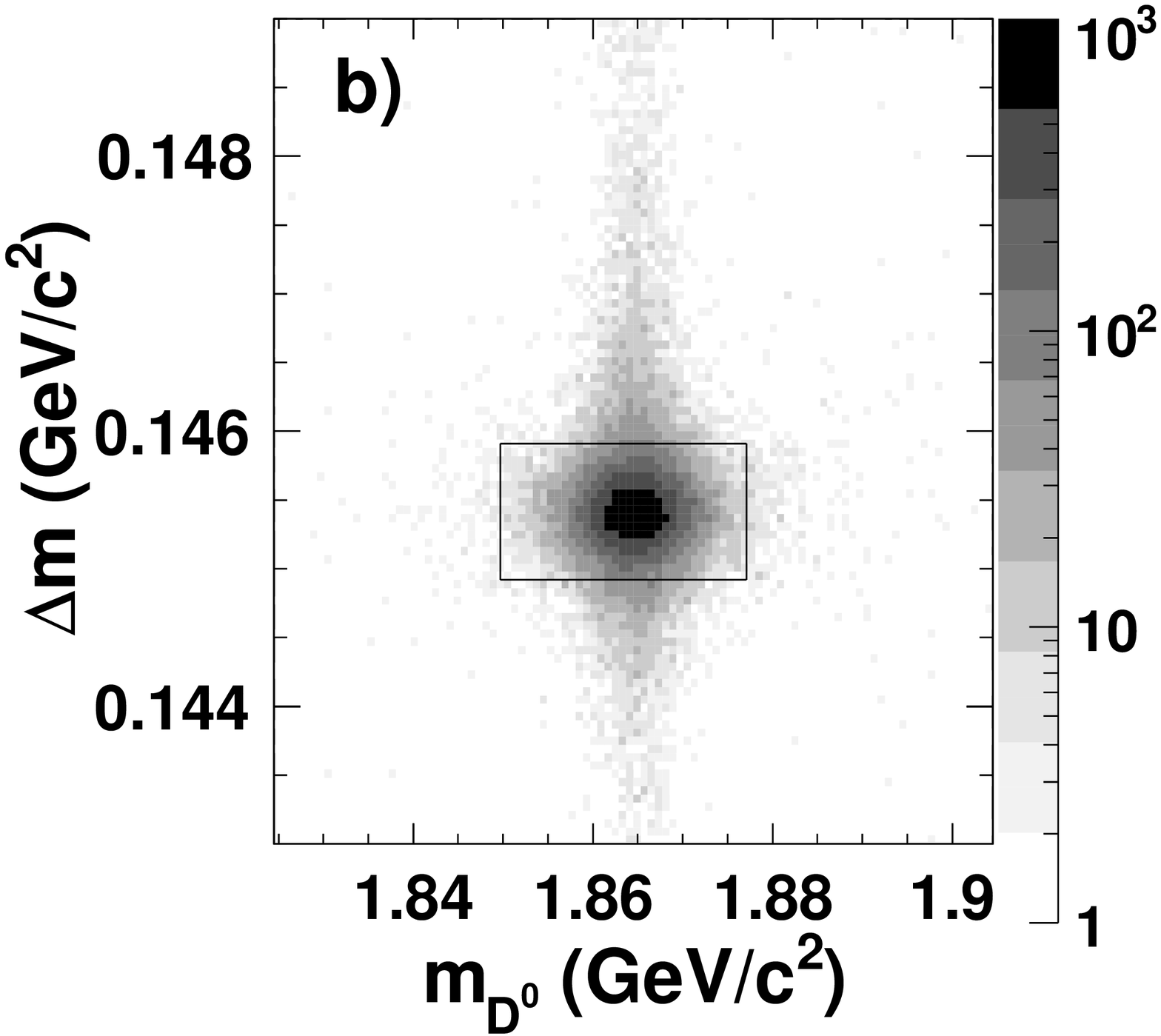}  
}
\caption{\babar\ distributions of $m_{\Dz}$ and $\DeltaM$ for the $\KS h^+h^-$
analysis, after all selection cuts. The shaded scale indicates the number 
density of events.  The boxes indicate the signal regions as used in the
analysis for the mixing fits. (a)~$\KS\pi\pi$.  (b)~$\KS K K $. 
Reprinted figures with permission from 
P. del Amo Sanchez {\it et al.}, {\it Phys. Rev. Lett.} 105, 081803 (2010). 
Copyright 2010 by the American Physical 
Society.\protect\cite{delAmoSanchez:2010xz}}
\label{fig:BaBar_Kspipi_mD_Deltam}
\end{center}
\end{figure}

\begin{figure}[ht]
\begin{center}
\hbox to\hsize{%
  \includegraphics[scale=0.35]{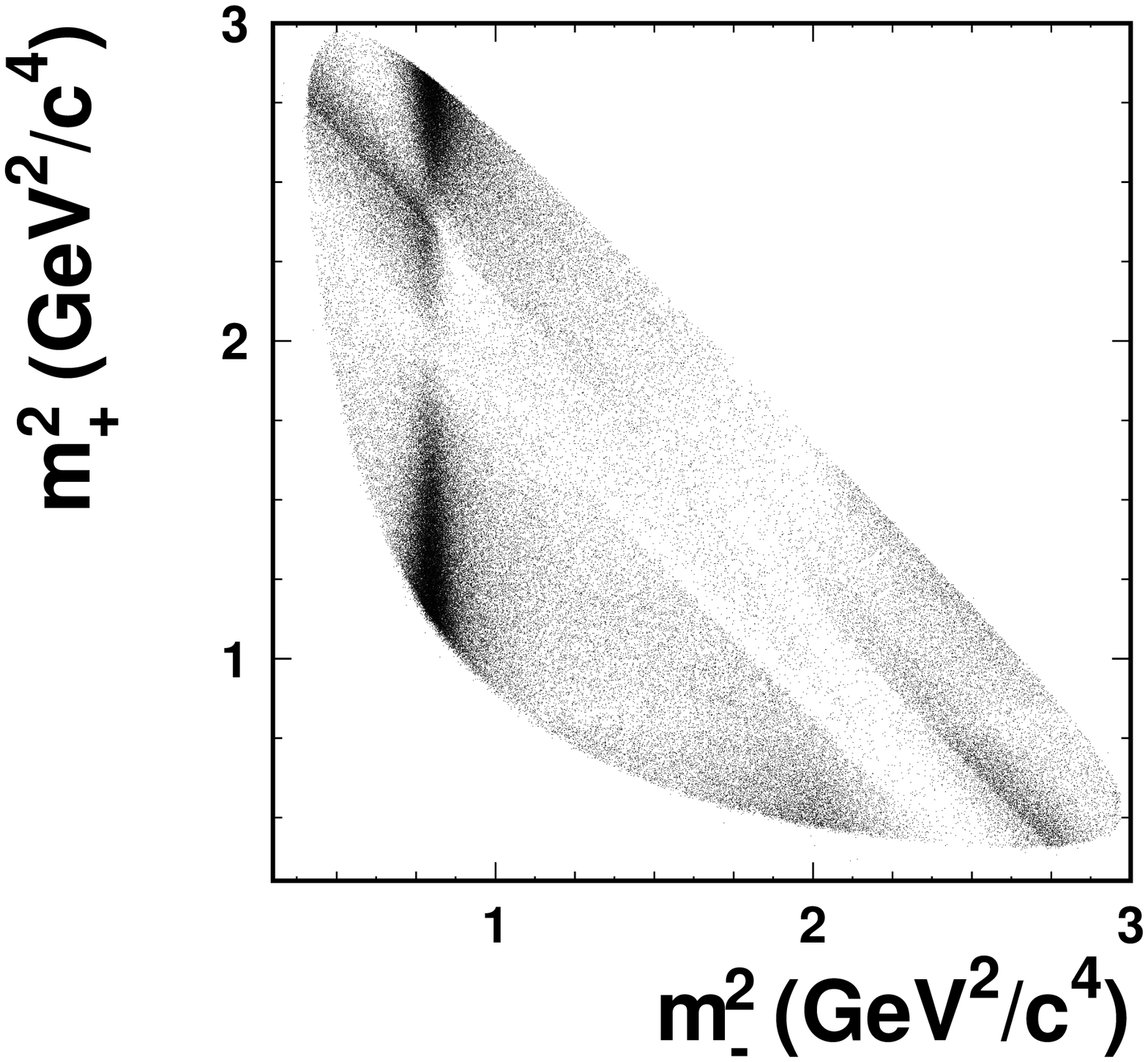}
  \hfil
  \includegraphics[scale=0.35]{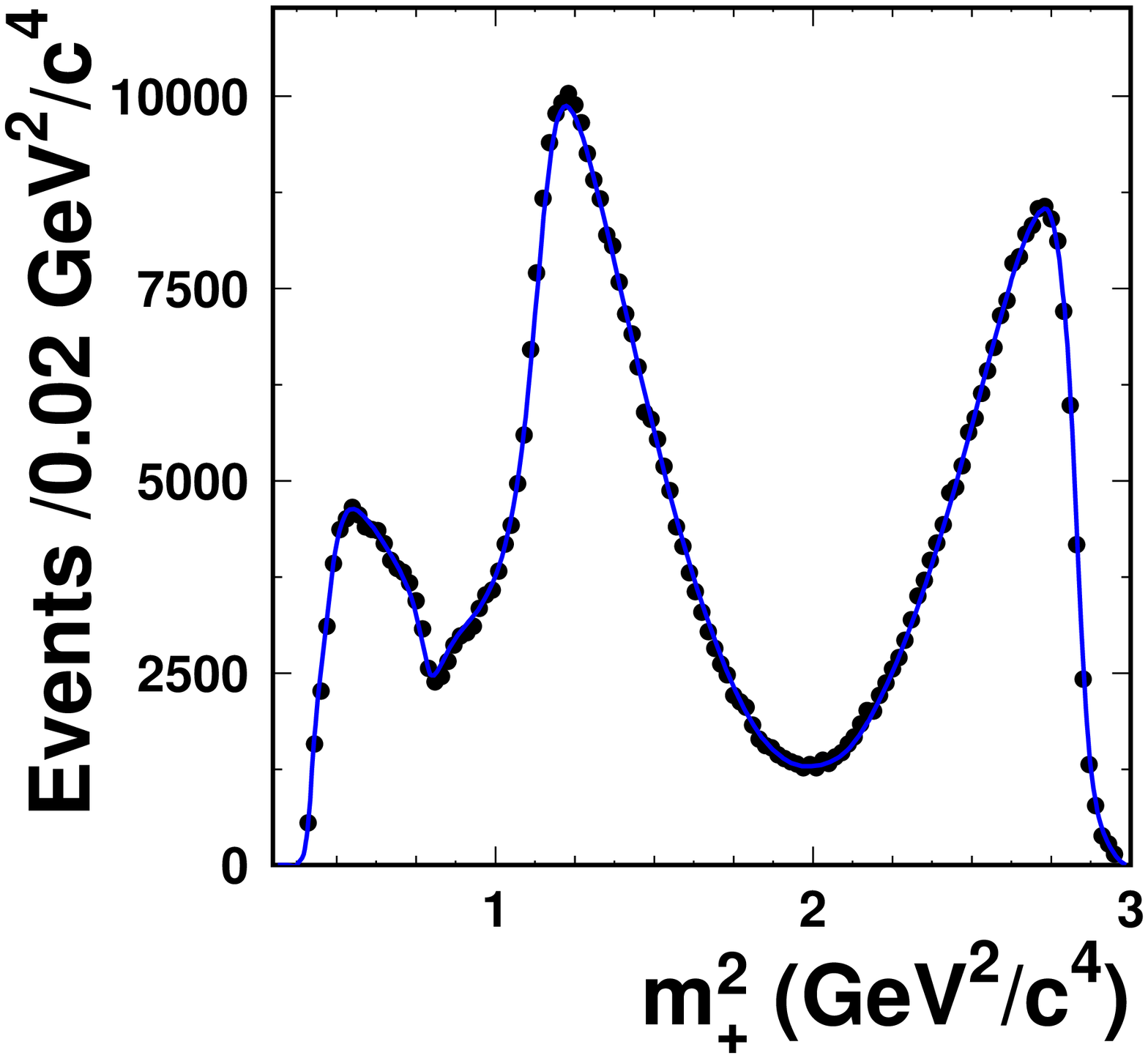}  
}
\hbox to\hsize{%
  \includegraphics[scale=0.35]{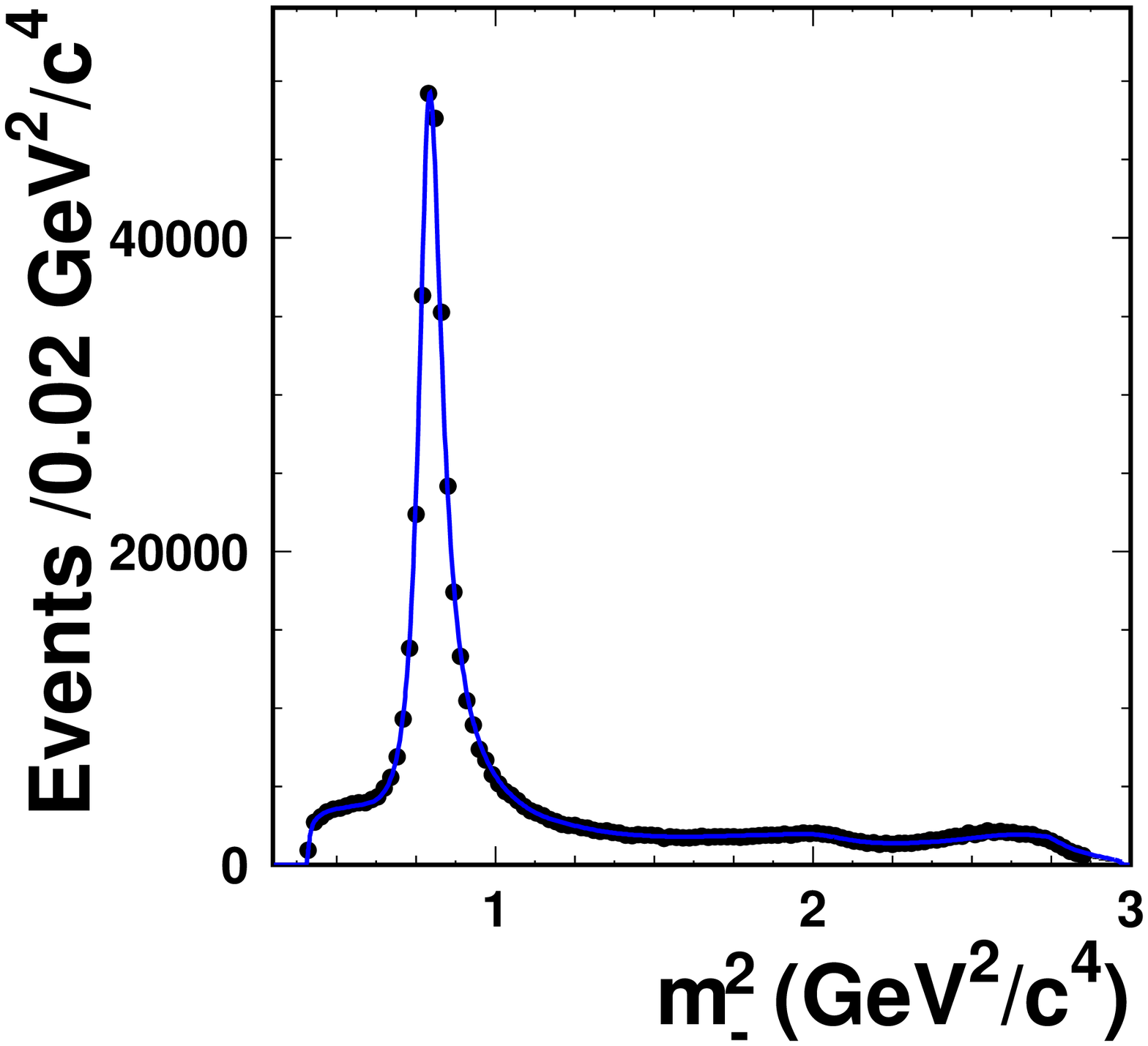}
  \hfil
  \includegraphics[scale=0.35]{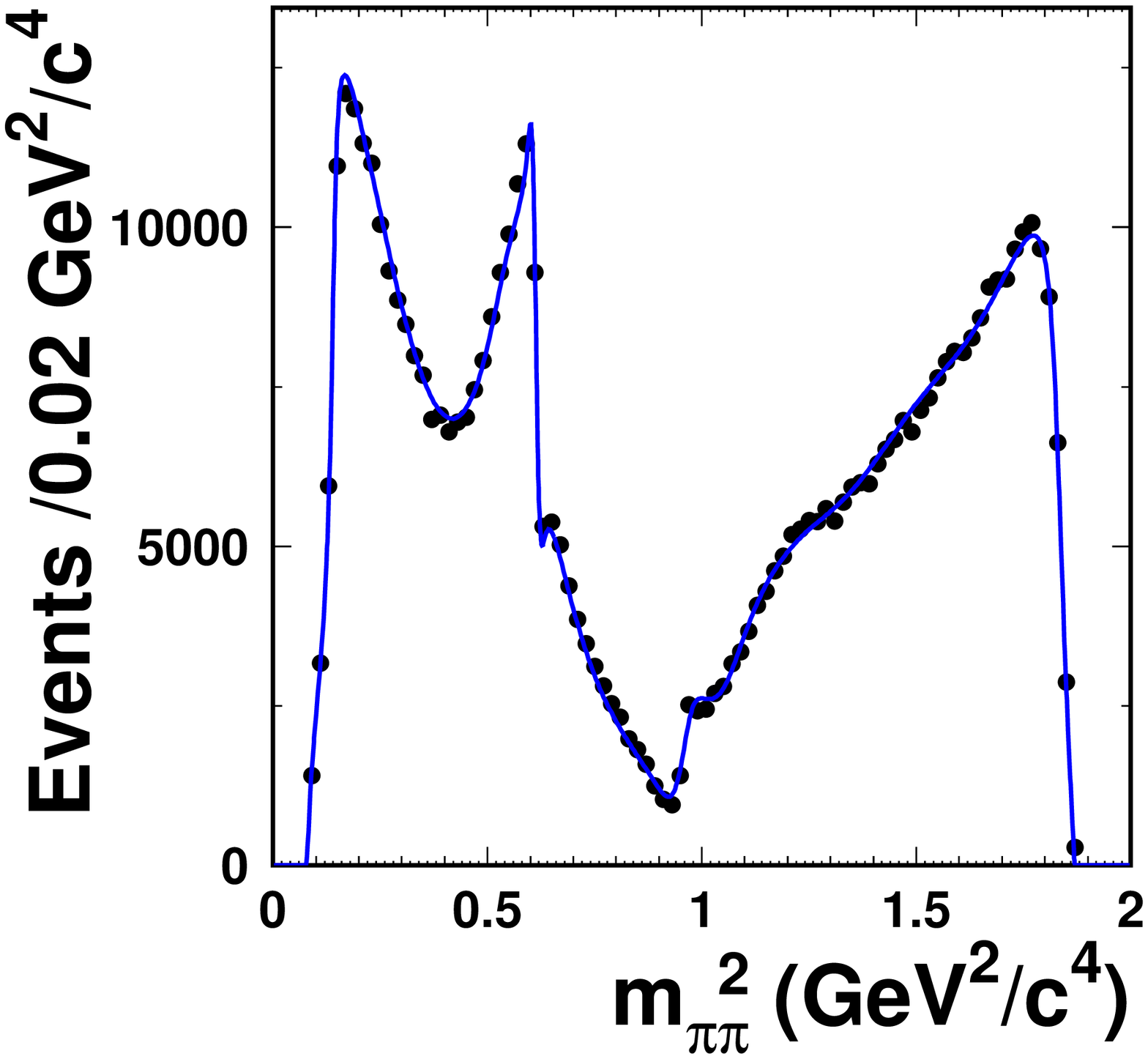}  
} 
\caption{Dalitz plot and fit projections for the Belle $\Dz\to\KS\pip\pim$
analysis. In the projections, the data are shown as points and the fit as
a solid line.  $m^2_{\pm} = m^2_{\KS\pipm}$
for \Dz decays, and  $m^2_{\pm} = m^2_{\KS\pimp}$ for \Dzb decays.
See Ref.~\protect\refcite{Abe:2007rd} for details of the 18 quasi-two-body
resonance plus non-resonant background Dalitz model and resulting fit.
Reprinted figure with permission from 
L. M. Zhang {\it et al.}, {\it Phys. Rev. Lett.} 99, 131803 (2007). 
Copyright 2007 by the American Physical Society.}
\label{fig:Belle_Kspipi_distributions}
\end{center}
\end{figure}

These analyses proceed in a manner similar to the two-body, time-dependent
analyses: they make use of the sign of the slow pion~$\pi_s$ from a 
\Dstarp decay to tag the neutral meson as \Dz or \Dzb at its creation.
After selecting appropriate-quality charged tracks, $\pip\pim$ 
pairs that have an invariant mass close to the \KS mass
(typically within 10 MeV) are selected, forming a $\KS$ candidate.  
Another set of
oppositely-charged tracks that share a common vertex are combined with
the $\KS$ candidate to form a $\Dz$ candidate.  This allows for the
$\KS$ decay vertex to be displaced from the $\Dz$ decay vertex.
A kinematic fit then provides
the \Dz decay position and its momentum vector~$p_{\Dz}$, which is required to point
back the the luminous interaction region.  The decay time~$t$ is calculated
using the $\Dz$ decay length~$l$ $ = \beta\gamma c t = c t p_{\Dz}/m_{\Dz}$, 
along with its error~$\terr$, on an 
event-by-event basis.  

Background sources include the random $\pi_s$ background, where an incorrect 
assignment between a low-momentum
pion and a good \Dz decay has been made, misreconstructed $\Dz$, and combinatoric background.
A few other sources of backgrounds (classification varies from experiment to
experiment) may also be included in the fit model as well to 
model specific non-signal decay modes.

The time-dependent analysis uses candidates from 
a two-dimensional signal region of $\m_{\Dz}$,
the reconstructed \Dz candidate mass, and
either $\DeltaM$ (\babar),
or $Q = m_{\KS\pi\pi\pi_s} - m_{\KS\pi\pi} - m_{\pi}$,
the available kinetic energy released in the \Dstarp decay (Belle). 
See Fig.~\ref{fig:BaBar_Kspipi_mD_Deltam}.
\babar\ (Belle) determines the yields of signal and background
in the signal box by fitting the $m_{\Dz}$ and $\DeltaM$ ($Q$)
to PDFs characterizing each background source over the full range in 
$\mDz$ and $\DeltaM$ (\babar) or $Q$ (Belle), and rescaling the 
component yields to the signal region.
Belle finds 534,410 signal candidates in 540~\invfb.\cite{Abe:2007rd}
\babar\ finds a signal yield of $540,800$ ($79,900$) $\KS\pi\pi$
($\KS\Kp\Km$) with purity 98.5\% (99.2\%) $\KS\pi\pi$ ($\KS\Kp\Km$)
in 468.5~\invfb of data.\cite{delAmoSanchez:2010xz}  
Fig.~\ref{fig:Belle_Kspipi_distributions} shows the 
Belle experiment's time-integrated distribution of \Dz decays 
and projections of the fit to the data where 
$m^2_{\pm} = m^2_{\KS\pipm}$
for \Dz decays and  $m^2_{\pm} = m^2_{\KS\pimp}$ for \Dzb
decays~\cite{Abe:2007rd}.

To determine the mixing parameters,
PDFs are defined that include the dependence of the signal
and background components on decay time $t$, $\terr$,
and location in the Dalitz plot.
Included in the signal PDF are the matrix elements
\begin{eqnarray}
{\cal M}(s_-, s_+, t) & = & 
   {\cal A}(s_-, s_+) g_+(t) + \qoverp {\cal \bar A}(s_-, s_+) g_-(t),\label{eq:KShh_matrix_element1}\\
{\cal \bar M}(s_-, s_+, t) & = & 
   {\cal \bar A}(s_-, s_+) g_+(t) + \poverq {\cal A}(s_-, s_+) g_-(t),\label{eq:KShh_matrix_element2}
\end{eqnarray}
which are convolved with a decay-time resolution function that depends
on position in the Dalitz plot.  Eqs.~\ref{eq:KShh_matrix_element1}
and \ref{eq:KShh_matrix_element2} are generalizations 
of Eqs.~\ref{eq:AmpEvol} and \ref{eq:AmpEvolb} to multi-body decays.
Different resolution functions are used for $\KS\pi\pi$
and $\KS K K$ distributions.  The decay-time resolution function is
a sum of Gaussians with widths that may scale with the event-by-event,
decay-time 
error~$\delta t$, and also depends weakly on position in the Dalitz plot.
Belle uses three Gaussians with different scale factors
and a common mean, which 
are allowed to vary in the fit.  \babar\ uses two Gaussians that scale with
$\delta t$, one of which is allowed to have a non-zero mean ($t_0$ offset),
and a third Gaussian which does not scale with $\delta t$.
The 
results of this procedure are discussed in section~\ref{sec:KS_h+h-_results}.
\par
\subsubsection{$\Dz\to\Kp\pim\piz$ Analysis}
\label{sec:Analysis_Techniques_Hadronic_Multi-body_Kpipiz}

As in the case of the two-body WS decay $\Dz\to\Kp\pim$, the three-body
WS decay $\Dz\to\Kp\pim\piz$ can occur through DCS decay or via $\Dz\to\Dzb$ mixing followed by the CF
decay $\Dzb\to\Kp\pim\piz$.  With a WS branching fraction of $( 3.04 \pm 0.17 ) \times 10^{-4}$ compared
with $(1.47 \pm 0.07 ) \times 10^{-4}$ for $\Dz\to\Km\pip$,\cite{Nakamura:2010zzi} the $K\pi\piz$ channel is competitive
in sensitivity to the two-body channel, despite the lower efficiency of
reconstructing the three-body final state.

\begin{figure}[ht]
\begin{center}
\hbox to\hsize{%
  \includegraphics[scale=0.3]{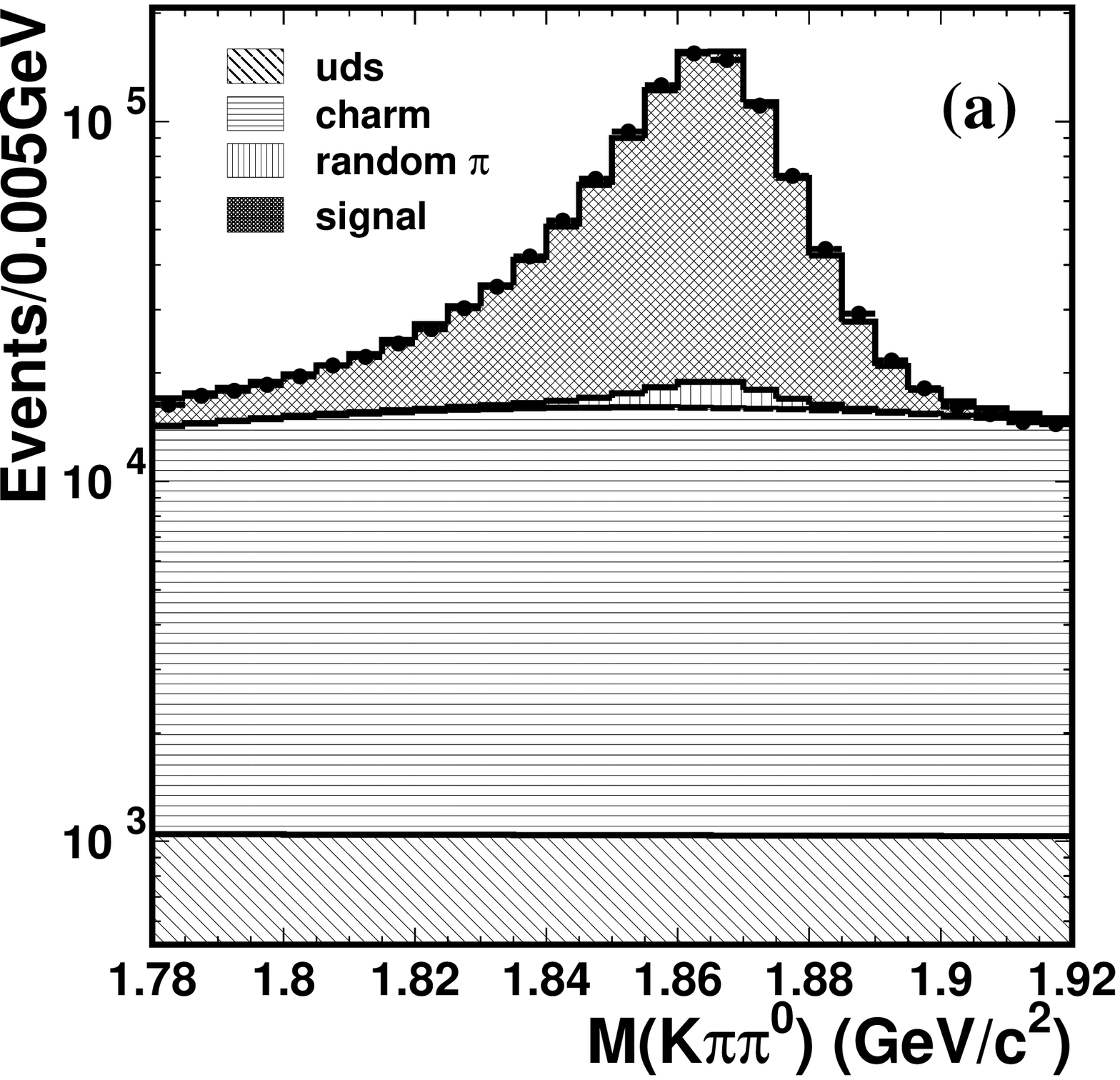}%
  \hfil
  \includegraphics[scale=0.3]{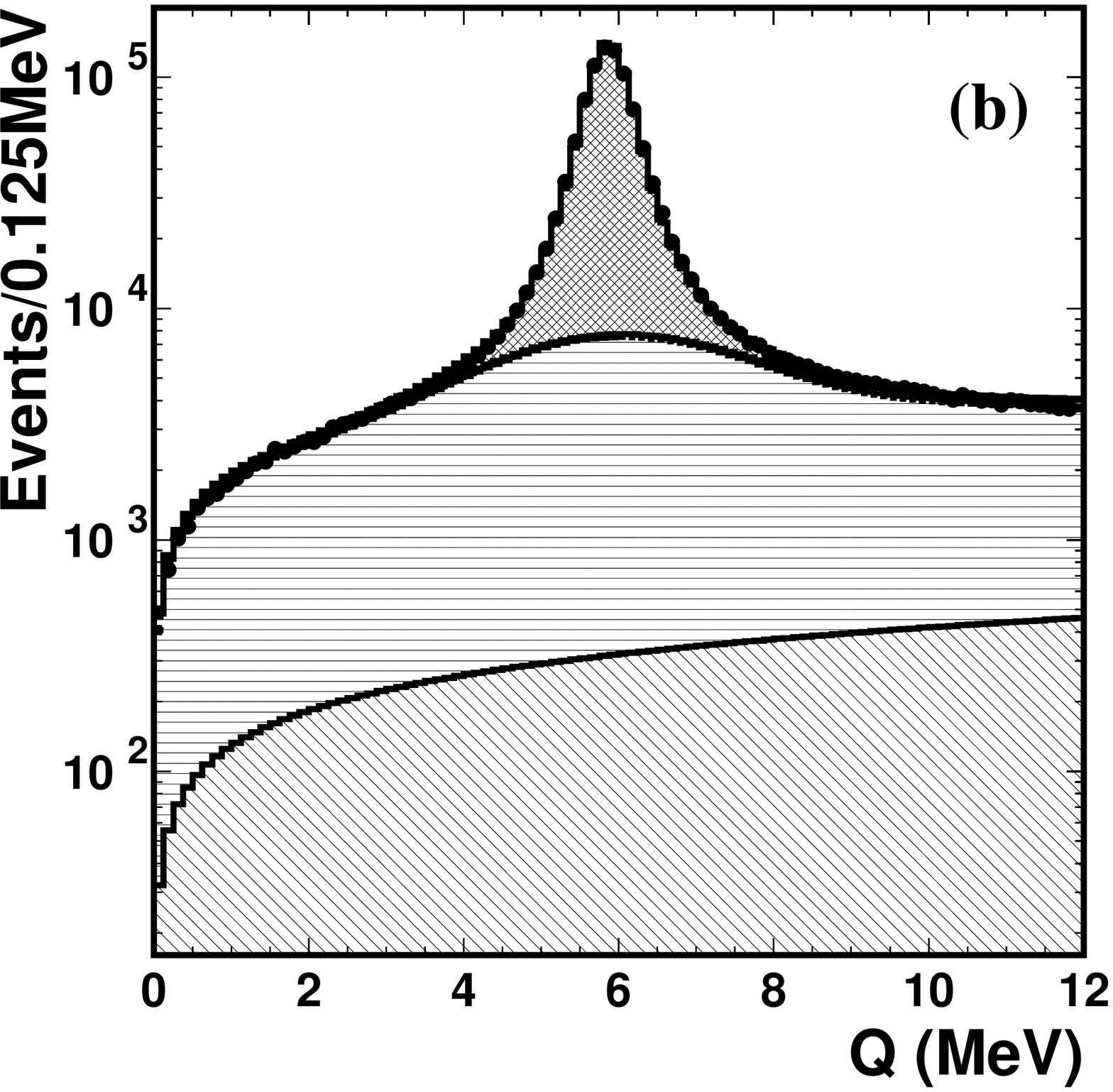}%
}
\hbox to\hsize{%
  \includegraphics[scale=0.3]{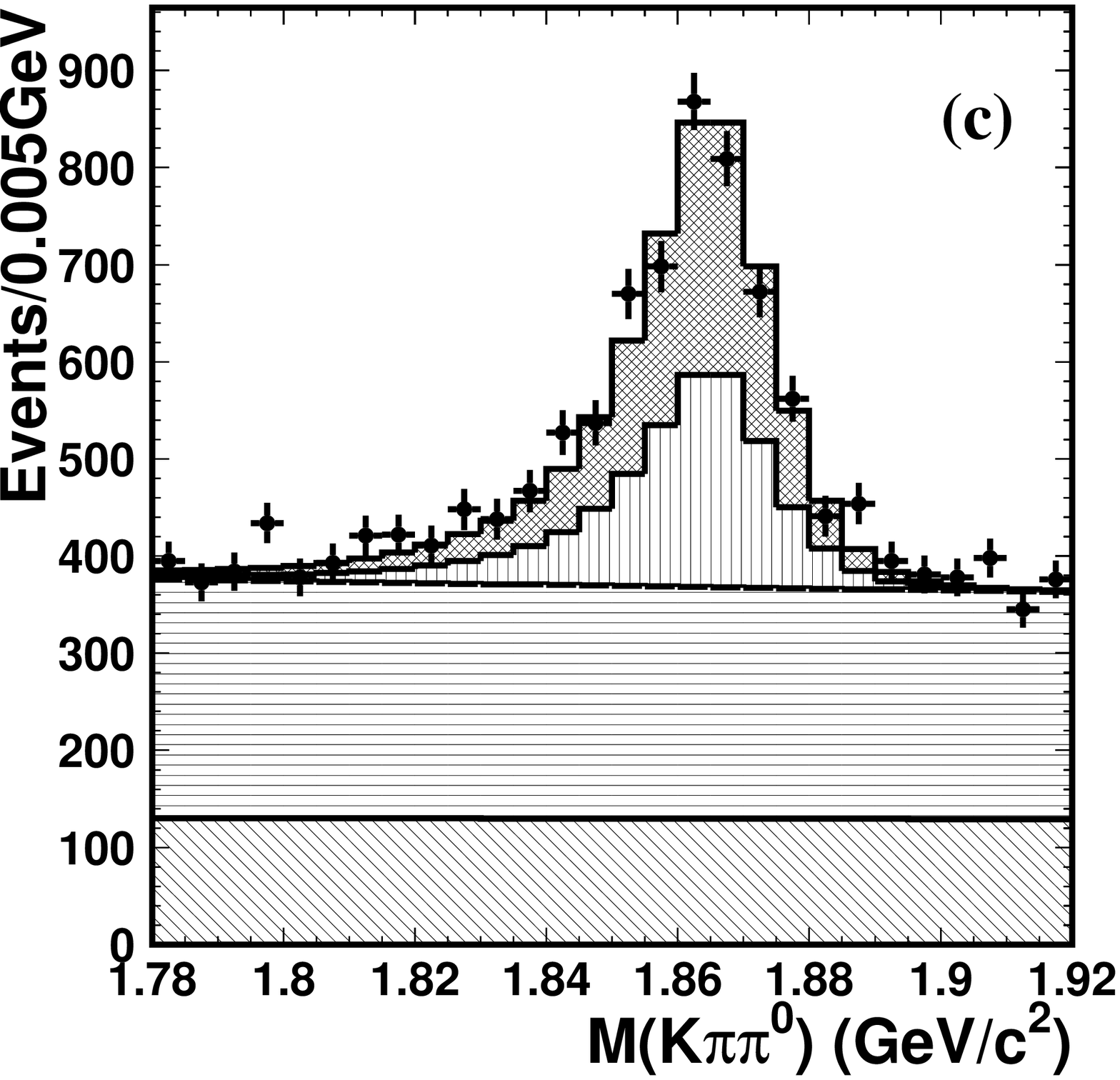}%
  \hfil
  \includegraphics[scale=0.3]{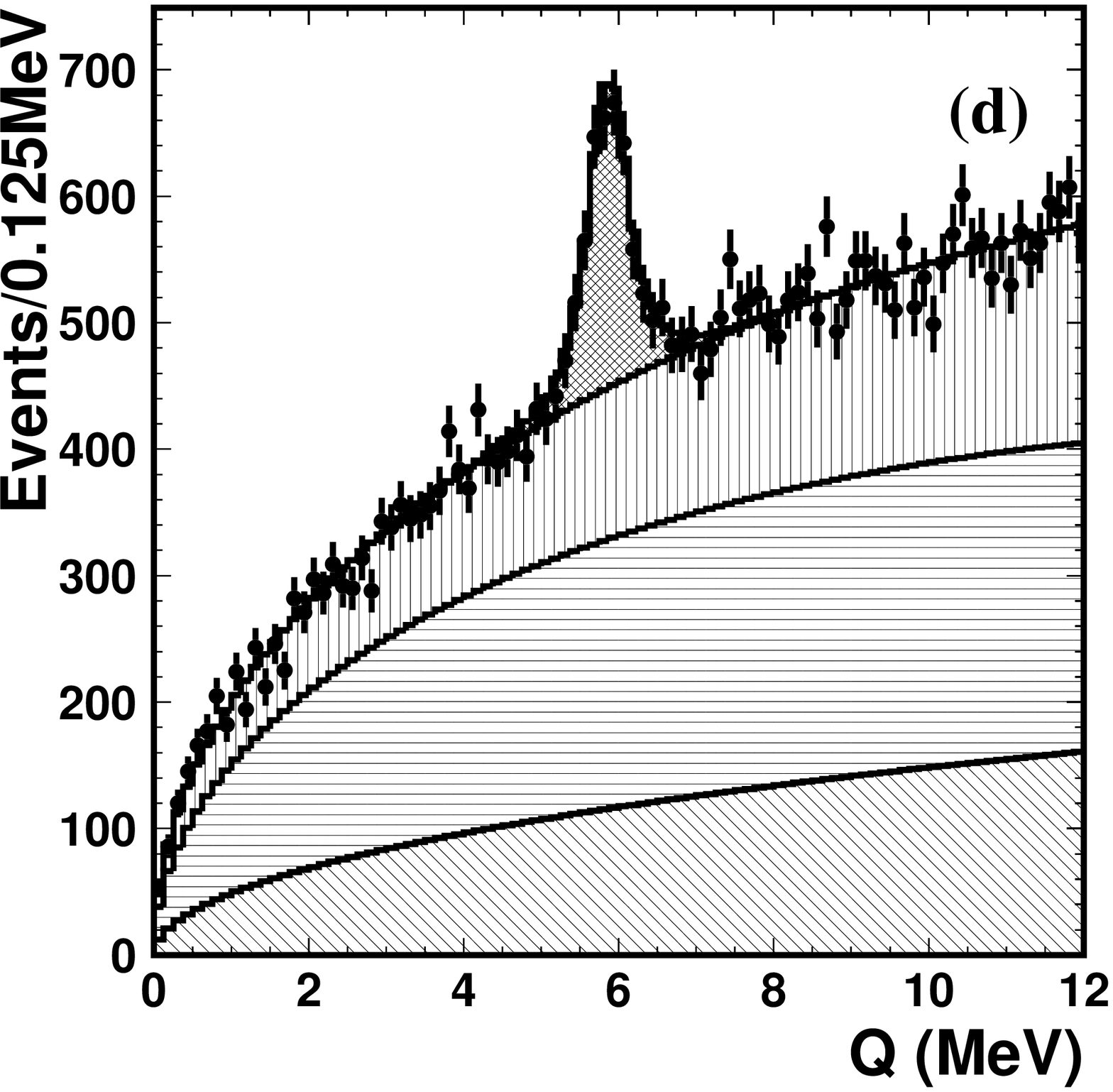}%
}
\caption{Belle $K\pi\piz$ reconstructed mass ($m_{K\pi\piz}$) 
and available kinetic energy ($Q$) distributions, showing signal and
background components as determined by a fit to the two-dimensional
($m_{K\pi\piz}$, $Q$) distribution.  Top row: RS; bottom 
row: WS.  
Reprinted figure with permission from 
X.C. Tian {\it et al.}, {\it Phys. Rev. Lett.} 95, 231801 (2005). 
Copyright 2005 by the American Physical Society.\protect\cite{Tian:2005ik}}  
\label{fig:Belle_Kpipi0_dists}
\end{center} 
\end{figure}

Reconstruction details of $\Dz\to K\pi\piz$ events
vary from experiment to experiment, but the basic selection
process is as follows.
The decay $\Dstarp\to\slowpi\Dz$ is used 
to tag the flavor of the neutral $D$ at production.
To form \Dz candidates, pairs of oppositely charged tracks 
originating from a common vertex
are combined with a \piz candidate
whose momentum in the laboratory is $\gsim 300~\mevc$.
\Dz candidates whose mass is within $\sim 60~\mevcc$ of
the nominal \Dz mass are retained.  
Particle identification requirements are imposed to 
reduce feedthrough of doubly misidentified, CF candidates
into the WS sample.
The momentum of each \Dz
candidate is required to point back to the interaction region
and its momentum in the CM system (\pstar) is required to satisfy $\pstar\gsim 2.5~\gevc$ to
suppress \Dz candidates from $B$ decay.  

Each \Dz candidate is paired with
a slow pion $\pi_s$ to form a \Dstarp candidate.  
\Dstarp candidates 
which have an appropriate value of $\DeltaM$ (or,
equivalently, $Q$; see Section~\ref{sec:Analysis_Techniques_Hadronic_Multi-body_Kshh}) and have sufficiently good $\chi^2$ per degree of freedom 
from the kinematic and/or vertex fits are retained.

Background sources considered are
random $\slowpi$ (an incorrectly associated \slowpi\ combined 
with good \Dz forming a \Dstarp candidate), incorrectly reconstructed charm decays,
and $uds$ combinatorial background.
Maximum likelihood fits to the two-dimensional ($m_{K\pi\piz}$, $\DeltaM$)
distribution are performed
to determine the yields of signal and background candidates in the RS and
WS samples.  
As an example, see Fig.~\ref{fig:Belle_Kpipi0_dists} for
the Belle fit to the \Dz candidate mass and $Q$ distributions.

\babar\ analyzed the $\Dz\to K\pi\piz$ mode using two
different methods.  The first, method~I,\cite{Aubert:2006kt}
uses the different decay-time dependence of
DCS and mixed decays and analyzes regions of phase space
chosen to 
optimize sensitivity to mixing. Although the rates of DCS and 
CF decays vary across the Dalitz plot, the 
mixing rate is the same  at all phase space points.

The time dependence of the WS-to-RS decay rate ratio can be expressed for
a given phase-space region (a tilde indicates integration of a quantity
over this region) as 
\begin{equation}
\frac{\Gamma_{WS}^{K\pi\piz}(t)}{\Gamma_{RS}^{K\pi\piz}(t)} = \tilde R_D^{K\pi\piz} + \alpha {\tilde y}^{\prime}
 \sqrt{\tilde R_D^{K\pi\piz}} (\Gamma t) + \frac{{\tilde x}^{\prime2} +
{\tilde y}^{\prime2}}{4} (\Gamma t)^2,
\label{eq:Kpipiz_WS_decay_rate}
\end{equation}
where $\alpha$ is an averaging factor that accounts for the variation of
the strong phase over the phase space region ($0 \leq \alpha \leq 1$).
$\tilde R_D$ is the DCS branching ratio,
${\tilde x}^{\prime}$ and ${\tilde y}^{\prime}$ are the mixing parameters
$x$ and $y$ rotated by an integrated strong phase $\tilde\delta$:
\begin{equation}
\begin{array}{rl}
{\tilde x}^{\prime}&=\phantom{-}x\cos{\tilde\delta}+y\sin{\tilde\delta},\\
{\tilde y}^{\prime}&=-x\sin{\tilde\delta}+y\cos{\tilde\delta}.\\
\end{array}
\label{eq:Kpipiz_tilde_primes}
\end{equation}
Note that $R_M = ({\tilde x}^{\prime 2} + {\tilde y}^{\prime 2} )/2 =
(x^2 + y^2)/2$ is independent of the integration region.

\begin{figure}[ht]
\begin{center}
\hbox to\hsize{%
  \includegraphics[scale=1.0]{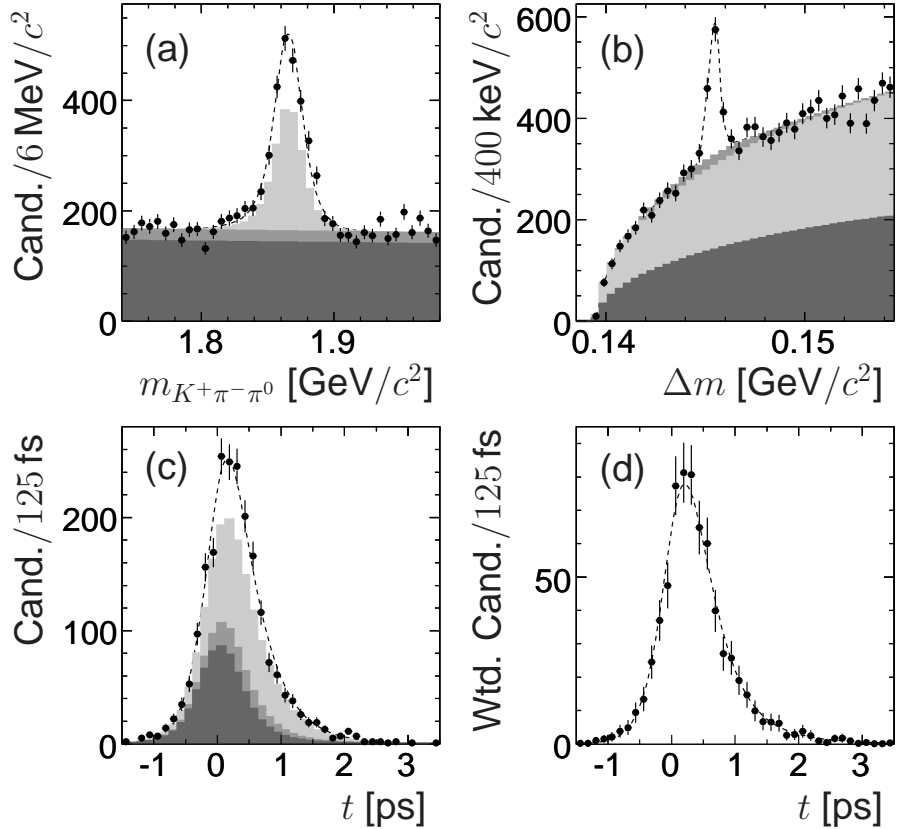}%
}
\caption{\babar\ fit projections for $K\pi\piz$ analysis method~I (see
text), showing data (points with error bars) and PDFs (dotted lines). (a) Reconstructed $m_{K\pi\piz}$ for candidates satisfying 
$0.1444 < \DeltaM < 0.1464$~\gevcc; (b) $\DeltaM$ for
candidates satisfying $1.85 < m_{K\pi\piz} < 1.88$~\gevcc;
(c) decay-time $t$ satisfying both mass selections in (a) and (b);
and (d) signal-enhanced version of (c) using a channel-likelihood
signal projection.\protect\cite{Pivk:2004ty,Condon:1974rh}
Reprinted figure with permission from 
B. Aubert {\it et al.}, {\it Phys. Rev. Lett.} 97, 221803 (2006). 
Copyright 2006 by the American Physical Society.\protect\cite{Aubert:2006kt}}  
\label{fig:BaBar_Kpipi0_method1_fits}
\end{center} 
\end{figure}

After signal and background yields are determined, a fit is performed to
the decay-time distribution.
The functional forms of the PDFs are 
determined based on MC studies, but all parameters are determined
by fitting the data.  
The large RS signal is used to determine
the resolution function used in the WS fit.
The observed decay-time dependence of the 
WS PDF is given by Eq.~\ref{eq:Kpipiz_WS_decay_rate}
convolved with the resolution function.  
Projections of
the reconstructed \Dz mass, $\DeltaM$, and decay-time fits are
shown in Fig.~\ref{fig:BaBar_Kpipi0_method1_fits}.

\begin{figure}[ht]
\begin{center}
\hbox to\hsize{%
  \includegraphics[scale=0.3]{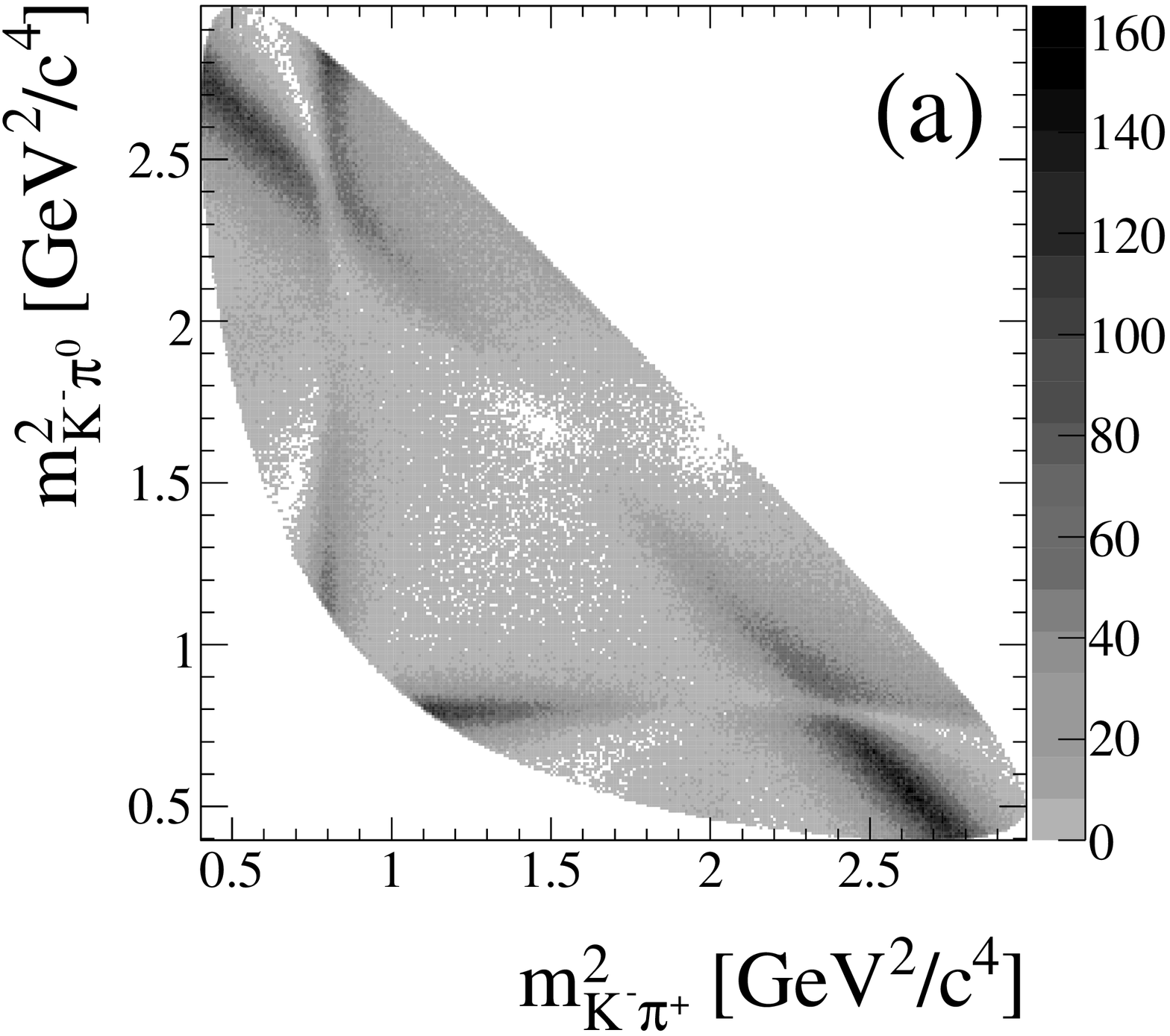}%
  \hfil
  \includegraphics[scale=0.3]{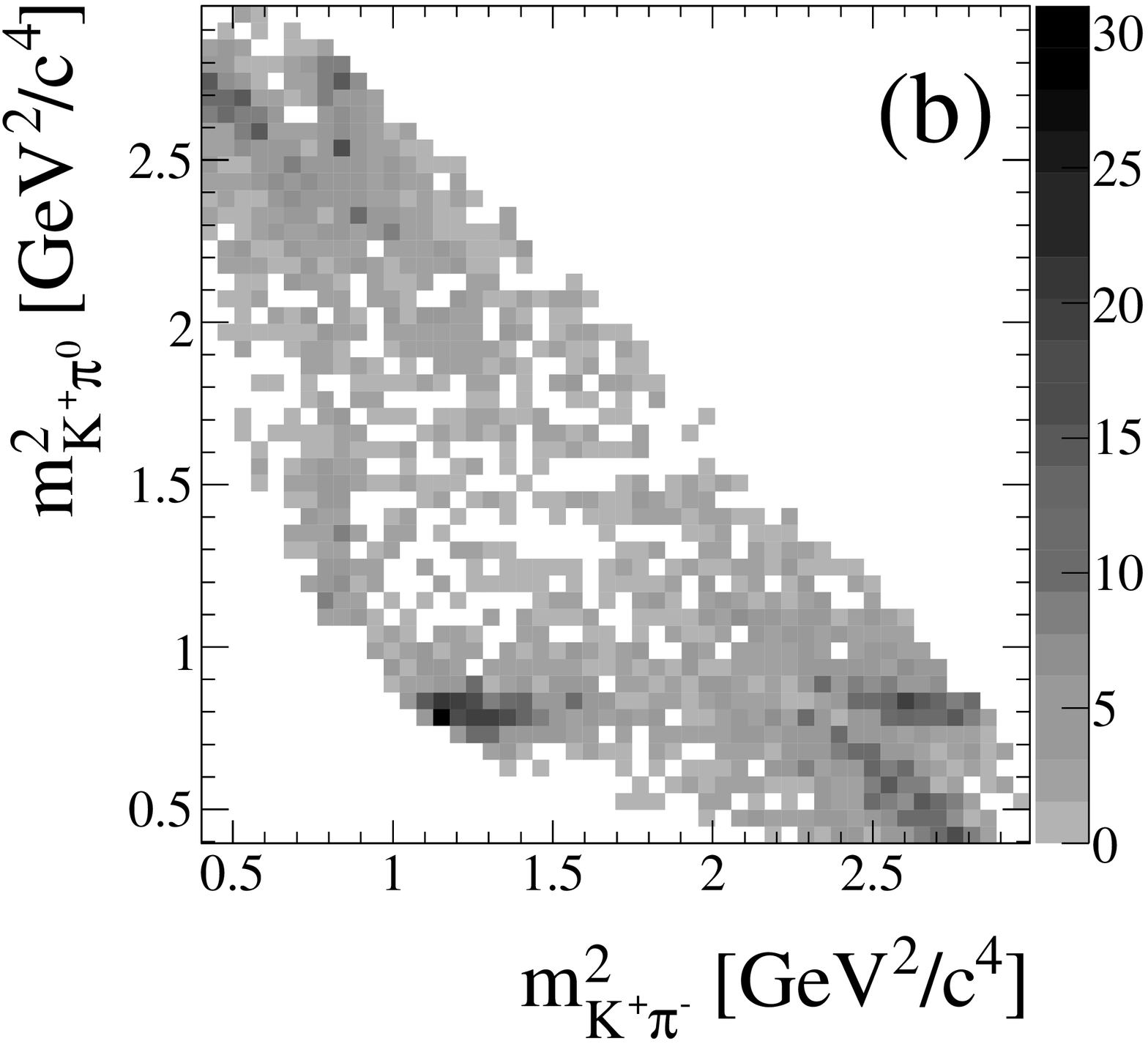}%
}
\hbox to\hsize{%
  \includegraphics[scale=0.3]{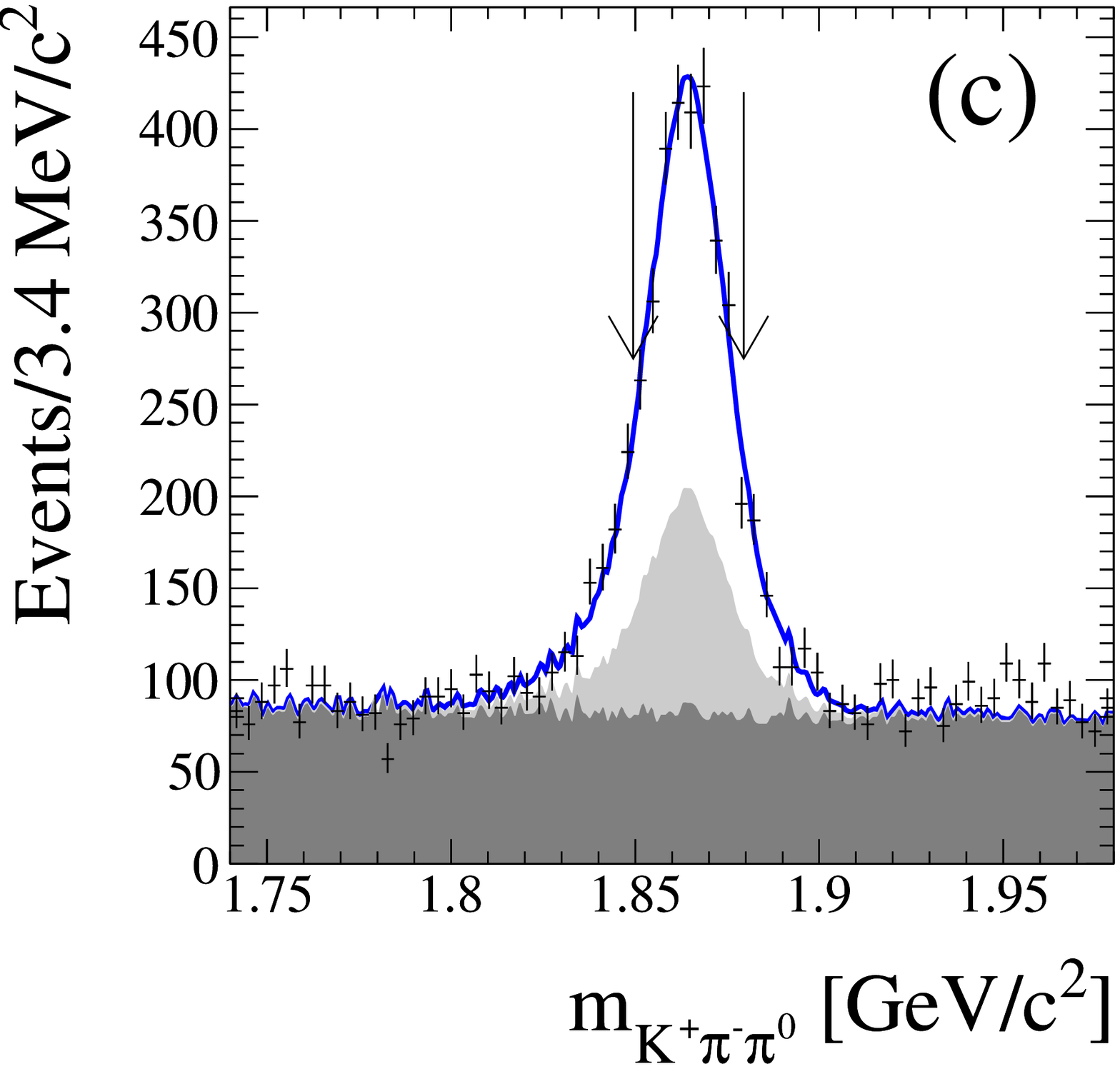}%
  \hfil
  \includegraphics[scale=0.3]{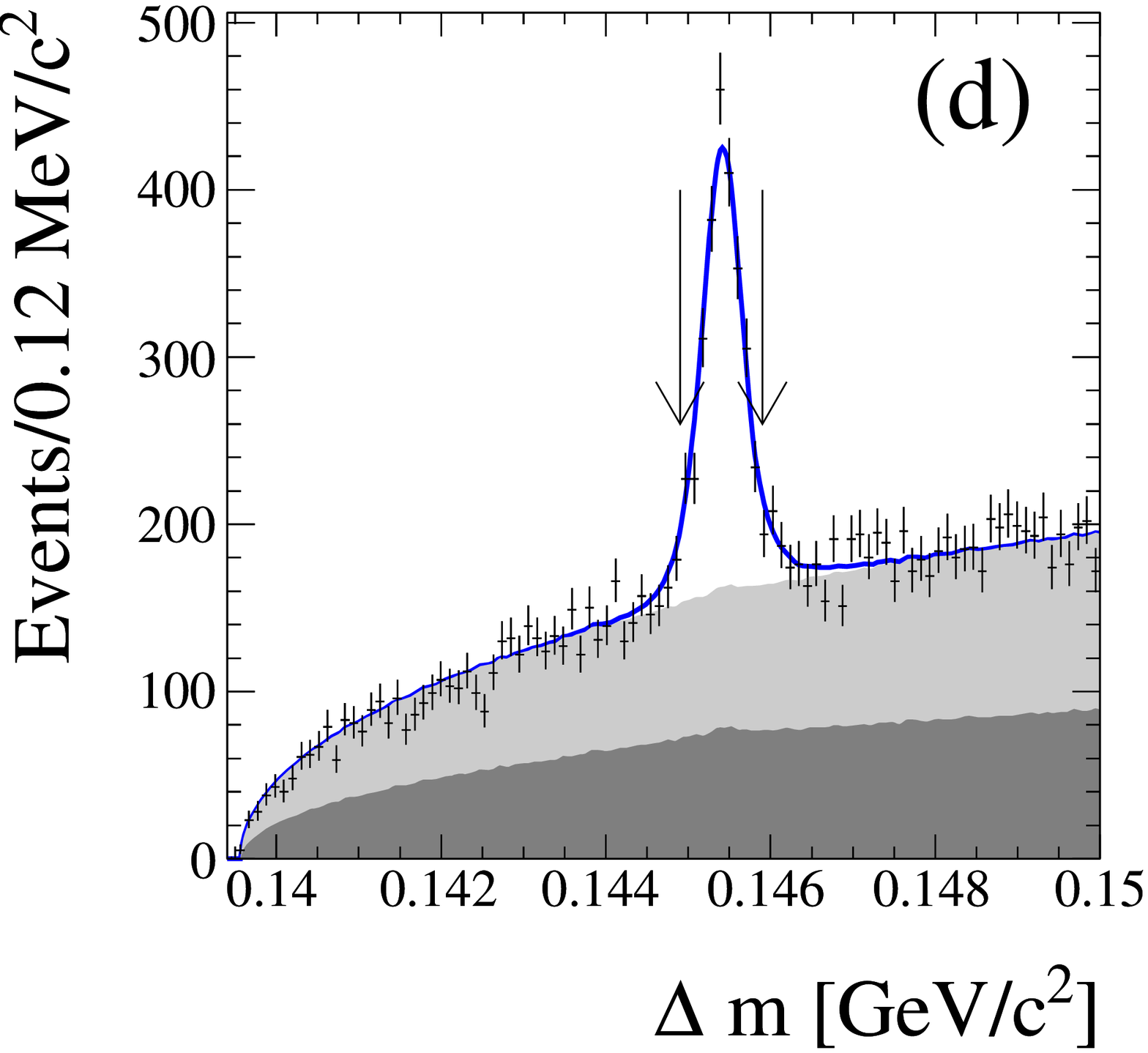}%
}
\caption{\babar\ RS (top left) and WS (top right) Dalitz distributions and reconstructed WS \Dz mass (bottom left)
and $\DeltaM$ (bottom right) distributions for $\Dz\to K\pi\piz$ using method~II.
Shaded regions indicate signal (white), mistag background (gray), and
combinatoric background (dark gray).
Reprinted figure with permission from 
B. Aubert {\it et al.}, {\it Phys. Rev. Lett.} 103, 211801 (2009). 
Copyright 2009 by the American Physical Society.\protect\cite{Aubert:2008zh}} 
\label{fig:BaBar_Kpipi0_dists}
\end{center} 
\end{figure}

The second method (method~II) used by \babar\cite{Aubert:2008zh}
to search for mixing in $\Dz\to K\pi\piz$
decays is a time-dependent, Dalitz-plot analysis that uses an 
isobar model\cite{Kopp:2000gv}
to describe the dynamics.
The time-dependent decay rate for WS decays to
a particular final state $f$ at a given point in the Dalitz plot
$(s_{12}, s_{13}) = (m_{\Kp\pim}^2, m_{\Kp\piz}^2)$, and assuming
$|x|$, $|y| \ll 1$, may be written
as
\begin{eqnarray}
\lefteqn{\frac{dN_f (s_{12}, s_{13}, t)}{ds_{12} ds_{13} dt} =  } \nonumber \\
& \qquad  &   e^{-\Gamma t} \left[
|\Af|^2 + |\Af| |\Abf| ( y \cos \delta_{f} - x \sin \delta_{f})\Gamma t
+ \frac{x^2 + y^2}{4} |\Abf|^2 (\Gamma t)^2
\right],
\label{eq:Kpipiz_method2_decay_rate}
\end{eqnarray}
where $f=\Kp\pim\piz$, the DCS amplitude is 
$\Af(s_{12},s_{13}) = \langle f|{\cal H}|\Dz\rangle$, 
the CF amplitude is $\Abf(s_{12},s_{13}) = \langle f|{\cal H}|\Dzb\rangle$,
and $\delta_f(s_{12},s_{13}) = \arg (A_f^*/\Abf) $.

Written in terms of normalized mixing parameters and
normalized amplitude distributions, the time dependence can be expressed
as:
\begin{eqnarray}
\lefteqn{\frac{dN_f (s_{12}, s_{13}, t)}{ds_{12} ds_{13} dt} \propto }
\nonumber \\
&&e^{-\Gamma t} r_0^2 \left[
|\Af^{\rm DCS}|^2 + |\Af^{\rm DCS}| |\Af^{\rm CF}| 
( {\hat y} \cos \delta_{f} - {\hat x} \sin \delta_{f})\Gamma t
+ \frac{{\hat x}^2 + {\hat y}^2}{4} |\Af^{\rm CF}|^2 (\Gamma t)^2
\right],\qquad
\label{eq:Kpipiz_method2_decay_rate_normalized}
\end{eqnarray}
where $\Af^{\rm DCS} = \Af/\sqrt{\int |\Af|^2 ds_{12} ds_{13} }$,
$\Af^{\rm CF} = \Abf/\sqrt{\int \Abf|^2 ds_{12} ds_{13} }$ are normalized
distributions, $r_0 = \sqrt{\int |\Af|^2 ds_{12} ds_{13} /  
\int |\Abf|^2 ds_{12} ds_{13} }$, and ${\hat x} = x/r_0$ and 
${\hat y} = y/r_0$ are normalized mixing parameters.

The isobar model parametrizes the amplitudes
$\Af$ and $\Abf$ as a
coherent sum of seven resonances plus a $K\pi$ $S$-wave
component derived from $K\pi$ scattering data,\cite{Aston:1987ir}
including a non-resonant component.
The high-statistics RS sample ($\sim 659,000$ candidates) is
used to determine the isobar model parameters for CF decays and the
decay-time resolution function for both the RS sample
and the WS sample ($\sim 3000$ candidates).  
See Fig.~\ref{fig:BaBar_Kpipi0_dists}.
Sensitivity to the mixing parameters arises primarily from the
interference terms (linear in $t$) in 
Eq.~\ref{eq:Kpipiz_method2_decay_rate} 
and Eq.~\ref{eq:Kpipiz_method2_decay_rate_normalized}.
PDFs expressing the dependence of the WS decay rate on Dalitz plot
position and decay time are convolved with the decay-time resolution and
a fit performed, determining the
DCS isobar model parameters (amplitudes and phases)
and the mixing parameters.

An unknown strong phase difference 
$\delta_{K\pi\piz}$
between the DCS decay $\Dz\to\rho^-\Kp$ and the CF decay
$\Dzb\to\Kp\rho^-$ cannot be determined in this analysis, so the
mixing parameters measured are 
\begin{equation}
\begin{array}{rl}
x^{\prime}_{K\pi\piz}&=\phantom{-}x\cos{\delta_{K\pi\piz}}+y\sin{\delta_{K\pi\piz}},\\
y^{\prime}_{K\pi\piz}&=-x\sin{\delta_{K\pi\piz}}+y\cos{\delta_{K\pi\piz}}.\\
\end{array}
\label{eq:Kpipiz_method_II}
\end{equation}
Results of the two methods are discussed in Section~\ref{sec:Kpipi0_results}.
 
\subsubsection{$\Dz\to\Kp\pim\pip\pim$ Analysis}
\label{sec:Analysis_Techniques_Hadronic_Multi-body_K3pi}

The $\Dz\to \Km\pip\pim\pip$ (CF $K3\pi$) 
decay has been used to study charm physics since
soon after the discovery of the \Dp and \Dz mesons.  An early search
for wrong-sign \Dz decays saw no significant signal, but did 
set limits on the WS rate.\cite{Goldhaber:1977qn}
E791 reported\cite{Aitala:1996fg}
a measurement attributed to DCS decay of 
$R_D^{K3\pi} = [0.25^{+0.36}_{-0.34} \pm 0.03]\%$ by analyzing
the distribution of wrong-sign \Dz decay times.
In a time-integrated measurement, CLEO reported 
evidence\cite{Dytman:2001rx} for
wrong-sign $\Dz\to\Kp\pim\pip\pim$ decays.  They found
a $3.9$ standard deviation result in the wrong-sign to right-sign
branching fraction:
$R_{WS}^{K3\pi} = [0.41^{+0.12}_{-0.11}{\rm (stat)} \pm 0.04 {\rm (syst)}
 \pm 0.10 {\rm (\hbox{phase space})}]\%$.

The $\Dz\to K3\pi$ decay offers some advantages over $\Dz\to K\pi$, and a 
couple of difficulties.  One advantage is that
the RS branching fraction for $\Dz\to K\pi\pi\pi$ of $\approx 8.1\%$
is twice that of $\Dz\to K\pi$ of $\approx 3.9\%$.  Another is that the
vertex resolution of the four-body $K3\pi$ decay is usually better than that
of the two-body $K\pi$ decay, which leads to an improved
decay-time resolution.  These advantages
are somewhat offset by the reduced
efficiency of reconstructing the four-body decay relative to the two-body
decay and by 
complications in determining the mixing parameters $x$ and $y$ due to
variations in the strong phase over the four-body phase space (the 
mixing rate $R_M$, however, is independent of position in phase space).
As in the case of 
$\Dz\to K\pi\piz$,
the strong phase $\delta_{K\pi\pi\pi}$ 
cannot
be determined in this analysis alone.
To date, no amplitude analysis of $\Dz\to K3\pi$ decays 
has been attempted.

\subsubsection{Analysis of Wrong-sign Semileptonic Decays}
\label{Analysis_Techniques_Semileptonic_Decays}

The WS semileptonic decays $\Dz\to K^{(*)+} e^-\bar\nu_e$ and 
$\Dz\to K^{(*)+} \mu^-\bar\nu_{\mu}$ offer unique features to searches
for \Dz-\Dzb mixing.  
One unique feature is that 
doubly Cabibbo-suppressed,
wrong-sign decays do not occur in the semileptonic mode
in the SM.  This simplifies
the time-dependence of the WS rate relative to the RS rate, Eq.~\ref{eq:Rws},
to:
\begin{equation}
\Rws(t)=\frac{\Rm}{2}(\Gamma t)^2\, .
\label{eq:Rws_semilep}
\end{equation}
The WS decay rate is
thus directly sensitive to the presence of mixing, as there 
is no contribution from either DCS decay or from interference
between DCS decay and mixing. 
On the other hand, semileptonic decays present a challenge not encountered
when analyzing 
hadronic decays: the presence of the unobserved neutrino in the final state
precludes exact determination of the \Dz candidate 
mass, leading to degraded decay-time and mass-difference resolutions
and higher backgrounds.

Distinguishing characteristics of
mixing in WS semileptonic decays include the quadratic time dependence of
Eq.~\ref{eq:Rws_semilep} and a peak in the available kinetic energy spectrum
$Q = m(Kl\nu\pi_s) - m(\Dz) - m(\pi)$ near 5.8~\mevcc (or, equivalently, in 
$\DeltaM = m(Kl\nu\pi_s) - m(Kl\nu)$ near 145~\mevcc).  
WS semileptonic decays share the peaking behavior in $Q$ (or $\DeltaM$)
with RS semileptonic decays, but have a time dependence modified by
the quadratic term
given in Eq.~\ref{eq:Rws_semilep} instead of the pure exponential 
decay-time distribution characterizing the RS decay.
Semileptonic decays are also susceptible 
to feed-through from RS $\Dz\to K\pi$ decays, where the kaon is
mis-identified as a lepton and the pion as a kaon.  This is particularly
a concern in the case of semi-muonic decays, since the kaons and pions
are more prone to mis-identification as muons than as electrons.

Many searches for mixing using semileptonic decays have been carried out.
The $K\mu\nu$ mode in particular has 
been used to search for and set limits on mixing since
shortly 
after the discovery of the \Dz meson.~\cite{Aubert:1981gx,Bodek:1981mi}
More recently, E791, CLEO, Belle, and 
\babar\ have reported measurements.\cite{Bitenc:2008bk,Aitala:1996vz,Cawlfield:2005ze,Aubert:2007aa,Aubert:2004bn}

The E791 analysis estimates the missing momentum of the neutrino by
using the measured decay vertex positions of the \Dstarp and the \Dz, the
kaon and lepton momenta, and attributing the \Dz mass to the secondary decay.
This results in a two-fold ambiguity, which is resolved by always
choosing the higher-momentum solution for the \Dz (motivated by MC
studies).  This choice results in some degradation in the decay-time resolution
which is accounted for as a systematic error.

The Belle analysis 
applies the following procedure to estimate the neutrino four-momentum 
and consequently, \DeltaM. 
Applying four momentum balance to the initial \epem system, the $Kl$ system,
the missing $\nu$, and the rest of the event, an approximation for the
missing momentum is obtained.  This value is refined by the use of two
additional constraints.
First, a \Dz mass constraint is applied to the $Kl\nu$ 
system, resulting in a scale factor that is used to produce a refined
$m(Kl\nu\pi_s)$ value, the $m(Kl\nu)$ mass having been fixed to $m(\Dz)$.
A second constraint on $m_{\nu}^2$ is applied, resulting in a correction
to the angle between the three-momentum of the $Kl$ system and that of the rest
of the event. 

\babar\  has published two semileptonic mixing analyses, one using
a single-tag ($\Dstarp\to\Dz\slowpip$)
method and the other using a double-tag method.  
The single-tag analysis includes
both $\Dz\to Ke\nu$ and $\Dz\to K^*e\nu$ decays, and treats
them essentially the same way.  No attempt is made to reconstruct the
$K^*$ explicitly; its kaon daughter is used directly in reconstructing the
\Dz, as if it were a \Dz daughter.  After selection cuts are imposed,
resulting $K$, $e$, and $\slowpi$ tracks, the position of the $K$-$e$ vertex,
and the event thrust axis are used to reconstruct the three components of 
the \Dz momentum vector by means of three neural net estimators.  These
estimators
have been trained using ${\cal O}(10^5)$ simulated signal events to 
reproduce the $\Dz$ momentum vector components.
Events are required to pass a neural net selection which discriminates
prompt charm from background events.  The majority of the remaining background
comes from charm events not from $B$ decays where a random charged pion has 
been combined
with a charged~$K$ daughter and an 
electron daughter from the charm decay, or with $K$ and
electron combinations not from a common parent.  Understanding the origin of
these backgrounds is important as they do not share exactly the same 
decay-time distribution as true charm decays, and this must be accomodated 
in the 
decay-time fit.  After performing an extended maximum likelihood fit to the large RS
data sample, which determines many of the PDF parameters describing the 
RS and WS $\DeltaM$ and decay-time PDFs, the mixing quantities are determined from a fit to the WS data, including the decay-time information.

\begin{figure}[ht]
\begin{center}
\hbox to\hsize{%
  \includegraphics[scale=0.40]{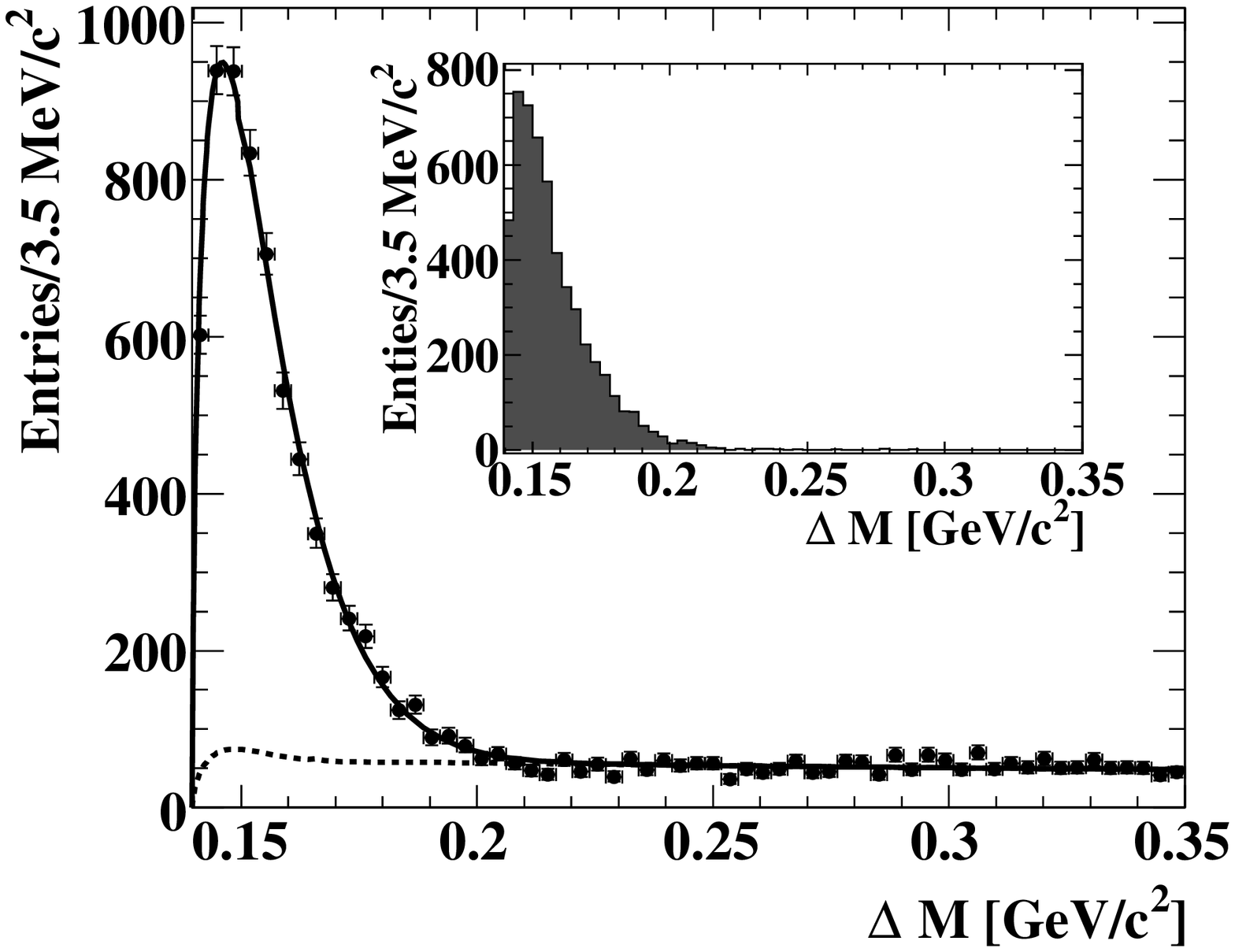}%
  \hfil
  \includegraphics[scale=0.30]{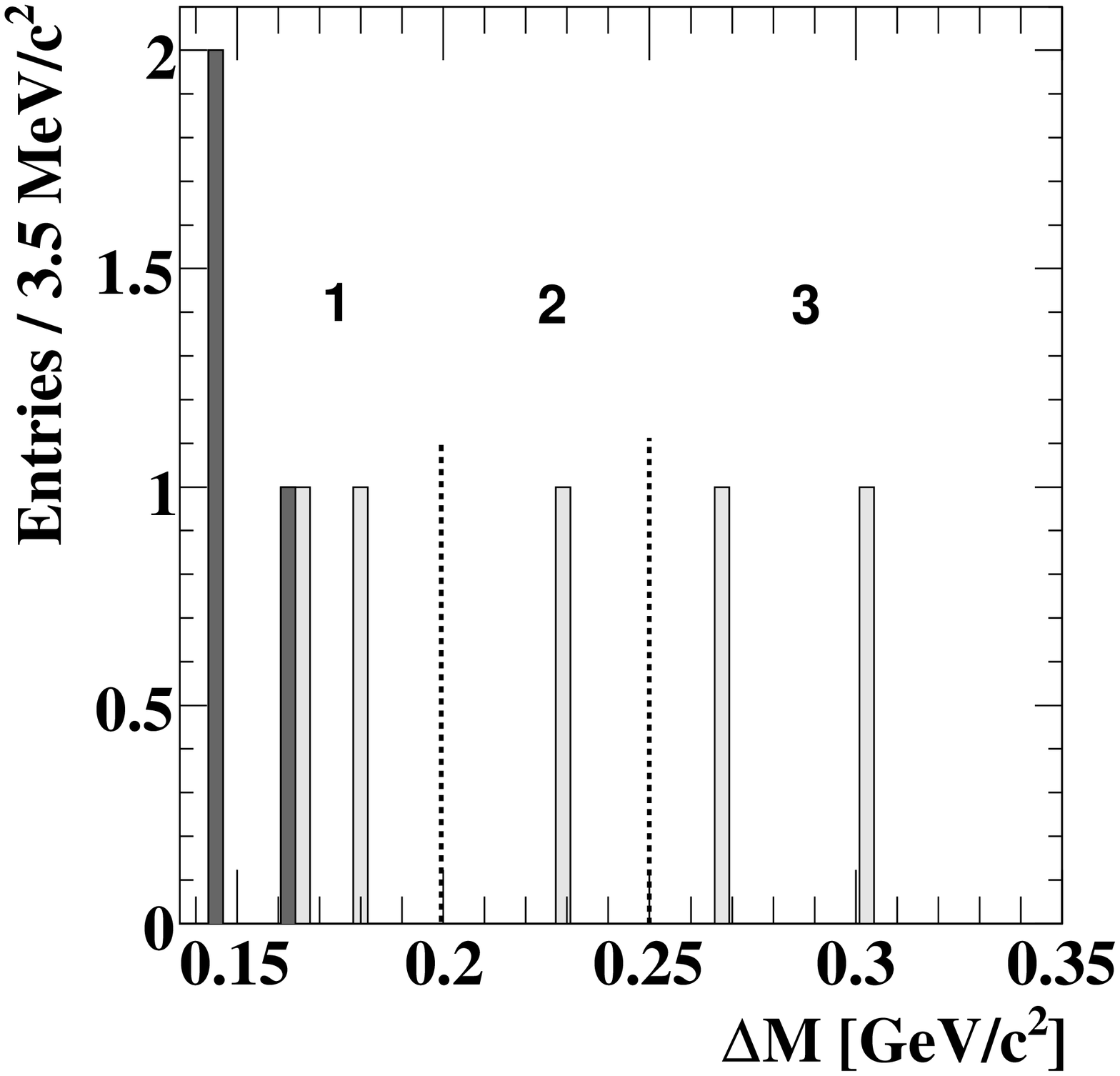}%
}
\caption{\babar\ semileptonic $\DeltaM$ distributions from 
Ref.~\protect\refcite{Aubert:2007aa}, showing both singly-tagged (white)
and doubly-tagged distributions (gray)
Left plot: RS data (points)
before the double-tag selection, along with the 
total fit projection (solid line), and the
background fit projection (dashed line). Inset: RS $\DeltaM$ distribution
after applying the double-tag selection.  Right plot: WS data 
events satisfying all
selection criteria (gray histogram) and all but the double-tag selection 
(white histogram).
Reprinted figures with permission from 
B. Aubert {\it et al.}, {\it Phys. Rev.} D~76, 014018 (2007). 
Copyright 2007 by the American Physical Society.}
\label{fig:BaBar_doubly_tagged_semilep_dm}
\end{center} 
\end{figure}

The \babar\  double-tagged analysis attempts to address the predominant background
present in the singly-tagged analysis: feedthrough into the WS sample 
from RS semileptonic 
\Dz decays, where the \Dz has been wrongly associated with a random slow pion~$\slowpi$.  In addition
to the \Dstarp tag, a second tag is constructed by requiring either a
fully reconstructed, high-momentum, hadronically decaying $\Dz$ or $\Dp$
in the the opposite hemisphere.
While greatly improving the purity of the tagged sample,
the 
selection
efficiency drops by more than a factor of ten.  Additional background
suppression criteria are imposed which bring the sensitivity of this
analysis to about the same as the single-tag analysis above.  Resulting
$\DeltaM$ distributions for RS and WS events are shown in 
Fig.~\ref{fig:BaBar_doubly_tagged_semilep_dm}.

%%------ CURRENT EXPERIMENTAL RESULTS --------
            
\section{Current Experimental Results}
\label{sec:Current_Experimental_Results}

\subsection{Time-independent Experiments}
\label{sec:Time-independent_results}

% Time-independent methods results

\subsubsection{Correlated decay results at 3.770~\gev}
\label{sec:correlated_decays_results}

\par
Using methods described in Section~\ref{sec:correlated_decays_method},
the CLEO Collaboration reported the first measurement of the 
strong phase difference \deltaKpi in 2008.\cite{Rosner:2008fq,Asner:2008ft}.  
From 281~\invpb of data 
collected at $\sqrt{s} = 3.770~\gev$ with the CLEO-c detector, 
the correlated analysis was performed using different sets of
external measurements as input.
These included:
measurements of two-body \Dz branching fractions;
the previous, plus measurements of the time-integrated WS rate
$\Rws \equiv \Gamma(\Dz\to\Kp\pim)/\Gamma(\Dz\to\Km\pip)$ 
and the mixing rate \Rm (the ``standard'' fit); and
the previous, plus measurements of $x$, $y$, 
\xPrimeSq, \yPrime, and $r^2$ (the ``extended'' fit).
Correlations between external inputs were incorporated in the fits.
In the standard fit, $x\sin\deltaKpi$ is fixed to zero.  In the extended
fit, this condition is relaxed.  

\begin{figure}[ht]
\begin{center}
\hbox to\hsize{%
  \includegraphics[scale=0.5]{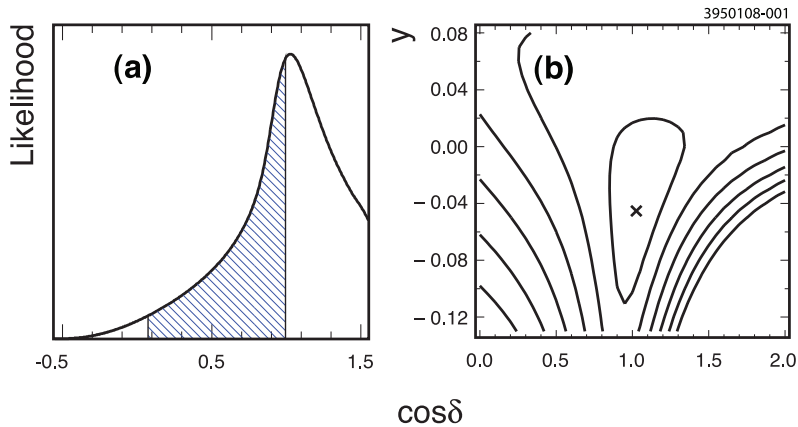}%
}
\caption{Fit likelihood from the CLEO standard fit, including statistical
and systematic uncertainties.  (a)~For $\cos\delta$; the hatched region
contains 95\% of the physically allowed area. (b)~Two-dimensional
likelihood for $\cos\delta$ and $y$.
Reprinted figure with permission from 
J. Rosner {\it et al.}, {\it Phys. Rev. Lett.} 100, 211801 (2008). 
Copyright 2008 by the American Physical Society.\protect\cite{Rosner:2008fq}}
\label{fig:CLEO_standard_fit}
\end{center}
\end{figure}

Systematic uncertainties accounted for in the analyses included
estimates of efficiencies for particle identification,
reconstructing tracks, and for reconstructing neutral 
neutral decays (\KS and \piz).  Other systematic uncertainties
included efficiencies for $\eta$ reconstruction, selection cuts,
fit model description, and detector and physics modeling. 
The largest systematics were for $\eta$ reconstruction (4.0\%) and
$\Delta E = E_{\Dz} - E_{{\rm beam}}$ 
selection (0.5--5.0\%).  Bias estimates on the fitting procedure
were obtained by studying a sample of simulated \DzdashDzb decays
fifteen times 
the size of the recorded dataset.  Biases from the fitting procedure were
less than one-half the size of the statistical errors on the 
fitted parameters.
\par
\begin{figure}[ht]
\begin{center}
\hbox to\hsize{%
  \includegraphics[scale=0.5]{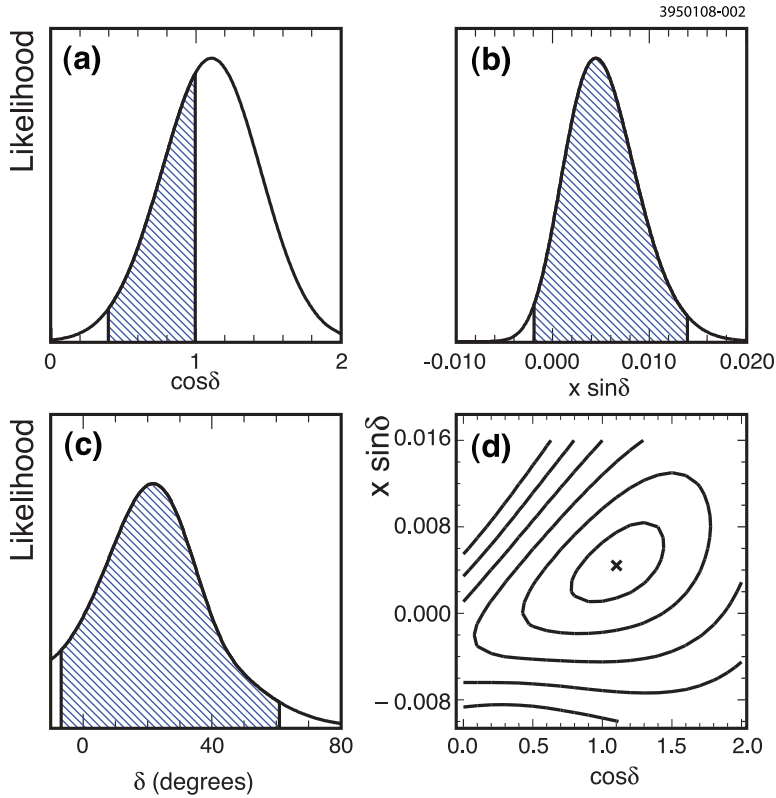}%
}
\caption{Fit likelihoods from the CLEO extended fits, including statistical
and systematic uncertainties.  Hatched regions
contain 95\% of the physically allowed area.
(a)~$\cos\delta$.
(b)~$x\sin\delta$. (c)~$\delta$. (d)~Two dimensional likelihood for
$\cos\delta$ and $x\sin\delta$.
Reprinted figure with permission from 
J. Rosner {\it et al.}, {\it Phys. Rev. Lett.} 100, 211801 (2008). 
Copyright 2008 by the American Physical Society.\protect\cite{Rosner:2008fq}}
\label{fig:CLEO_extended_fit}
\end{center}
\end{figure}

Results 
from the standard and extended
fits are given in 
Table~\ref{tab:CLEOResults2008}.  Likelihoods from the standard fit
are shown in Fig.~\ref{fig:CLEO_standard_fit} and from the extended fit
are shown in Fig.~\ref{fig:CLEO_extended_fit}.
The final result for \deltaKpi, including asymmetric errors estimated
from the shape of the likelihood function shown in 
Fig.~\ref{fig:CLEO_standard_fit}, is
$\cos \deltaKpi = 1.03^{+0.31}_{-0.17}\pm0.06$.  Limiting to the
region $|\cos\deltaKpi|<1$, $|\deltaKpi| < 75^\circ$ at the 95\% confidence
level.  From the extended fit, they obtain 
$\cos\deltaKpi = 1.10 \pm 0.35 \pm 0.07$ and 
$x\sin\deltaKpi = (4.4^{+2.7}_{-1.8}\pm2.9)\times 10^{-3}$ and
$\deltaKpi = (22^{+11}_{-12}{}^{+9}_{-11})^\circ$.  In both cases,
the statistical errors were obtained by inspection of the log likelihood.

\begin{table}[ht]
\tbl{CLEO mixing and $\delta_{K\pi}$ strong phase difference measurements.
From Refs.~\protect\refcite{Rosner:2008fq} and~\protect\refcite{Asner:2008ft}.
See text for fit descriptions.}
{\label{tab:CLEOResults2008}
\begin{tabular}{lcc}
\toprule
Parameter & Standard fit & Extended fit \\
\colrule
$y$                & $(-45\pm59\pm15)\times 10^{-3}     $ & $(6.5\pm0.2\pm2.1)\times 10^{-3}$ \\
$x^2$              & $(-1.5\pm3.6\pm4.2)\times 10^{-3} $ & $(0.06\pm0.01\pm0.05)\times 10^{-3}$ \\
$r^2$              & $(8.0\pm6.8\pm1.9)\times 10^{-3}  $ & $(3.44\pm0.01\pm0.09)\times 10^{-3}$ \\
$\cos \deltaKpi$   & $1.03^{+0.31}_{-0.17}\pm0.06               $ & $1.10\pm0.35\pm0.07$ \\
$x \sin \deltaKpi$ & Fixed at 0                          & $(4.4^{+2.7}_{-1.8}\pm2.9)\times 10^{-3}$\\ 
$\deltaKpi$ & --- & $(22^{+11}_{-12}{}^{+9}_{-11})^{\circ}$ \\
\botrule 
\end{tabular}} 
\end{table}

In a recent preliminary analysis,\cite{Sun:2010zz}  
CLEO extended its quantum correlated coherent decay analysis
to measure the strong phase differences in $\Dz\to\Kp\pim$, 
$\Dz\to\Kp\pim\piz$,  $\Dz\to\Kp\pim\pip\pim$ and $\Dz\to K^0_{S,L}h^+h^-$,
$h=\pi,K$, using the full dataset ($818\invpb$)
together with additional single- and double-tag modes. This analysis makes
direct measurements of $r_{K\pi}^2$ and $\sin\deltaKpi$, resulting in 
approximately a factor of two smaller 
(and more symmetric) statistical uncertainties on $\cos\deltaKpi$.
In the near future, BES-III will likely produce strong phase difference
measurements with improved statistical precision.

\subsection{Results from Time-dependent Analyses of Two-body Decays}
\label{sec:Two-body_Decays_results}

\subsubsection{$\Dz\to\Kp\pim$ Wrong-sign Decay Results}
\label{sec:Kpi_results}
\par
Several experiments, E691,\cite{Anjos:1987pw} E791,\cite{Aitala:1996fg} 
FOCUS,\cite{Link:2004vk} CLEO,\cite{Godang:1999yd} 
\babar,\cite{Aubert:2003aea} and Belle\cite{Zhang:2006dp} have set upper limits
on \DzdashDzb mixing by analyzing the time dependence of WS $\Dz\to\Kp\pim$ decays outlined in Sec.~\ref{sec:Analysis_Techniques_Hadronic_Two-body_Decay_Modes}.
Of these, the Belle limit, based on 400\invfb, is the most stringent.
Assuming \CP conservation, they find: $\xPrimeSq<0.72 \times 10^{-3}$ and $-9.9 \times 10^{-3}<\yPrime<6.8 \times 10^{-3}$ at the $95\%$ confidence level.
\par
\begin{figure}[h!]
\begin{center}
  \includegraphics[width=9cm]{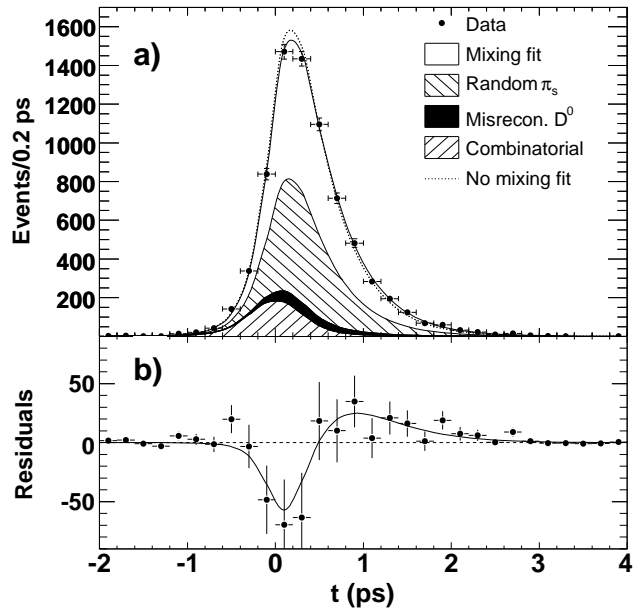}
\caption{\babar: (a) Projections of the reconstructed decay-time distribution 
of \Dz and \Dzb WS 
candidates (points with error bars) 
and fit result integrated over the region $1.843<\mKpi<1.883\gevcc$ and 
$0.1445<\DeltaM<0.1465\gevcc$. Assuming \CP conservation, the results of fitting with mixing 
and without mixing are shown as the solid and dashed curves, respectively. (b) 
The points with error bars represent the difference between the data and the 
average value of the no-mixing fit in each data bin. The solid curve shows the difference between
the fit with and without mixing. If there were no mixing, the data points would scatter randomly 
around the dashed horizontal line.
Reprinted figure with permission from 
B. Aubert {\it et al.}, {\it Phys. Rev. Lett.} 98, 211802 (2007). 
Copyright 2007 by the American Physical Society.\protect\cite{Aubert:2007wf}}
\label{fig:KpiWSResidual}
\end{center}
\end{figure}

In 2007 the \babar\ Collaboration reported evidence for \DzdashDzb mixing
from $(4030\pm90)$ WS signal candidates and $(1141500\pm1200)$ RS signal candidates in a 
$384\invfb$ data sample.~\cite{Aubert:2007wf} 
The reconstructed decay-time distribution for WS data and the
fit results with and without mixing (assuming \CP conservation) are shown in 
Fig.~\ref{fig:KpiWSResidual}. The fit with mixing provides a substantially better description
of the data than the fit with no mixing.
The parameters obtained from fitting the \babar\ data assuming \CP conservation are listed 
in Table~\ref{tab:KpiResults}.
\par
\begin{figure}[h!]
\begin{center}
  \includegraphics[width=9cm]{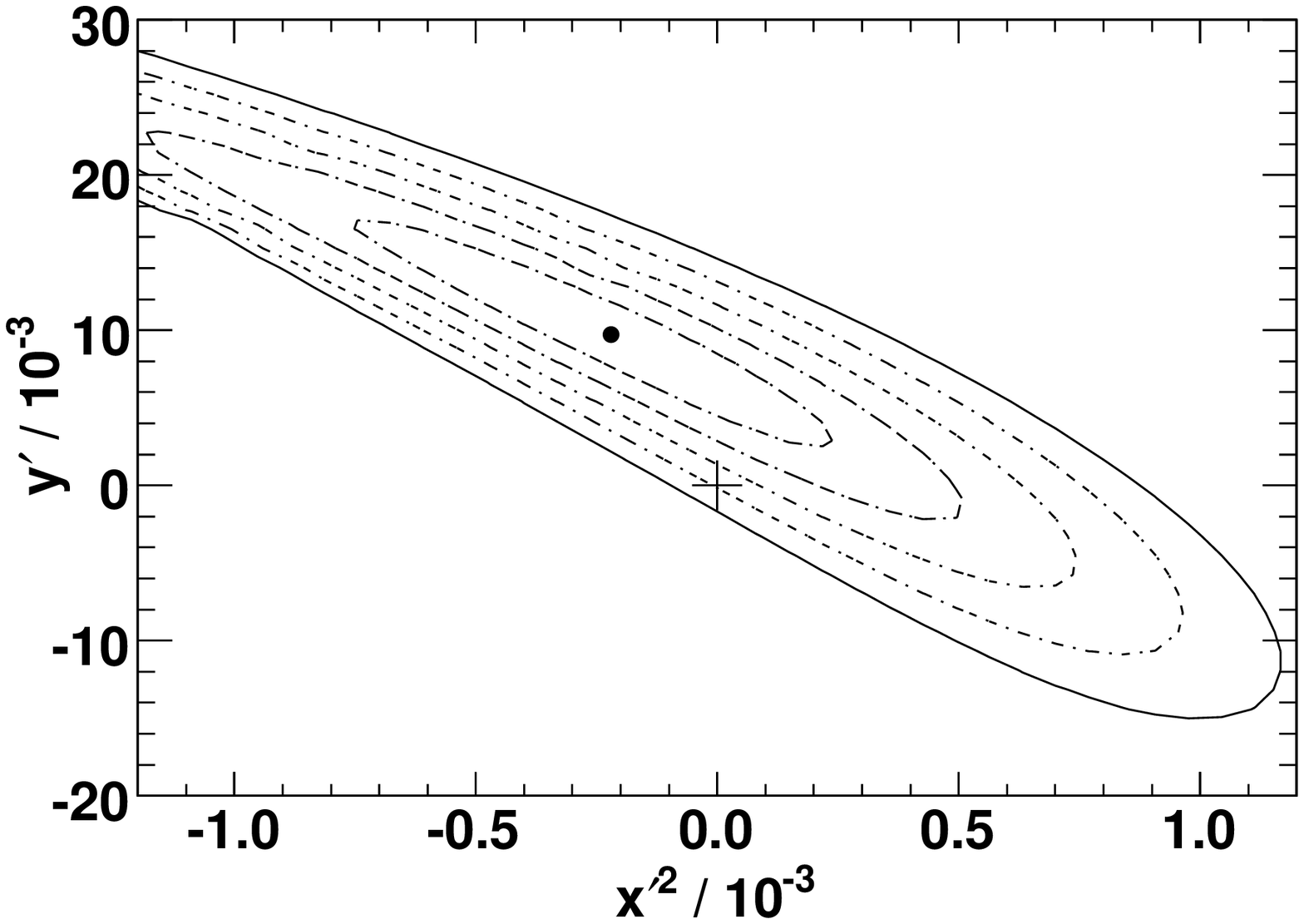}
\caption{\babar: The best fit value 
for the mixing parameters $(\xPrimeSq,\yPrime)$ ($\bullet$), the no-mix point $(0,0)$ $(+)$,
and confidence level (C.L.) contours evaluated for $(1-\hbox{C.L.})=0.317\thinspace(1\sigma)$, $4.55\times10^{-2}\thinspace(2\sigma)$, 
$2.70\times10^{-3}\thinspace(3\sigma)$, $6.33\times10^{-5}\thinspace(4\sigma)$ and
$5.73\times10^{-7}\thinspace(5\sigma)$, using the change $-2\Delta\log\Like{}$
from the joint estimation of two parameters. 
The contours include the estimated systematic uncertainty.
Reprinted figure with permission from 
B. Aubert {\it et al.}, {\it Phys. Rev. Lett.} 98, 211802 (2007). 
Copyright 2007 by the American Physical Society.\protect\cite{Aubert:2007wf}}
\label{fig:KpiWSContours}
\end{center}
\end{figure}

 
In the \babar\ measurement,
the significance of the mixing signal is estimated from the change in the log likelihood, 
\changeL, with respect to its value at the global minimum. Fig.~\ref{fig:KpiWSContours} shows
the confidence-level contours calculated using the change in log likelihood from the joint
estimation of two parameters. 
The best fit value of the $(\xPrimeSq,\yPrime)$ parameters to the \babar\ data is at the unphysical
value of $(\xPrimeSq=-2.2\times 10^{-4},\yPrime=9.7\times 10^{-3})$. As can be seen
from Fig.~\ref{fig:KpiWSContours}, 
the two parameters
are highly correlated with each other. Constraining the fit region to 
$\xPrimeSq\ge0$ yields $(\xPrimeSq=0,\yPrime=6.4\times 10^{-3})$, and corresponds to
$\changeL=0.7$. The no-mix point corresponds to $\changeL=23.9$ statistical units.
The maximum log likelihood is denoted as $\log\Like(\xPrimeSq,\yPrime)$. 
Each systematic variation is included one at a time into the fit and a new log liklihood
$\log\Like(\xPrimeSq_i,\yPrime_i)$
is obtained. The significance of the $i^{th}$ systematic variation is 
$s_i^2=2\left[\log\Like(\xPrimeSq,\yPrime)-\log\Like(\xPrimeSq_i,\yPrime_i)\right]/2.3$, 
where the factor
of 2.3 is the 68\% confidence level for two degrees of freedom.
Reducing 
\changeL by $1+\sum_is_i^2=1.3$ everywhere to account for systematic uncertainties results in a significance
equivalent to 3.9 standard deviations.  
Predominant systematic uncertainties on the
mixing parameters arise
from modeling the long decay times of other charm decays populating
the signal region and to a non-zero mean in the proper decay-time resolution
function.  

\begin{figure}[h!]
\begin{center}
  \includegraphics[width=8cm]{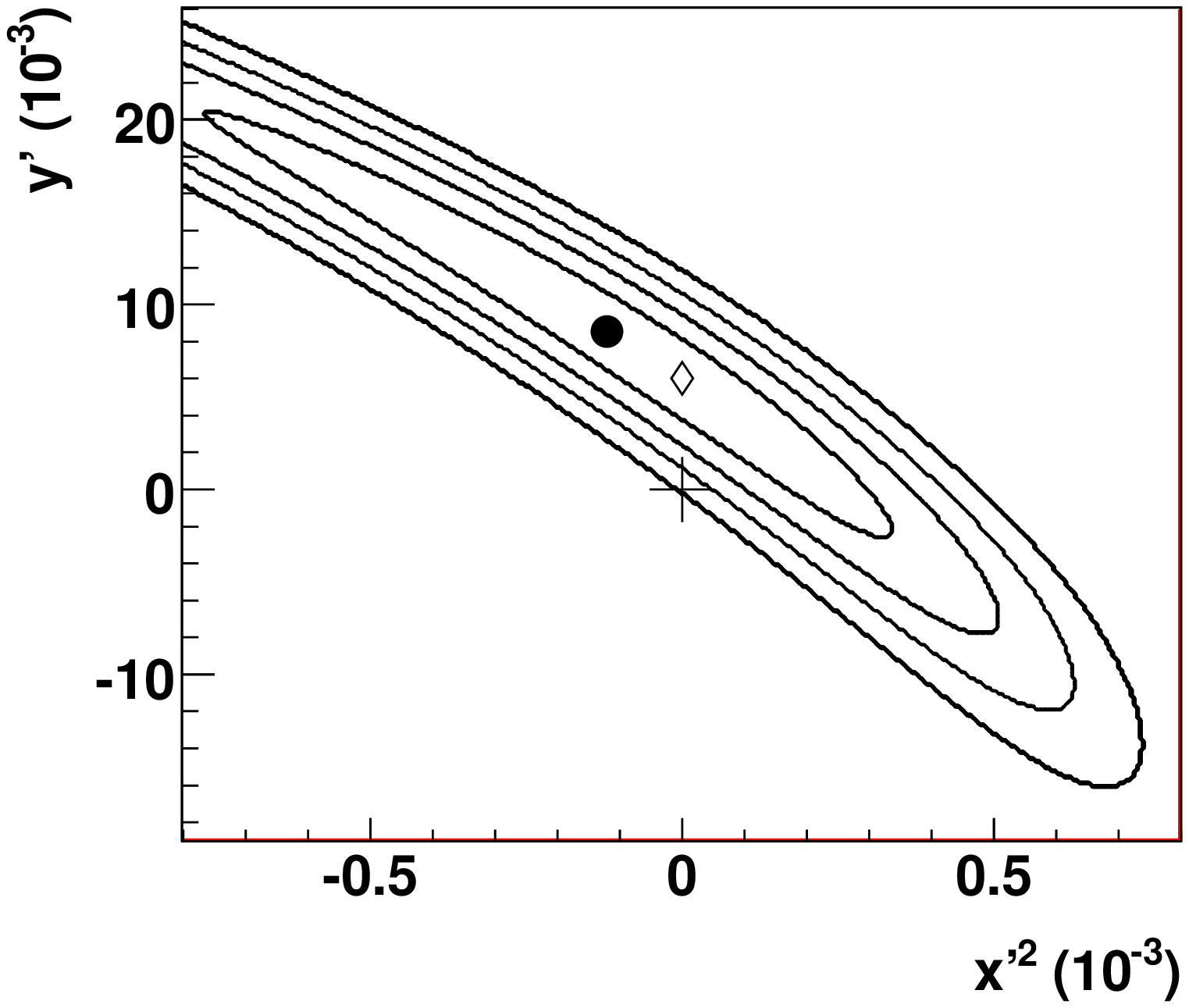}
\caption{CDF: The unconstrained best fit value
for the mixing parameters $(\xPrimeSq,\yPrime)$ 
($\bullet$),  requiring $(\xPrimeSq\ge0)$ ($\Diamond$), the $(0,0)$ (no-mix) point ($+$) and  
Bayesian probability contours corresponding to one through four equivalent Gaussian
standard deviations.
Reprinted figure with permission from 
T. Aaltonen {\it et al.}, {\it Phys. Rev. Lett.} 100, 121802 (2008). 
Copyright 2008 by the American Physical Society.\protect\cite{Aaltonen:2007uc}}
\label{fig:KpiCDFWSContours}
\end{center}
\end{figure}

\par
To allow for \CPV, the \babar\ analysis 
fits the \Dz and \Dzb samples separately 
to determine the parameters $(\RdcsP,\xPrimePSq, \yPrimeP)$ and $(\RdcsM,\xPrimeMSq, \yPrimeM)$, 
respectively. 
From these fitted values, the parameters 
$\Rdcs=\sqrt{\RdcsP\RdcsM}$ and $\AD=(\RdcsP-\RdcsM)/(\RdcsP+\RdcsM)$ are computed. 
The systematic component of the error on the \babar\ measurement of $A_D$ is mainly due to
uncertainties in modeling the slight asymmetry between the interactions of 
\Kp and \Km mesons in the detector. 
\par
The CDF Collaboration, using a $1.5\invfb$ data sample of $\bar pp$ collisions at $\sqrt{s}=1.96\tev$
has shown evidence for mixing in the $\Dz\to\Kp\pim$ channel.\cite{Aaltonen:2007uc}
Since the CDF experiment was not running on the $\Upsilon(4S)$ as \babar\ and Belle were,
removal of $B\to D$ decays was considerably more challenging than applying a simple center of mass momentum cut,
as was done in the $B$-factory measurements.  On the other hand, due to the much larger 
average boost, the average flight distance in the lab is greater than in the $B$-factory experiments.
Despite the vastly different 
environment,
the central
values of the mixing parameters shown in Table~\ref{tab:KpiResults} and the $(\xPrimeSq, \yPrime)$ C.L. contours 
shown in Fig.~\ref{fig:KpiCDFWSContours}, both from the CDF experiment, 
agree remarkably well with the corresponding \babar\ results.
There is no evidence for \CPV from any of the reported measurements.
\begin{table}[ht!]
  \tbl{\DzdashDzb mixing and \CPV parameters from $\Dz\to\Kp\pim$ decays. For results with 
two reported uncertainty components, the first is statistical and the second is systematic. 
The results with a 
single uncertainty component include both statistical and systematic uncertainties.} 
  {  \begin{tabular}{lcccc}\toprule
     Fit type & Parameter & \multicolumn{3}{c}{Fit Results ($/10^{-3}$)}  \\
              &           & \babar\cite{Aubert:2007wf} & CDF\cite{Aaltonen:2007uc} & Belle\cite{Zhang:2006dp} \\
    \colrule
    No \CPV or mixing 
    & $\Rdcs$         &   $3.53\pm0.08\pm0.04$   &$4.15\pm0.10$           &$3.77\pm0.08\pm0.05$\\
    \colrule
    No \CPV
    &  $\Rdcs$        &   $3.03\pm0.16\pm0.10$   &$3.04\pm0.55$ &$3.64\pm0.17$\\
    &  $\xPrimeSq$    &  $-0.22\pm 0.30\pm0.21$   &$-0.12\pm0.35$ &$0.18^{+0.21}_{-0.23}$\\
    &  $\yPrime$      &    $9.7\pm 4.4\pm3.1$    &$8.5\pm7.6$   &$0.6^{+4.0}_{-3.9}$\\
    Significance     &&     $3.9$          & $3.8$        & $2.0$ \\
    \colrule
    \CPV allowed
    & $\Rdcs$         &   $3.03\pm 0.16\pm0.10$   &$-$&$-$ \\
    & $\AD$           &    $-21\pm 52\pm15$     &$-$&$23\pm47$\\
    & $\AM$           &    $-$             &$-$&$670\pm1200$\\
    & $\xPrimePSq$    &  $-0.24\pm 0.43\pm0.30$   &$-$&$-$ \\
    & $\yPrimeP$      &  $ 9.8 \pm 6.4\pm4.5$    &$-$&$-$\\
    & $\xPrimeMSq$    &  $-0.20\pm 0.41\pm0.29$   &$-$&$-$\\
    & $\yPrimeM$      &  $ 9.6 \pm 6.1\pm4.3 $   &$-$&$-$\\
    & $\xPrimeSq$     &        $-$         &$-$&$<0.72$ (95\% C.L.)\\
    & $\yPrime$       &        $-$         &$-$&$-28<\yPrime<21$ (95\% C.L.)\\
\botrule
  \end{tabular}\label{tab:KpiResults}}
\end{table}

\subsubsection{$\Dz$ Lifetime Ratio Results}
\label{sec:KK_pipi_results}

\begin{figure}[h]
\begin{center}
\hbox to\hsize{%
  \includegraphics[scale=0.35]{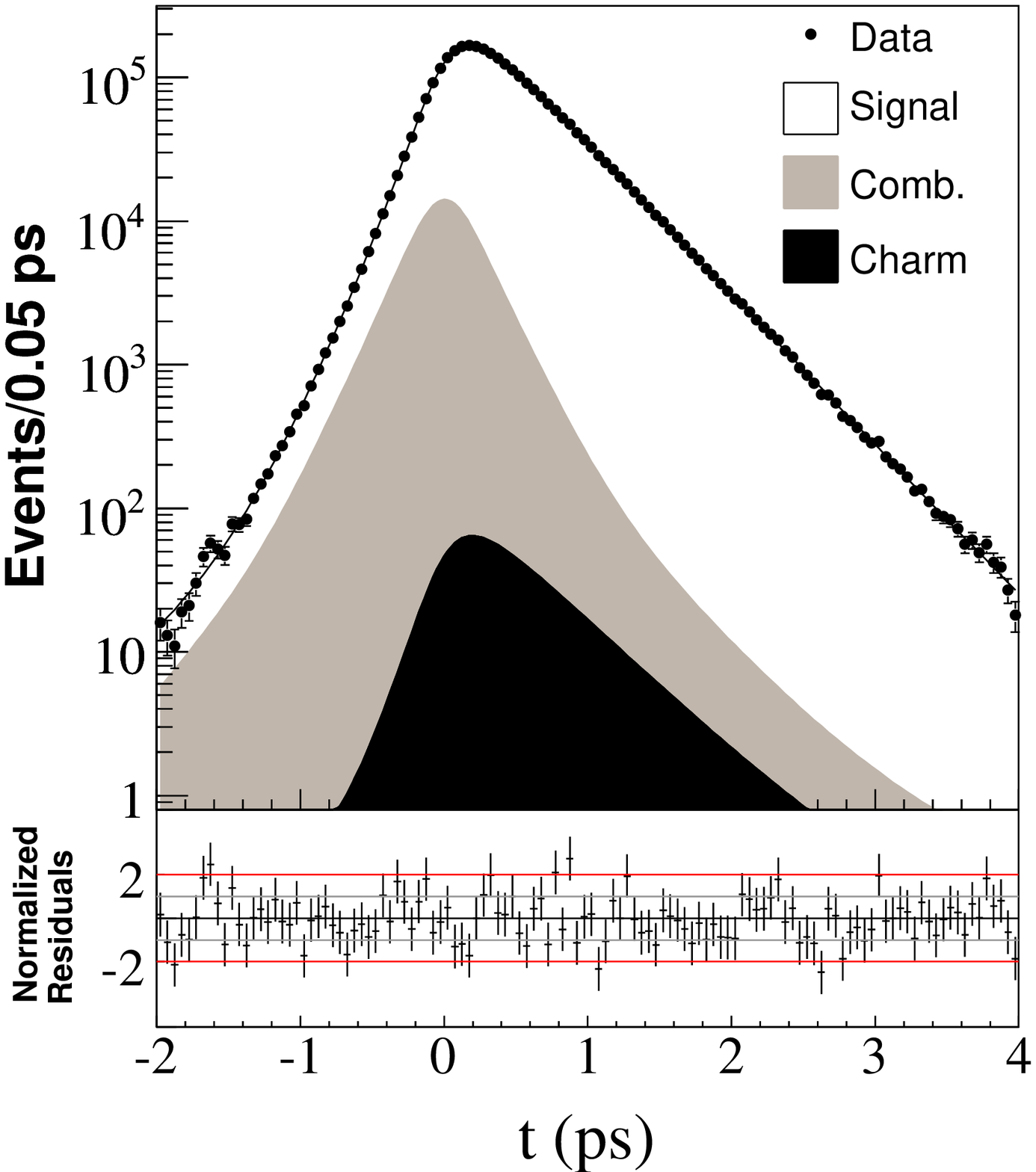}% 
  \hfil
  \includegraphics[scale=0.35]{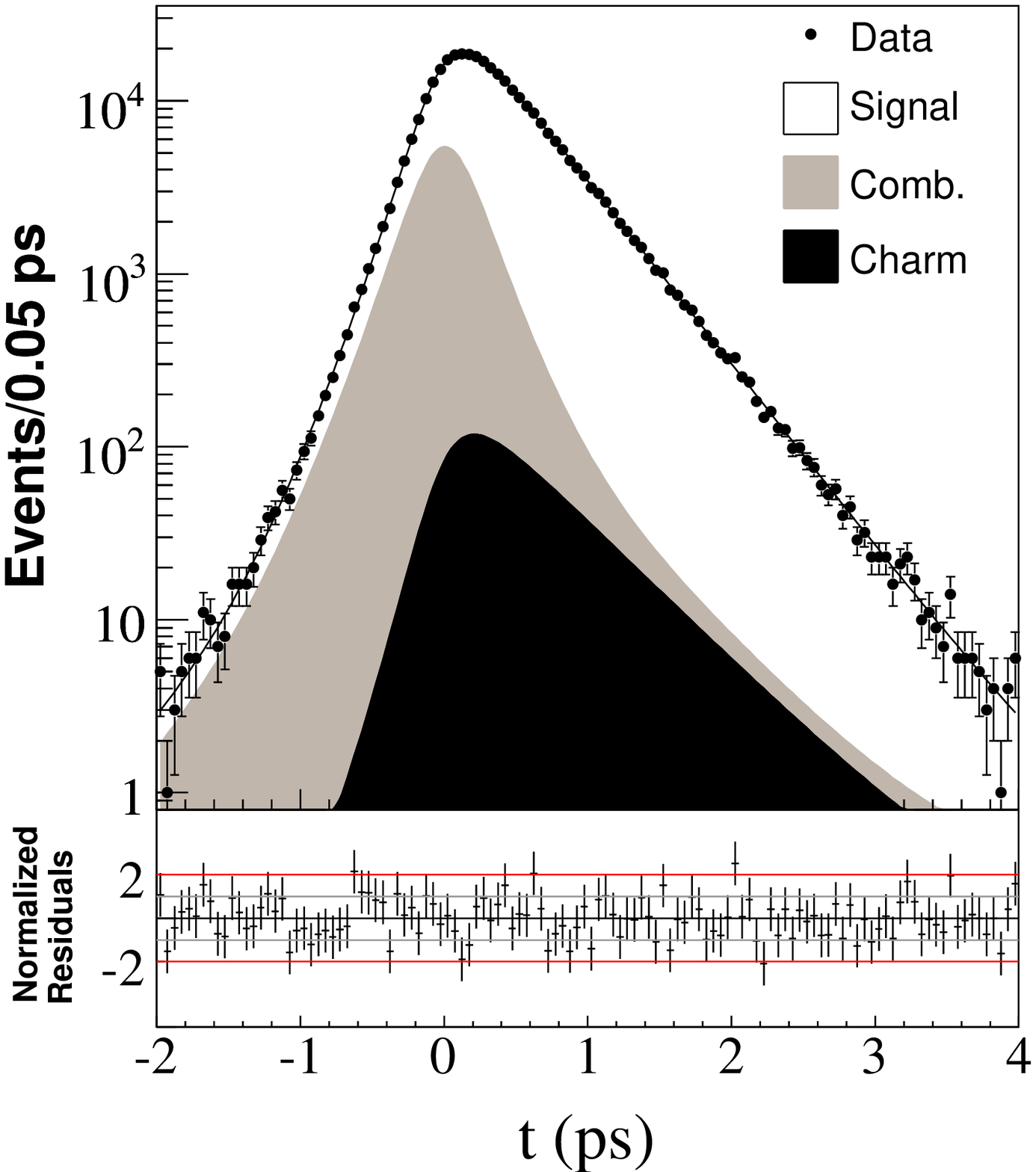}%
}

\caption{\babar\ decay-time distributions of $D^0~\to~K^-~\pi^+$~(left), and $D^0 \to K^+ K^-$~(right), from untagged \Dz decays.  In each plot, the total fit
is shown as a solid line.  Fit components are signal (white), combinatorics (gray), and charm background (black).  
Reprinted figures with permission from 
B. Aubert {\it et al.}, {\it Phys. Rev.} D~80, 071103(R) (2009). 
Copyright 2009 by the American Physical Society.\protect\cite{Aubert:2009ck}}
\label{fig:DztoKpiKKLifetime}
\end{center}
\end{figure}

\begin{figure}[h]
\begin{center}
\hbox to\hsize{%
  \includegraphics[scale=0.35]{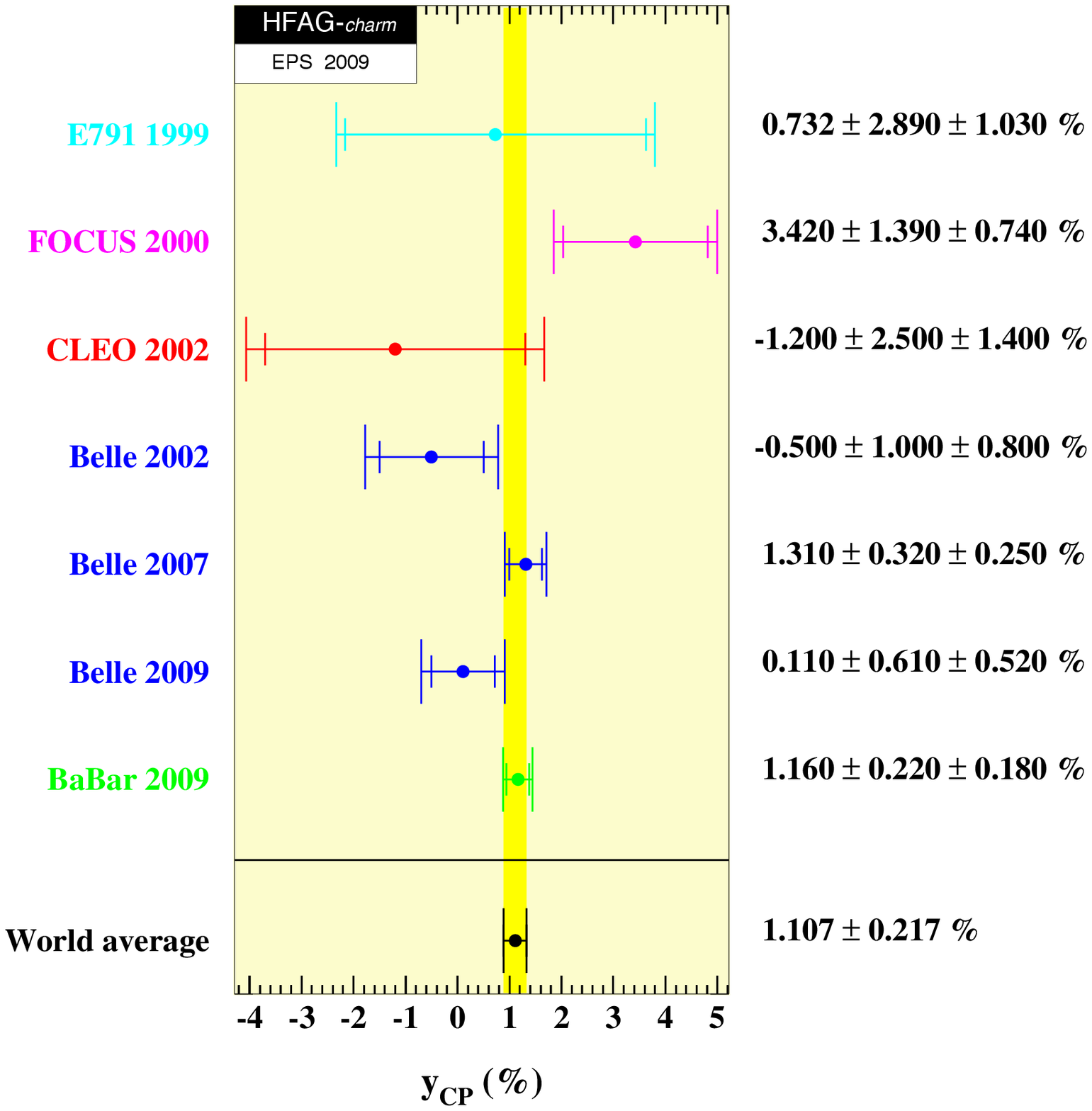}%
  \hfil
  \includegraphics[scale=0.35]{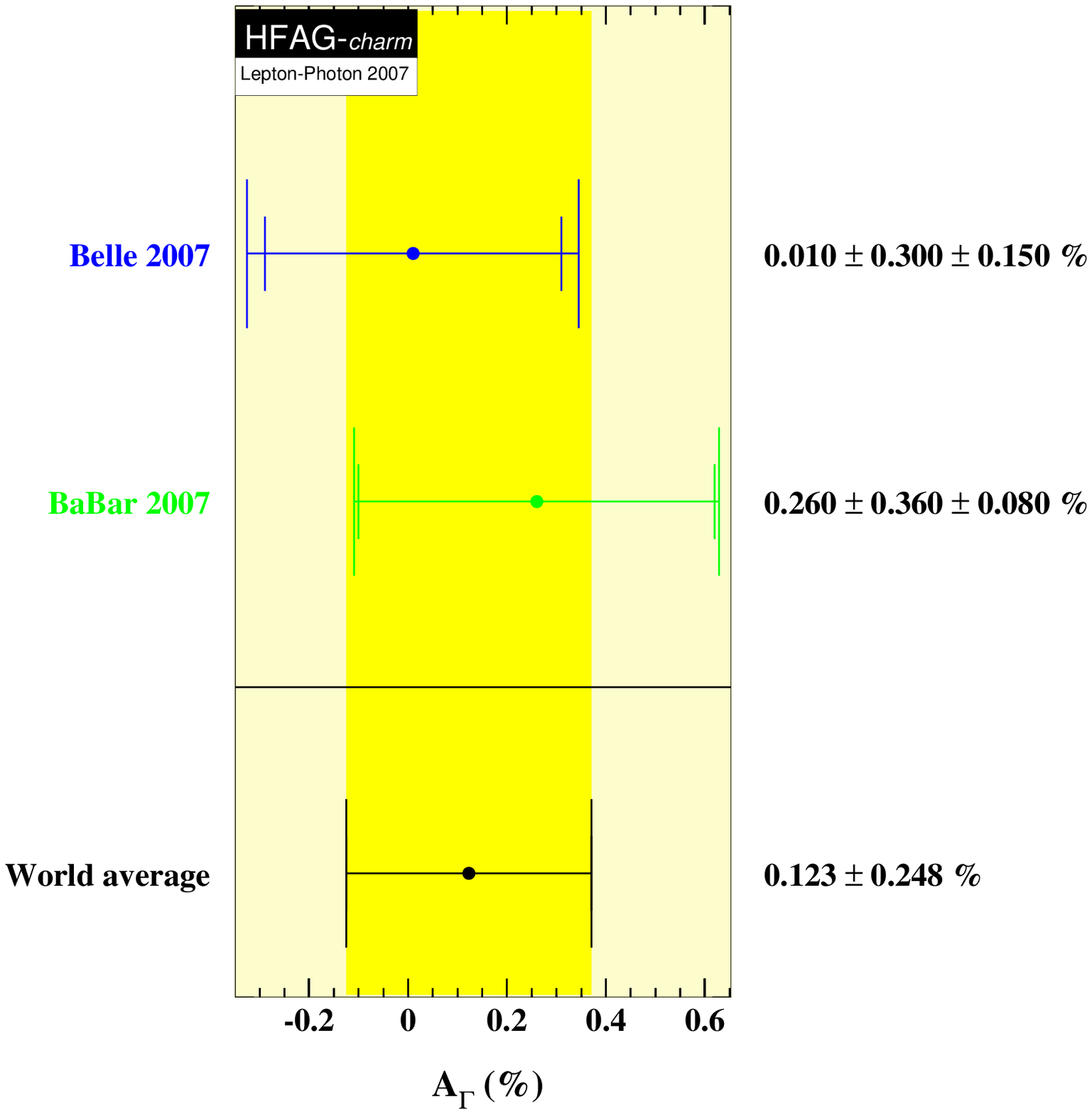}% 
}

\caption{HFAG compendium of  $y_{CP}$ (left) and $A_{\Gamma}$ (right) measurements, along with world average values. From Refs.~\protect\refcite{Asner:2010qj,HFAGwebsite}.}
\label{fig:yCP_Agamma}
\end{center}
\end{figure}

E791, FOCUS, CLEO, Belle, and \babar\ have published $y_{CP}$ results 
\cite{Staric:2007dt,Aitala:1999dt,Aubert:2007en,Aubert:2009ck,Link:2000cu,Csorna:2001ww}
from lifetime measurements of $\Dz\to K^+K^-$ and $\Dz\to \pip\pim$ decays relative to that from $\Dz\to K^-\pip$ decays. 
Fits to the proper time distributions from the \babar\  untagged analysis\cite{Aubert:2009ck} are shown in Fig.~\ref{fig:DztoKpiKKLifetime}.
The lifetime measurement is determined from a fit using 
the decay time and decay-time error of candidates
in a signal region, as described in Sec.~\ref{sec:Analysis_Techniques_Lifetime_Ratio}. 
The charm background component shape and yield in the signal region are obtained from MC
simulated events. The combinatorial background component shape in the signal region is estimated from sideband
data. A fit to the \Dz data mass distribution is performed over the full mass range to estimate the total background and signal yields in the signal region. 
The combinatorial yield in the signal region is then obtained by subtracting the charm background yield from the total background yield there.

The results from the Belle and \babar\ experiments on $A_{\Gamma}$ show no evidence for \CP violation. These measurements
are shown at the bottom of Table~\ref{tab:yCP_Results}.
Results for $y_{CP}$ are shown in Fig.~\ref{fig:yCP_Agamma}, and are given in Table~\ref{tab:yCP_Results}.
The Heavy Flavor Averaging Group (HFAG) world average $y_{CP}$ value\cite{Asner:2010qj,HFAGwebsite} is more than four standard deviations away from the no-mixing hypothesis ($y_{CP}=0$).

\begin{table}[ht]
\tbl{Results from $y_{CP}$ ($\Delta\Gamma$) and $A_{\Gamma}$ measurements from E791, FOCUS, CLEO, Belle, and \babar\ 
experiments.\protect\cite{Staric:2007dt,Aitala:1999dt,Aubert:2007en,Aubert:2009ck,Link:2000cu,Csorna:2001ww}
Measurement uncertainties are given as statistical (first) and systematic (second).  The world-average uncertainty is statistical and systematic combined.}
{\label{tab:yCP_Results}
\begin{tabular}{lccc}
\toprule
Experiment & Parameter & Result (\%) & data sample \\
\colrule
E791\protect\cite{Aitala:1999dt}  & $\Delta\Gamma$ & $ 0.04 \pm 0.14 \pm 0.05$ & 500 \gev $\pi$N interactions ($2\times 10^{10}$ events) \\ \hline
FOCUS\protect\cite{Link:2000cu} & $y_{CP}$ & $3.42 \pm 1.39 \pm 0.74 $ & $\gamma$N interactions ($1\times 10^6$ reconstr. $D\to K n(\pi)$ \\ \hline 
CLEO\protect\cite{Csorna:2001ww}  & $y_{CP}$ & $ −1.2 \pm 2.5 \pm 1.4$ & $9.0$ \invfb near $\Upsilon (4S)$ resonance ; untagged \\ \hline
Belle\protect\cite{Staric:2007dt} & $y_{CP}$ & $−0.5 \pm 1.0 ^{+0.7}_{-0.8} $ & $23.4$ \invfb near $\Upsilon (4S)$ resonance; untagged  \\
Belle\protect\cite{Staric:2007dt} & $y_{CP}$ & $1.31 \pm 0.32 \pm 0.25$ & $540$ \invfb near $\Upsilon (4S)$ resonance ; $D^*$ tagged  \\
Belle\protect\cite{Staric:2007dt} & $y_{CP}$ & $0.11 \pm 0.61 \pm 0.52$ & $673$ \invfb near $\Upsilon (4S)$ resonance ; $D \to (K^+K^−) K^0$ \\ \hline
\babar\protect\cite{Aubert:2007en} & $y_{CP}$ & $1.03 \pm 0.33 \pm 0.19$ & $384$ \invfb near $\Upsilon (4S)$ resonance ; $D^*$ tagged\\ 
\babar\protect\cite{Aubert:2009ck} & $y_{CP}$ & $1.12 \pm 0.26 \pm 0.22$ & $384$ \invfb near $\Upsilon (4S)$ resonance ; untagged \\
\babar\protect\cite{Aubert:2009ck} & $y_{CP}$ & $1.16 \pm 0.22 \pm 0.18$ &   $D^*$ tagged + untagged combined \\ \hline
HFAG\protect\cite{HFAGwebsite}  & $y_{CP}$ & $1.107 \pm 0.217$ & World Average \\ \botrule
 & & & \\ 
Belle\protect\cite{Staric:2007dt} & $A_{\Gamma}$ & $0.01 \pm 0.30 \pm 0.15$ & $540$ fb$^−1$ near $\Upsilon (4S)$ resonance ; $D^*$ tagged  \\ \hline
\babar\protect\cite{Aubert:2007en} & $A_{\Gamma}$ & $0.26 \pm 0.36 \pm 0.08$ & $384$ fb$^−1$ near $\Upsilon (4S)$ resonance ; $D^*$ tagged  \\ \hline
HFAG\protect\cite{HFAGwebsite}  & $A_{\Gamma}$ & $0.123 \pm 0.248$  &  World Average \\ 
\botrule
\end{tabular}} 
\end{table} %

\subsection{Results from Time-dependent Analyses of Multi-body Decays}
\label{sec:Multi-body_Decays_results} 

\subsubsection{$\Dz\to\KS\hp\hm$ Analysis Results}
\label{sec:KS_h+h-_results}

\begin{figure}[ht]
\begin{center}
\hbox to\hsize{%
  \hfil
  \includegraphics[scale=0.30]{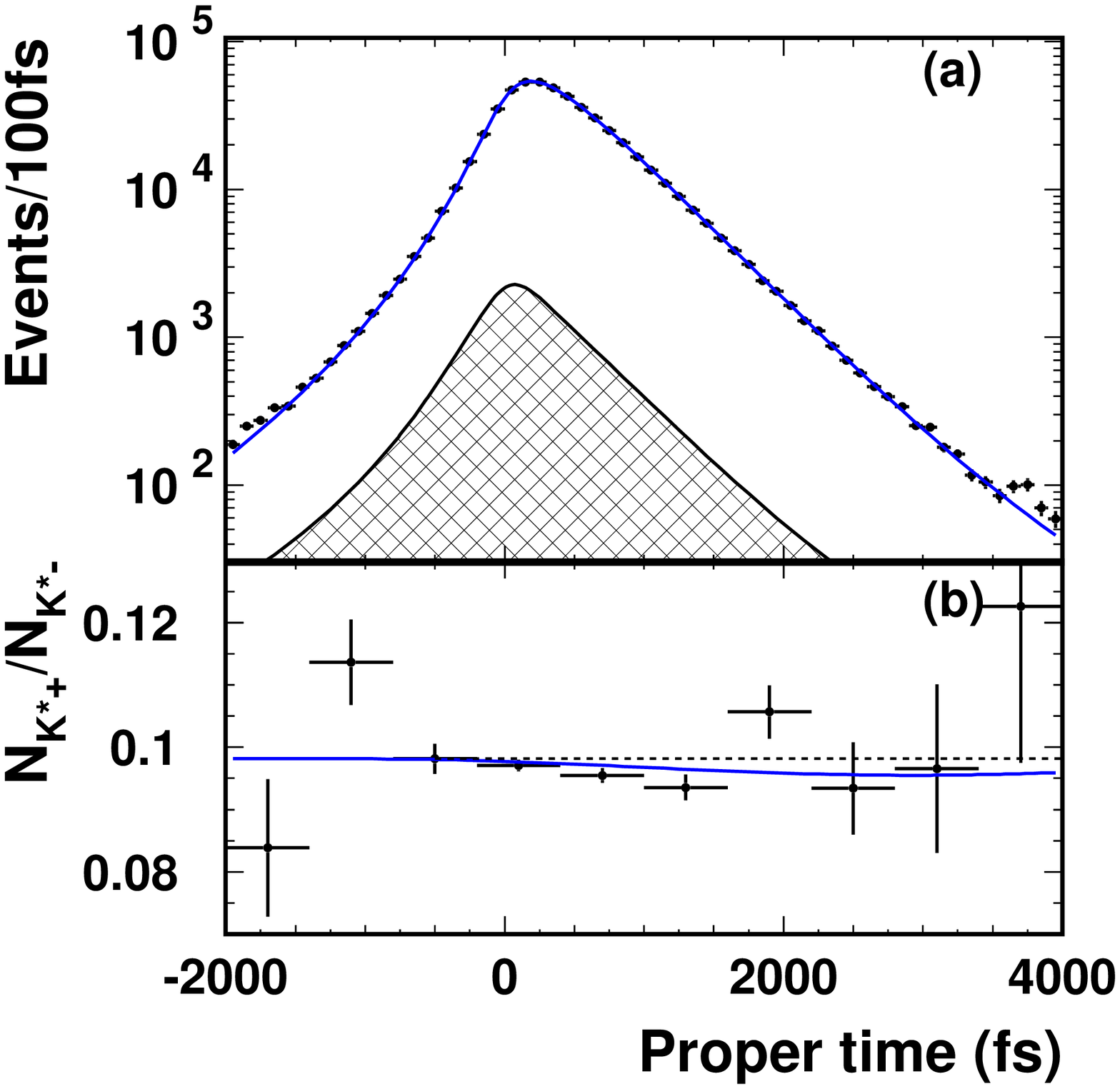}
  \hfil
}
\hbox to\hsize{%
  \includegraphics[scale=0.30]{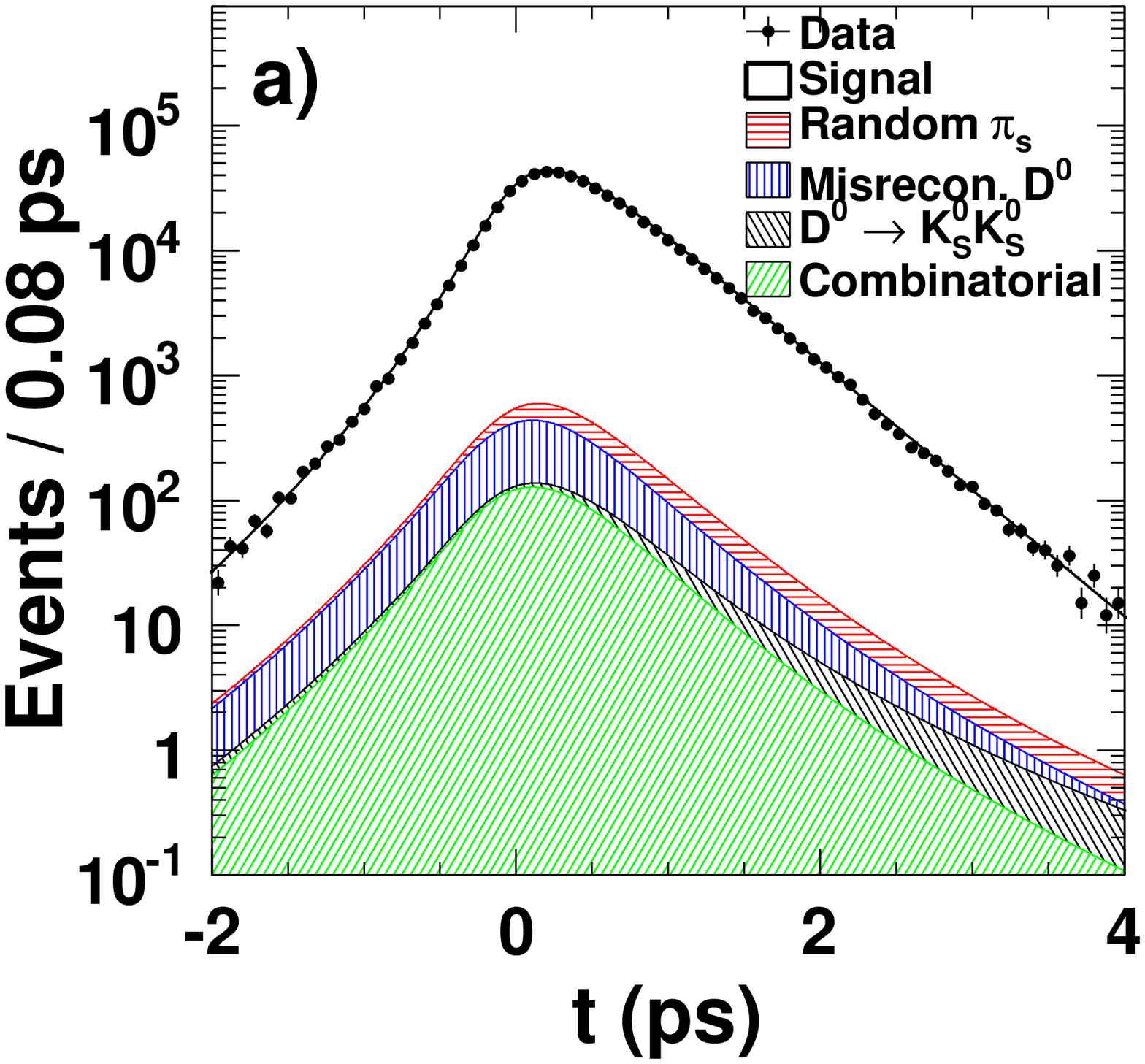}  
  \hfil
  \includegraphics[scale=0.30]{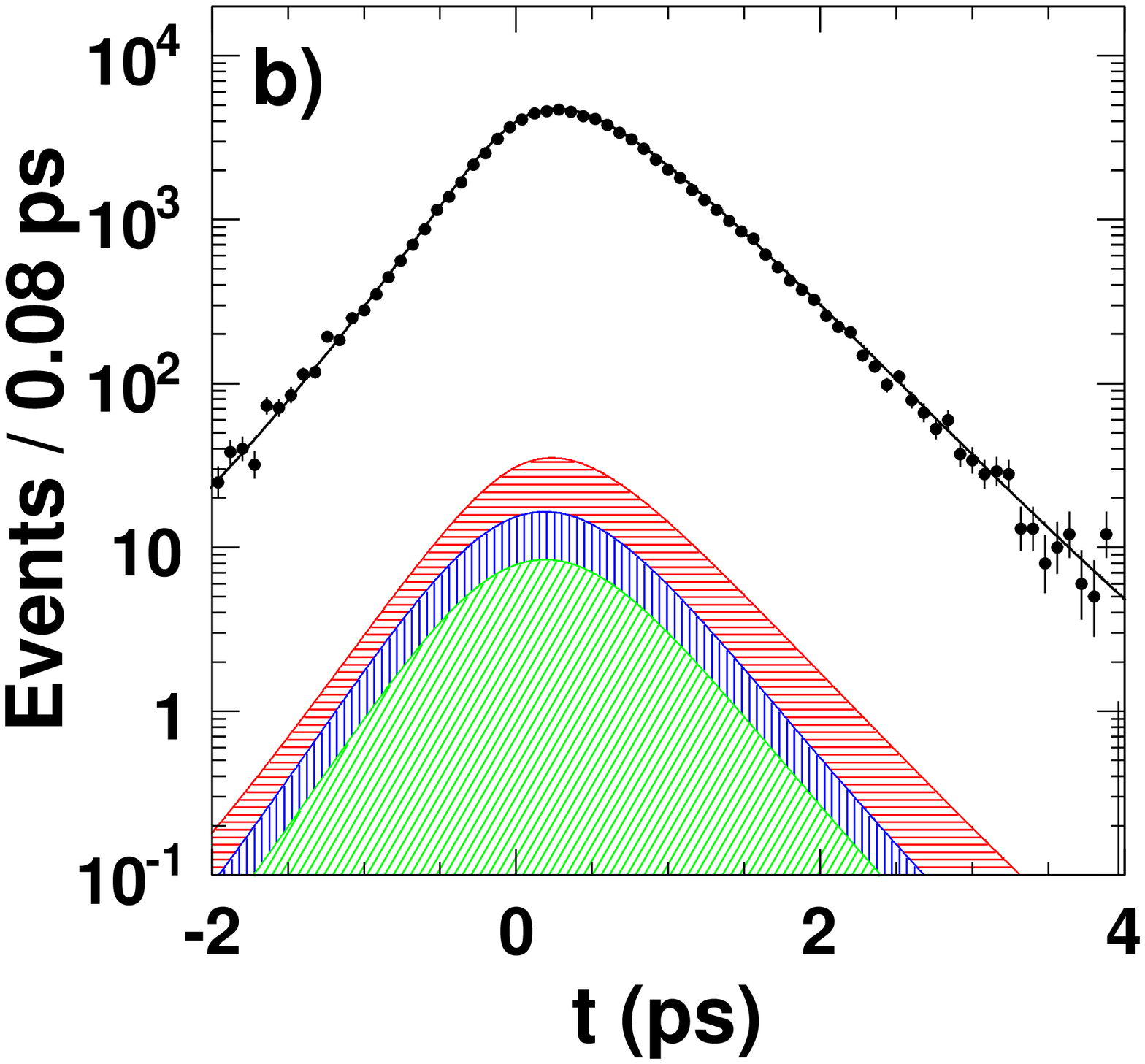}  
}
\caption{Reconstructed $\KS h^+h^-$ decay-time distributions.
Top: Belle $\KS\pip\pim$. Bottom left: \babar, $\KS\pip\pim$.  Bottom right:
\babar, $\KS \Kp\Km$.  
Reprinted figures with permission from 
L.M. Zhang {\it et al.}, {\it Phys. Rev. Lett.} 99, 131803 (2007) 
and
P. del Amo Sanchez {\it et al.}, {\it Phys. Rev. Lett.} 105, 081803 (2010) 
Copyright 2007 and 2010 by the American Physical 
Society.\protect\cite{delAmoSanchez:2010xz,Abe:2007rd}}
\label{fig:Kshh_time_dists}
\end{center}
\end{figure}

\begin{figure}[ht]
\begin{center}
\hbox to\hsize{%
  \includegraphics[scale=0.40]{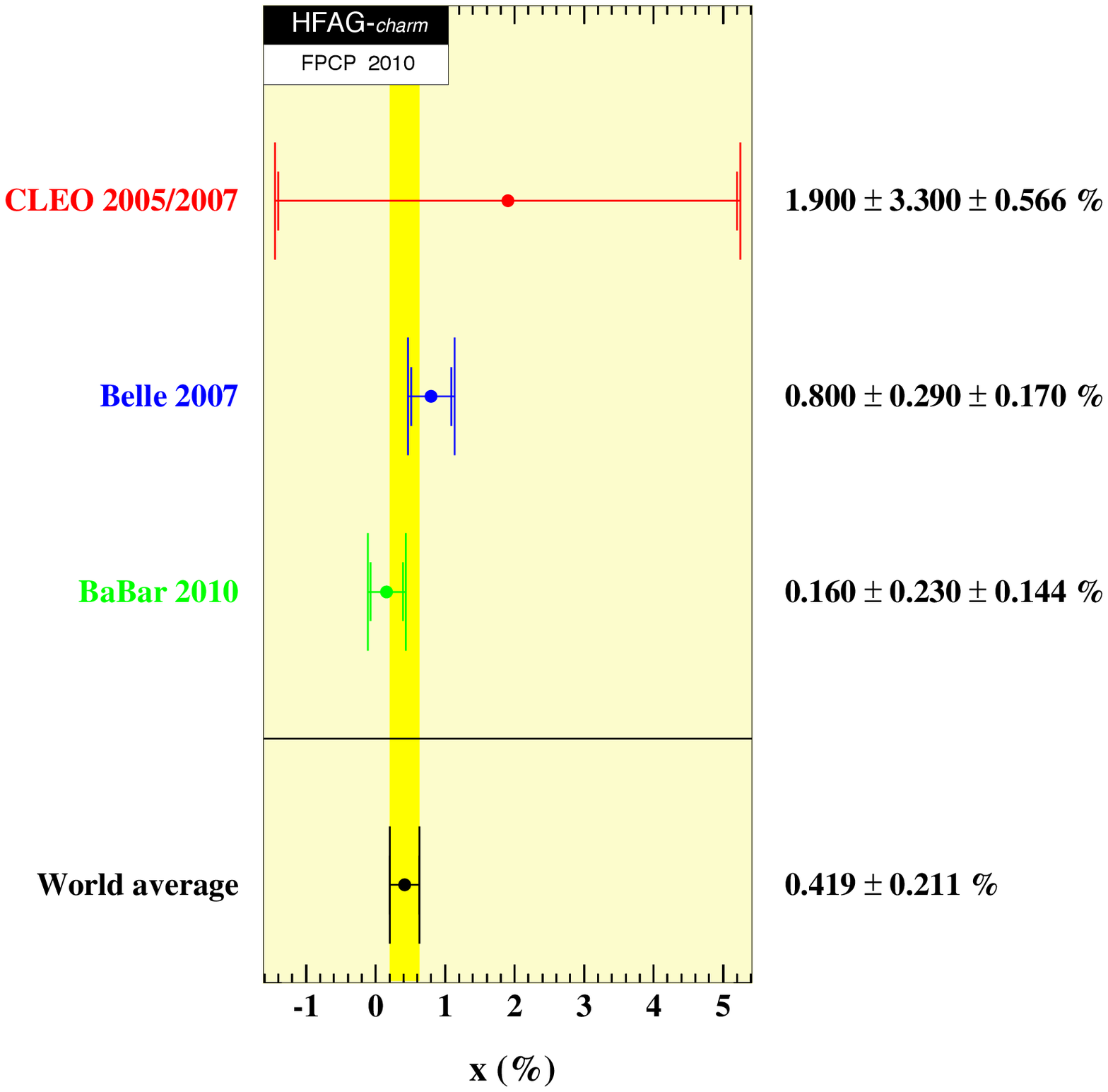}
  \hfil
  \includegraphics[scale=0.40]{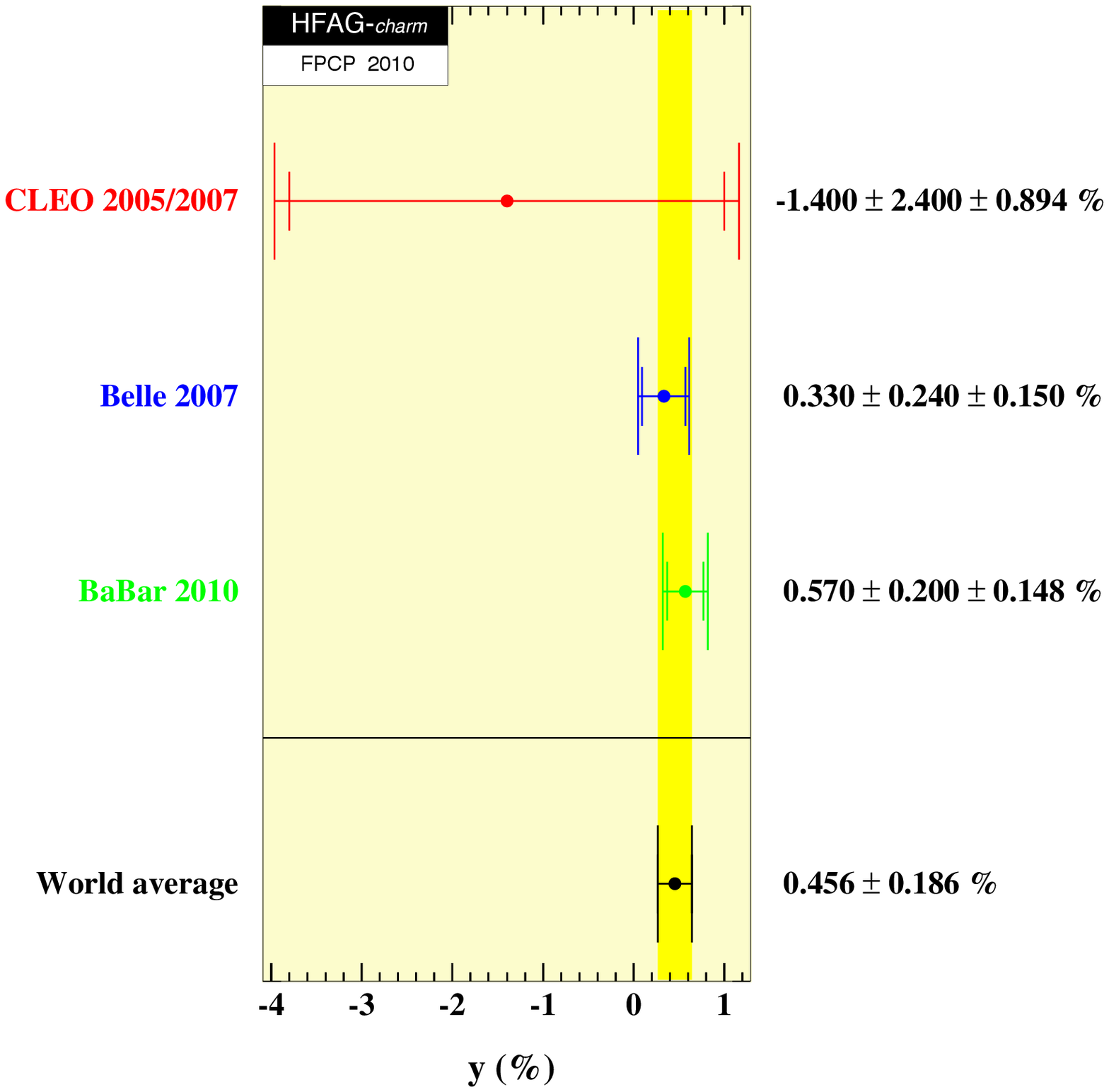}
}
\caption{$\KS h^+ h^-$ world averages for $x$ and $y$, assuming no \CP
violation.
From Refs.~\protect\refcite{Asner:2010qj,HFAGwebsite}.}
\label{fig:Kshh_x_and_y_avg}
\end{center}
\end{figure}

CLEO\cite{Asner:2005sz}, Belle\cite{Abe:2007rd,Zupanc:2009sy}, 
and \babar\cite{delAmoSanchez:2010xz}
have published results from time-dependent Dalitz-plot
analyses of $\Dz\to\KS\pip\pim$ decays and 
$\Dz\to\KS\Kp\Km$ (Belle, \babar).
The CLEO Collaboration, which
pioneered this technique, set 95\% confidence level (CL) limits
on the mixing parameters $x$ and $y$.  
Results for $x$ and $y$ are given in Table~\ref{tab:Kshh_Results} and
world averages in Fig.~\ref{fig:Kshh_x_and_y_avg}. 
No evidence for \CP violation has been
seen by any of the experiments.  

\begin{table}[ht]
\tbl{Results for $x$, $y$, and
$y_{CP}$ from $\Dz\to\KS\pip\pim$ and $\Dz\to\KS\Kp\Km$ 
time-dependent analyses.  All results are from Dalitz-plot analyses except 
the Belle $y_{CP}$ result.   Uncertainties on $x$ and $y$ are statistical,
experimental systematic, and resonance decay model systematic, respectively.
From Refs.~\protect\refcite{delAmoSanchez:2010xz,Asner:2005sz,Abe:2007rd}.}
{\label{tab:Kshh_Results}
\begin{tabular}{lcccc}
\toprule
Experiment & Fit Type & Parameter & Result  & 95\% C.L. Limit \\
\colrule
CLEO\protect\cite{Asner:2005sz}  & No CPV & $x$ (\%) & $ 1.8^{+3.4}_{-3.2} \pm 0.4 \pm 0.4$ & $(-4.7, 8.6)$ \\
CLEO\protect\cite{Asner:2005sz}  & No CPV & $y$ (\%) & $-1.4^{+2.5}_{-2.4} \pm 0.8 \pm 0.4$ & $(-6.3, 3.7)$ \\
CLEO\protect\cite{Asner:2005sz}  & CPV    & $x$ (\%) & $ 2.3^{+3.5}_{-3.4} \pm 0.4 \pm 0.4$ & $(-4.5, 9.3)$ \\
CLEO\protect\cite{Asner:2005sz}  & CPV    & $y$ (\%) & $-1.5^{+2.5}_{-2.4} \pm 0.8 \pm 0.4$ & $(-6.4, 3.6)$ \\
Belle\protect\cite{Abe:2007rd} & No CPV & $x$ (\%) & $0.80 \pm 0.29^{+0.09}_{-0.07}{}^{+0.10}_{-0.14}$ & $(0.0, 1.6)$ \\
Belle\protect\cite{Abe:2007rd} & No CPV & $y$ (\%) & $0.33 \pm 0.24^{+0.08}_{-0.12}{}^{+0.06}_{-0.08}$ & $(-0.34, 0.96)$\\
Belle\protect\cite{Abe:2007rd} & CPV    & $x$ (\%) & $0.81 \pm 0.30^{+0.10}_{-0.07}{}^{+0.09}_{-0.16}$ & $|x| < 1.6 $\\
Belle\protect\cite{Abe:2007rd} & CPV    & $y$ (\%) & $0.37 \pm 0.25^{+0.07}_{-0.13}{}^{+0.07}_{-0.08}$ & $|y| < 1.04$ \\
Belle\protect\cite{Abe:2007rd} & CPV    & $|q/p|$  &  $0.86^{+0.30}_{-0.29}{}^{+0.06}_{-0.03}\pm0.08$ & --- \\
Belle\protect\cite{Abe:2007rd} & CPV    & $\arg(q/p)$ (${}^{\circ}$) & $14^{+16}_{-18}{}^{+5}_{-3}{}^{+2}_{-4}$ & --- \\
Belle\protect\cite{Zupanc:2009sy} & No CPV & $y_{CP} (\%) $ & $0.11 \pm 0.61 \pm 0.52 $ & --- \\
\babar\protect\cite{delAmoSanchez:2010xz} & No CPV & $x$ (\%) & $0.16 \pm 0.23 \pm 0.12 \pm 0.08$ & --- \\
\babar\protect\cite{delAmoSanchez:2010xz} & No CPV & $y$ (\%) & $0.57 \pm 0.20 \pm 0.13 \pm 0.07$ & --- \\
\colrule
World average\protect\cite{HFAGwebsite} & No CPV & $x$ (\%) & $0.419 \pm 0.211 $ & \\
              &  No CPV & $y$ (\%) & $0.456 \pm 0.186 $  & \\ 
\botrule
\end{tabular}} 
\end{table} 

Experimental systematics include variations
in background and PDF models, efficiencies, event selection
criteria, and experimental resolution effects.  
In addition to these, systematics from 
the chosen resonance decay model are evaluated as well.

Additional cross-checks are performed.  Fitted values of background 
fractions, \Dz lifetimes, and decay-time scale factors are determined to be
consistent with expectations or previous results.  
Decay-time distributions are shown in Fig.~\ref{fig:Kshh_time_dists}.
Belle, CLEO, and \babar\ perform additional mixing fits to 
check for \CP violation. 
\babar\  performs separate fits to \Dz and \Dzb decays, and 
no evidence for \CP violation in mixing is seen.
CLEO performs separate fits to \Dz and \Dzb samples.  Belle incorporates
additional \CP-violating parameters in their mixing fit.
These additional fits yield results consistent with
the nominal fitted values.  No evidence for \CP violation is seen in 
any of the measurements.

\subsubsection{$\Dz\to K\pi\piz$ Analysis Results}
\label{sec:Kpipi0_results}

The first observation of the WS decay mode $\Dz\to \Kp\pim\piz$ was 
reported by CLEO in 2001.\cite{Brandenburg:2001ze}  Using a 9~\invfb
dataset of \epem\ collisions 
near the $\Upsilon(4S)$ resonance, they observed the decay with 
a 4.9 standard deviation significance and reported a wrong-sign rate of 
$R_{WS}^{K\pi\piz} = [0.43^{+0.11}_{-0.10}\pm0.07 ]\%$.  
Using 281~\invfb of \epem colliding-beam data near the $\Upsilon(4S)$,
Belle (2005) reported~\cite{Tian:2005ik} a WS branching fraction of 
$R_{WS}^{K\pi\piz} = [0.229\pm0.015^{+0.013}_{0.009}]\%$.  
In 2006 \babar\ reported~\cite{Aubert:2006kt} a measurement of $R_{WS}^{K\pi\piz} = [0.214\pm0.008\pm0.008]\%$
No evidence for \CP violation was observed in these studies.

\babar\ analyzed the $\Dz\to K\pi\piz$ decay mode using two
different methods (see Sec.~\ref{sec:Analysis_Techniques_Hadronic_Multi-body_Kpipiz}).
Method~I measured the time-integrated mixing rate 
$R_M = [0.023^{+0.018}_{-0.014} \pm 0.004]\%$ with a 95\% CL upper limit
of $R_M < 0.054\%$ assuming \CP conservation.  The result is compatible
with the no-mixing hypothesis at the 4.5\%~CL.  Additional results
are given in Table~\ref{tab:BaBar_Kpipi0_MethodIandII_Results}.  These include
measurements allowing for \CP violation.  

This study estimated systematic uncertainties by varying selection cuts,
changing background PDF shapes and the decay-time resolution model, 
varying the \Dz lifetime, and changing efficiency corrections, and
by performing the fit over the full Dalitz-plot phase space.  
Method~I results are statistics-limited.

\begin{table}[ht]
\tbl{Mixing and \CP violation results from \babar\  analyses of $\Dz\to K\pi\piz$ using method~I and method~II (see text for description of parameters).}
{\label{tab:BaBar_Kpipi0_MethodIandII_Results}
\begin{tabular}{ll}
\toprule
\CP conservation assumed  & \CP violation allowed  \\
\colrule
\multicolumn{2}{c}{\babar\  method~I\cite{Aubert:2006kt}} \\
$R_M = (0.023^{+0.018}_{-0.014} \pm 0.004)\%$ & 
$R_M = (0.010^{+0.022}_{-0.007} \pm 0.003)\%$ \\
$\alpha {\tilde y}^{\prime} = -0.012^{+0.006}_{-0.008} \pm 0.002$ &  
$\alpha {\tilde y}^{\prime} \cos {\tilde\phi} =  -0.012^{+0.006}_{-0.007} \pm 0.002$ \\ 
& $\beta {\tilde x}^{\prime} \sin {\tilde\phi} = 0.003^{+0.002}_{-0.005}\pm 0.000$\\ 
& $|p/q| = 2.2^{+1.9}_{-1.0}\pm 0.1$ \\ 
\colrule
\multicolumn{2}{c}{\babar\  method~II\cite{Aubert:2008zh}}\\
$x^{\prime}_{K\pi\piz}   =  \phantom{-}0.0261^{+0.0057}_{-0.0068} \pm 0.0039 $ & 
$x^{\prime+}_{K\pi\piz}  = \phantom{-}0.0253^{+0.0054}_{-0.0063} \pm 0.0039 $\\
$y^{\prime}_{K\pi\piz}   = -0.0006^{+0.0055}_{-0.0064} \pm 0.0034 $ & 
$y^{\prime+}_{K\pi\piz}  = -0.0005^{+0.0063}_{-0.0067} \pm 0.0050 $\\
&$x^{\prime-}_{K\pi\piz} = \phantom{-}0.0355^{+0.0073}_{-0.0083} \pm 0.0065 $\\
&$y^{\prime-}_{K\pi\piz} = -0.0054^{+0.0040}_{-0.0116} \pm 0.0041 $\\
\multicolumn{2}{c}{$r_0^2 = (5.25^{+0.25}_{-0.31} \pm 0.12)\times 10^{-3} $} \\
\botrule
\end{tabular}} 
\end{table} 

Method~II measured the mixing rate parameters $x^{\prime}_{K\pi\piz}$
and $y^{\prime}_{K\pi\piz}$ assuming \CP conservation, and 
$x^{\prime\pm}_{K\pi\piz}$ and $y^{\prime\pm}_{K\pi\piz}$ allowing for
\CP violation, where the $+(-)$ sign denotes measurement using only 
\Dz(\Dzb) candidates.  
The method~II results are inconsistent with the
no-mixing hypothesis at a significance of 3.2~standard deviations.
Mixing parameter values are 
given in Table~\ref{tab:BaBar_Kpipi0_MethodIandII_Results}.

The fit method is validated by generating Monte Carlo datasets using values
for the PDF parameters and amplitudes taken from fits to the data.
Toy Monte Carlo datasets are generated over the range $[-0.6, 0.6]$ in
$[x^{\prime}_{K\pi\piz}/r_0, [y^{\prime}_{K\pi\piz}/r_0]$, and fits to them
reconstruct 
the generated values to within an offset of 30\% of the statistical error.
Studies show that increased Monte Carlo statistics reduce this offset.  The
final results include a correction for this effect.

Systematic tests performed in the method~II  analysis include setting the 
decay-time resolution mean to zero (fitted value $4.2\pm 0.7$~fsec); 
changing the isobar model by varying the masses and widths of the 
included resonances within their errors; varying the numbers
of signal and background events in each category; and changing the definition
of the signal-region and candidate-selection criteria.

\subsubsection{$\Dz\to \Kp\pim\pip\pim$ Analysis Results}
\label{sec:K3pi_results}

\babar\  has reported a preliminary result\cite{Aubert:2006rz} using
a time-dependent analysis of 
the $\Dz\to K3\pi$ mode to set a limit on the mixing rate of 
$\Rm < 0.048\%$ at
the 95\% confidence level ($\Rm = [0.019^{+0.016}_{-0.015}\pm0.002]\%$),
and estimates that this is consistent with a no-mixing hypothesis at
the 4.5\% confidence level.  
This preliminary result, based on a 230~\invfb 
data sample, used tagged \Dstarp events
to determine the production flavor of the $\Dz$ candidate via the 
charge of the slow pion~$\pi_{\rm s}$ and reconstructed
the $\Dz$ candidate mass $m_{K\pi\pi\pi}$, the mass difference $\Delta m$,
the decay time $t$, and its uncertainty $\terr$. 

\begin{figure}[h!]
  \begin{center}
  \includegraphics[scale=0.7]{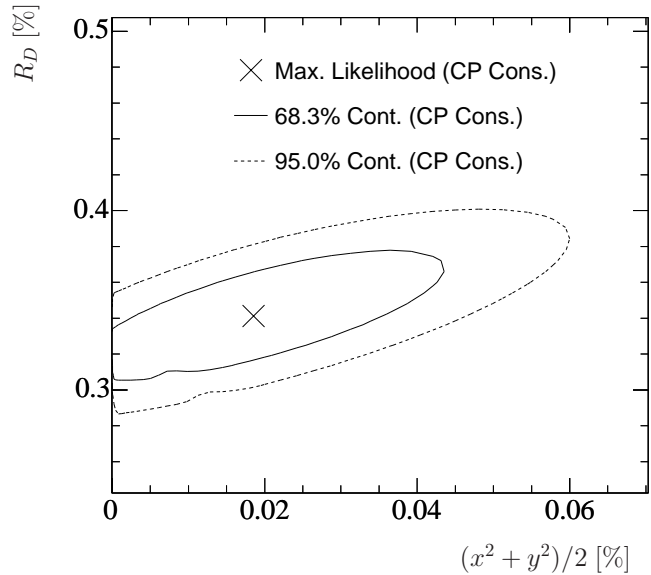}%
  \end{center}
  \caption{Likelihood contours from the \babar\ $K3\pi$ analysis for the $CP$ conserving fit for $\tilde R_D$
     vs. the mixing rate $R_M = (x^2+y^2)/2$.  Solid line:  $\Delta\ln {\cal L}
     = 1.15$; dotted line:   $\Delta\ln {\cal L} = 3.0$. 
     From Ref.~\protect\refcite{Aubert:2006rz}.}
  \label{fig:k3pi:nllrmix}
\end{figure}

\begin{figure}[h!]
  \begin{center}
  \includegraphics[scale=0.7]{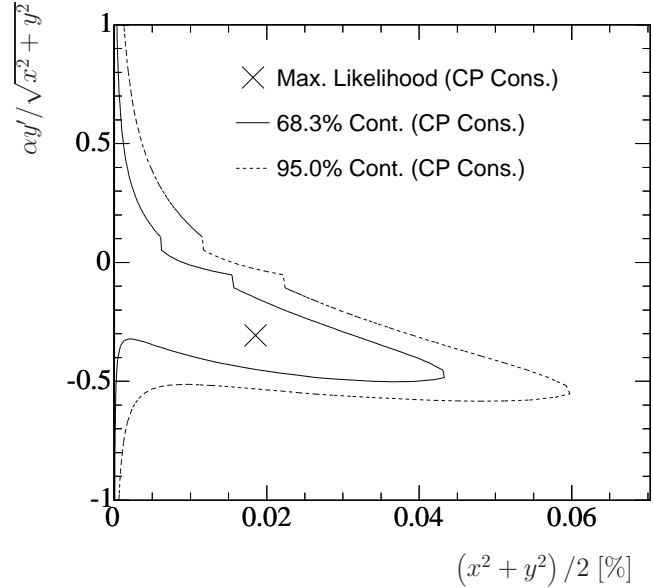}%
  \end{center}
  \caption{Likelihood contours from the \babar\ $K3\pi$ analysis for the $CP$ conserving fit for the interference
     term vs. the mixing rate $R_M = (x^2+y^2)/2$.  
     Solid line:  $\Delta\ln {\cal L}
     = 1.15$; dotted line:   $\Delta\ln {\cal L} = 3.0$.
     From Ref.~\protect\refcite{Aubert:2006rz}.} 
  \label{fig:k3pi:nllint}
\end{figure}

First, an unbinned extended maximum-likelihood fit is performed simultaneously
to the RS and WS two-dimensional distributions $(m_{K\pi\pi\pi}, \Delta m)$.
Approximately $3.5\times 10^5$ RS signal
candidates are found for $\Dz$ and $\Dzb$, respectively,
and about 1100 WS signal candidates for each.
The much larger RS sample effectively determines the
resolution function parameters and \Dz lifetime used in the WS mixing
fit.
Backgrounds include mis-reconstructed \Dstarp decays, mis-reconstructed
charm decays, and combinatorics.

Second, fits to the decay-time distributions are performed.   
The wrong-sign time-dependence is fitted to 
\begin{equation}
\frac{\Gamma_{\rm WS}(t)}{\Gamma_{\rm RS}(t)} =
  \tilde R_D + \alpha \tilde y^{\prime}\sqrt{\tilde R_D} (\Gamma t)
  + \frac{(x^2 + y^2)}{4} (\Gamma t^2),
\label{eq:k3pi:ws:time}
\end{equation}
where a tilde accent indicates quantities that are integrated 
over a region of phase space. 
The factor $\alpha$ accounts for strong-phase variation over the region;
A $CP$-conserving fit (which considers $\Dz$ and $\Dzb$ candidates
simultaneously) and a fit that is sensitive to $CP$ violation
(which treats them separately) are performed.  
The substitutions
\begin{eqnarray}
\alpha\tilde y^{\prime} &\to& |p/q|^{\pm} ( \alpha\tilde y^{\prime}   
     \cos \tilde\phi \pm \beta\tilde x^{\prime} \sin \tilde\phi ), \\         
x^2 + y^2 &\to& |p/q|^{\pm2} (x^2 + y^2)
\label{eq:k3pi:cpv}
\end{eqnarray}
are made in Eq.~\ref{eq:k3pi:ws:time}, choosing the ``$+$'' (``$-$'')
sign for $\Dz$ ($\Dzb$) candidate decays, respectively.
The factor $\beta$ accounts for $\phi$ variation over the phase space
region.

Systematic uncertainties are estimated by varying the selection criteria,
the PDF parametrization of the decay-time
resolution function, the background PDF shapes, and the measured $\Dz$ 
lifetime value.
The combined systematics are smaller than the statistical
errors by a factor of five.

Assuming $CP$ conservation, the \babar\ preliminary analysis finds
$R_M \equiv (x^2+y^2)/2$ and the effective mixing parameter
$\alpha \tilde y^{\prime}$ to be
\begin{eqnarray}
R_M &=& [0.019 {}^{+0.016}_{-0.015}({\rm stat.})\pm0.002 ({\rm syst.})]\%,\\
\alpha \tilde y^{\prime} &=& -0.006 \pm {0.005}({\rm stat.}) \pm 0.001({\rm syst.}),
\label{eq:k3pi:rm}
\end{eqnarray}
which are consistent with the no-mixing hypothesis at the
4.3\% C.L.  
Allowing for $CP$ violation, the analysis finds
\begin{eqnarray}
R_M &=& [0.017 {}^{+0.017}_{-0.016}({\rm stat.})\pm0.003 ({\rm syst.})]\%,\\
|p/q| &=& 1.1  {}^{+4.0}_{-0.6}({\rm stat.})\pm0.1({\rm syst.}),\\
\alpha \tilde y^{\prime}\cos\tilde\phi &=& -0.006^{+0.008}_{-0.006}({\rm stat.}) \pm 0.006({\rm syst.}),\\
\beta \tilde x^{\prime}\cos\tilde\phi &=& 0.002^{+0.005}_{-0.003}({\rm stat.}) \pm 0.006({\rm syst.}).
\label{eq:k3pi:pq}
\end{eqnarray}

Two-dimensional coverage probabilities of 68.3\% and 95.0\% 
($\Delta\log{\cal L} = 1.15$, $3.0$, respectively) are shown in 
Fig.~\ref{fig:k3pi:nllrmix} for the doubly Cabibbo-suppressed rate
$\tilde R_D$ vs. the mixing rate $R_M$, and
in Fig.~\ref{fig:k3pi:nllint} for the interference term 
$\alpha y^{\prime}/\sqrt{x^2+y^2}$ vs. $R_M$.

\subsubsection{Semileptonic Decays Analysis Results}
\label{Semileptonic_Decays_results}

\begin{figure}[ht]
\begin{center}
\includegraphics[scale=0.5]{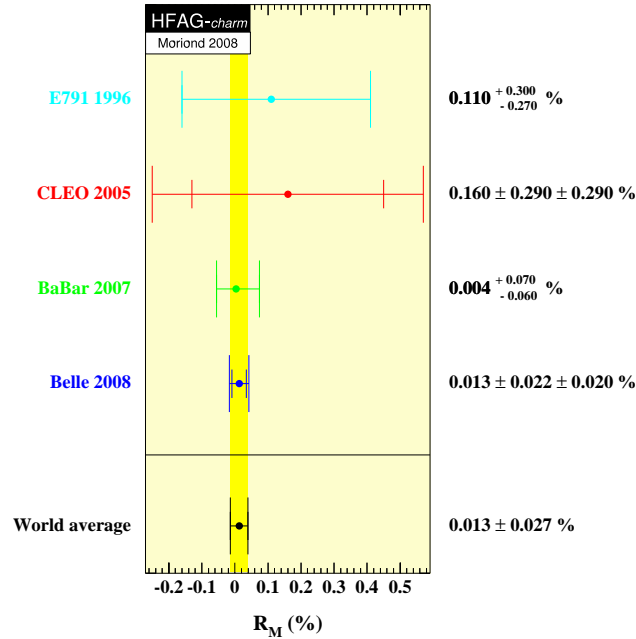}%
\caption{World average of semileptonic mixing results for \Rm in 
Table~\protect\ref{tab:Semileptonic_Mixing_Results}.  Due to possible 
correlations with the \babar\ doubly-tagged result, the \babar\ singly-tagged
result is not included in the average. 
From Ref.~\protect\refcite{Asner:2010qj}.} 
\label{fig:semilep_WA}
\end{center} 
\end{figure}

Results from five analyses using semileptonic \Dz decays are summarized in 
Table~\ref{tab:Semileptonic_Mixing_Results}.  These include results from
$\pim N$ collisions (E791\cite{Aitala:1996vz} at FNAL, $2 \times 10^{10}$ events) and
\epem interactions near the $\Upsilon(4S)$ resonance 
(CLEO II.V,\cite{Cawlfield:2005ze} 9.0~\invfb;
Belle,\cite{Bitenc:2008bk} 492~\invfb;
\babar\ singly-tagged,\cite{Aubert:2004bn} 87~\invfb; and \babar\ doubly-tagged,\cite{Aubert:2007aa} 344~\invfb).
The world average is shown in Fig.~\ref{fig:semilep_WA}.  Results from
both $K^{(*)}e\nu$ and $K^{(*)}\mu\nu$ decay modes are included.

\begin{table}[ht]
\tbl{Mixing results using semileptonic \Dz decay modes.  Uncertainties are statistical (first) and
systematic (second), except as noted.}
{\label{tab:Semileptonic_Mixing_Results}
\begin{tabular}{lll}
\toprule
Experiment & \Dz modes & Results \\
\colrule
E791\protect\cite{Aitala:1996vz} & $Ke\nu$ & $\Rm = (0.16^{+0.42}_{-0.37})$\% \\ 
& $K\mu\nu$ & $\Rm = (0.06^{+0.44}_{-0.40})$\% \\
& Combined &  $\Rm = 0.11^{+0.30}_{-0.27}$\% \\ 
&                     &  $\Rm < 0.50\%$ at 90\% CL \\
\colrule
CLEO II.V\protect\cite{Cawlfield:2005ze} & $Ke\nu$ & $\Rm = (1.10\pm 0.76)$\% (stat.+syst. combined)\\
& $K^*e\nu$ & $\Rm = (\phantom{0}0.0 \pm 0.31)$\% (stat.+syst. combined)\\
& Combined & $\Rm = (0.16 \pm 0.29)$\% (stat.+syst. combined)\\
\colrule
Belle\protect\cite{Bitenc:2008bk} & $K^{(*)}e\nu$ & $\Rm = (-0.6 \pm 2.7^{+1.8}_{-2.1})\times 10^{-4}$ \\
& $K^{(*)}\mu\nu$ & $\Rm = (5.9 \pm 3.7^{+3.9}_{-4.5})\times 10^{-4}$ \\
& Combined & $\Rm = (1.3 \pm 2.2 \pm 2.0) \times 10^{-4}$ \\
&          & $\Rm < 6.1 \times 10^{-4}$ at 90\% CL \\
\colrule
BaBar singly tagged\protect\cite{Aubert:2004bn} & $K^{(*)}e\nu$ & $\Rm = 0.0023 \pm 0.0012 \pm 0.0004$ \\
& & $\Rm < 0.0042$ at 90\% CL \\
\colrule
BaBar doubly tagged\protect\cite{Aubert:2007aa} & $K^{(*)}e\nu$ & $\Rm = 0.4 \times 10^{-4}$ (central value) \\
& & $\Rm$ in $(-2.2,2.8) \times 10^{-4}$ at 68\% CL \\
& & $\Rm$ in $(-13,12) \times 10^{-4}$ at 90\% CL \\
\botrule
\end{tabular}}
\end{table} 

Since the E791 analysis uses identical selections for both RS and WS
candidates, systematic uncertainties in the mixing rate measurement largely 
cancel.  Two sources of systematic uncertainty that were investigated
are the decay-time resolution modeling and feedthrough of hadronic
decays into the semileptonic sample.  The decay-time determination
is subject to detector effects and to the ambiguity from the missing
neutrino.  Decay times were estimated to be uncertain with a Gaussian
smearing of about 15\%.  This affects the final mixing result by only
about 10\% of its statistical uncertainty, and is not significant.
Feed-through of hadronic events via a hadron mis-identified as a 
lepton could increase the number of either RS or WS events.  The 
former would overestimate the size of the RS signal and cause
an incorrect estimate of the sensitivity to mixing, while the latter
would cause a false WS signal.  Since RS feed-through was estimated to
be very small (about 3\%) and, since no WS signal was seen, no 
corrections were made.  Evaluation of the fit modeling systematic
uncertainty (performed by adding 10 to 50 simulated mixed events
to the WS sample) showed a bias of 10--15\% toward a larger
mixing rate.  Since the final result is an upper limit, no correction
was applied.

In the Belle semileptonic result, the main systematics include uncertainty
in the signal and background $\Delta m$ distributions, the amount of RS and WS 
backgrounds, RS and WS efficiencies, and modeling of the decay-time 
distribution.  These are estimated separately for each of four subsamples,
which are categorized by whether the candidate contains an electron or 
a muon, and which of two silicon vertex detector 
configurations was used to record the event.
The overall muon sample systematics are about double that
of the electron samples, due in part to larger backgrounds in the signal
regions. 

The CLEO semileptonic analysis uses simulated events to model the background
and signal shapes used in the fit.  The largest systematic comes from 
the statistics of the simulation.  The second largest systematic is the
shape of the decay-time distribution, also obtained from simulation.  Other
sources of systematic uncertainties are 
the $Q$ shape ($Q = {}$energy released in the \Dstarp
decay), electron identification, and fit modeling.

The \babar\ singly-tagged semileptonic result systematics include contributions
from signal and background PDF shapes, the decay-time resolution model,
and decay-time PDFs for background charm decays.
Other possible contributions to the systematic on 
\Rm are evaluated and shown to provide no significant contribution. 
The total systematic error on \Rm is about 1/3 of the 
statistical uncertainty.

The \babar\ doubly-tagged semileptonic analysis finds three mixing signal candidates
where 2.85 background events are expected.  A 50\% systematic
to the background rate is assigned by comparing ten background control
samples with corresponding MC simulations.  
Other contributions to systematic uncertainties
are ignored in comparison to this 50\% error.  Confidence levels
are then calculated for \Rm using a frequentist method.

\subsection{World Average Results}
\label{sec:WA_results}
\begin{figure}[h!]
\begin{center}
\hbox to\hsize{%
  \includegraphics[width=0.49\hsize]{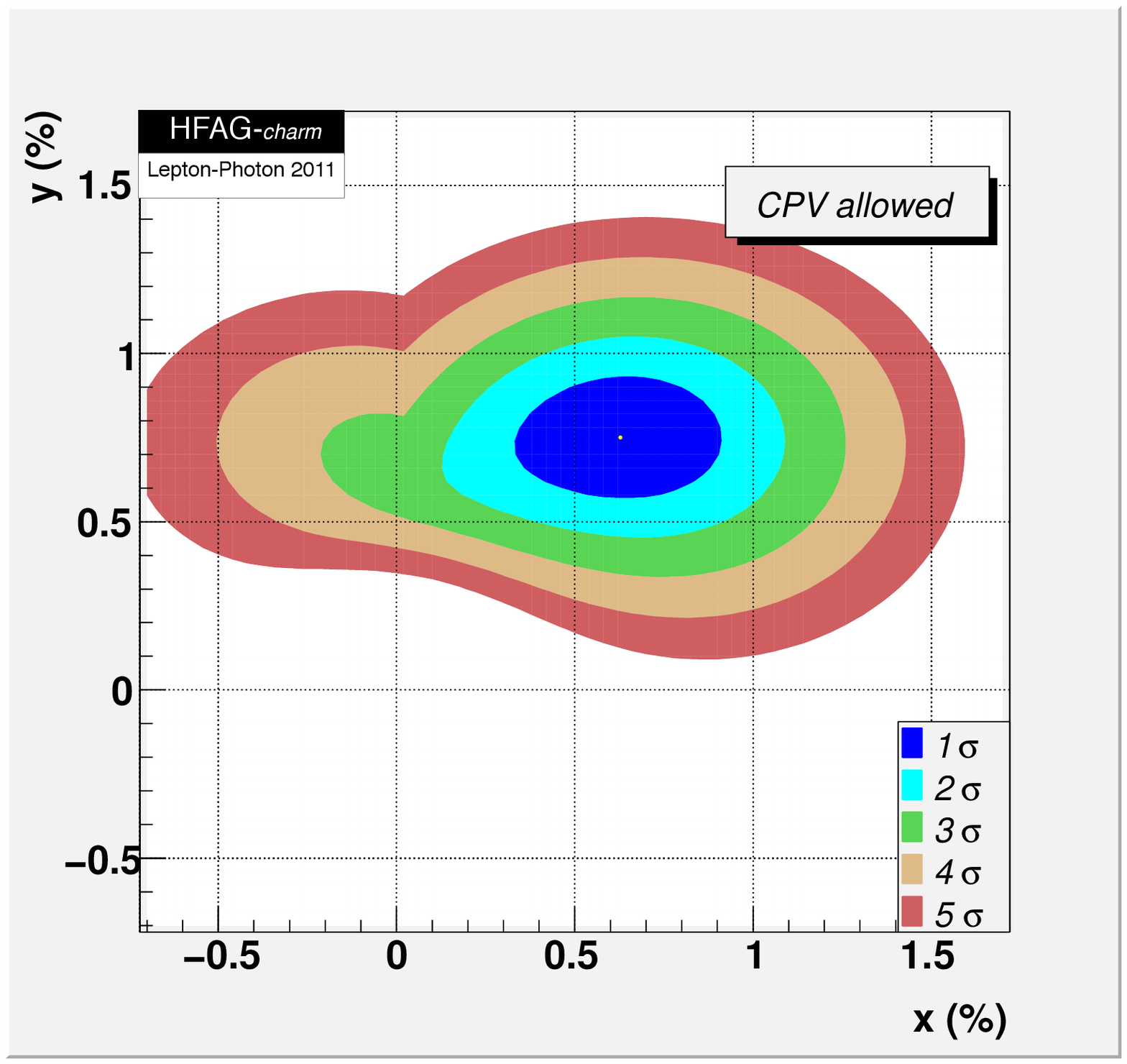}
  \hfil  
  \includegraphics[width=0.49\hsize]{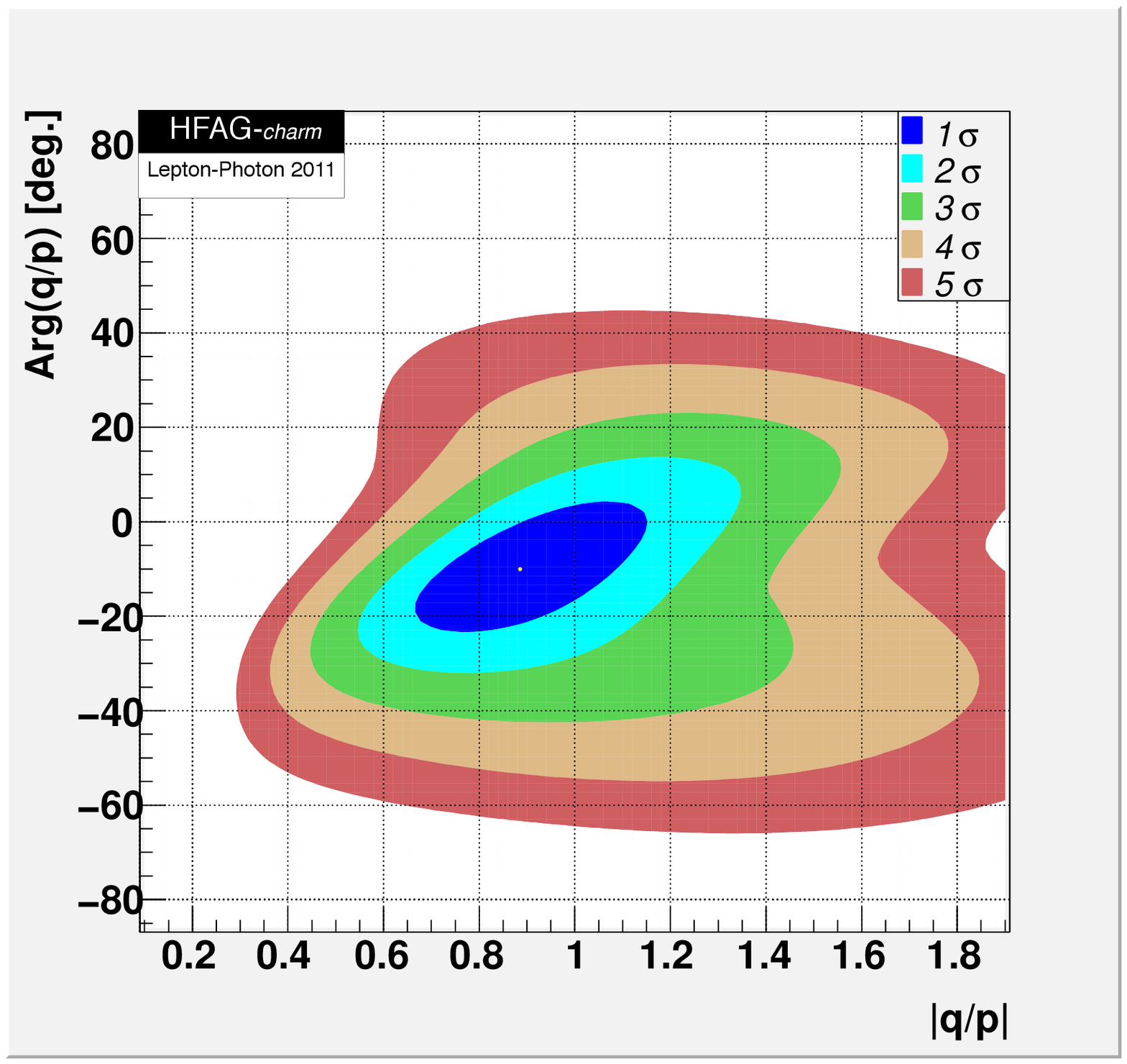}}%
\caption{HFAG world average probability contours\protect\cite{HFAGwebsite} for
 the \DzdashDzb mixing parameters $(x$, $y)$\ (left) and for the \CPV
parameters
$(|q/p|$, $\varphi=\arg(q/p))$ (right).} 
\label{fig:fig_plot_xyn2d}
\end{center}
\end{figure}

From HFAG, the world average values for the \DzdashDzb mixing and 
\CPV parameters are shown in 
Tab.~\ref{tab:WA_mixing}.
\begin{table}[ht]
\tbl{HFAG world average mixing and \CPV parameter values.\protect\cite{HFAGwebsite}}
{\label{tab:WA_mixing}
\begin{tabular}{cccc}
\toprule
Parameter & No \CPV & No direct \CPV & \CPV-allowed \\
\colrule
$x\ (\%)$ & $0.65^{+0.18}_{-0.19}$ & $0.63\pm0.19$ & $0.63^{+0.19}_{-0.20}$ \\
$y\ (\%)$ & $0.74\pm0.12$        & $0.75\pm0.12$ & $0.75\pm0.12$ \\
$|q/p|$   & ---                  & $1.02\pm0.04$ & $0.89^{+0.17}_{-0.15}$ \\
$\varphi\space(^\circ)$ & ---     & $-1.05^{+1.89}_{-1.94}$ & $-10.1^{+9.4}_{-8.8}$ \\
\botrule
\end{tabular}}
\end{table}
The probability contours, including both statistical and systematic uncertainties and allowing for \CPV, are shown in
Fig.~\ref{fig:fig_plot_xyn2d} for the mixing parameters $(x,y)$ and
\CPV parameters $(|q/p|$, $\varphi=\arg(q/p))$. 
The world average $(x$, $y)$ excludes the no-mixing point $(x=0$, $y=0)$
by $10.1$ standard deviations. To date, however, no single measurement
exceeds five standard deviations.  The no-\CPV point 
$(|q/p|=0$, $\varphi=0)$ lies within one standard deviation of the world average
$(|q/p|$, $\varphi)$ value. The recent LHCb measurements of mixing and direct 
$CPV$\cite{Charles:2011nx,Aaij:2011ad} are not included in these averages.

%%------ FUTURE PROSPECTS --------

\section{What's Next?}
\label{sec:Whats_Next}

Our understanding of charm physics has made great progress 
since 1975, when the first evidence for charm mesons was observed.
The fact that the no-mixing hypothesis has been excluded by 10 standard
deviations,
combined with improved understanding of the mechanisms leading to
\DzdashDzb mixing and \CPV,
leaves us in a position to make considerable progress in
the next few years in the charm sector in both experimental accuracy
and theoretical interpretation.  Here we survey a few
experiments that are likely to further 
our knowledge of charm physics and mixing
over the next few years.

\subsection{BES-III}
\label{sec:BES-III}

With well over two years of data-taking at the time of this writing, the 
BES-III experiment at BEPC-II has already surpassed both CLEO-c and BES-II
in recorded luminosity.\cite{Zheng:2011yu}  With over 200 million
$J/\psi$ and 100 million $\psi^{\prime}$ events, the experiment has
about $4\times$ the data samples of BES-II and CLEO-c.  Performance 
of the machine is good, with peak luminosities of order 
$10^{32}$~cm$^{-2}$sec$^{-1}$.

Of interest in the context of charm mixing is the machine's 
performance near the $\psi(3770)$, where it has reached a peak luminosity of 
$5.6\times 10^{32}$~cm$^{-2}$sec$^{-1}$ and recorded over 1~\invfb of
data in less than a year. 
BES-III plans to increase the $\psi(3770)$ dataset to 2.5~\invfb in the
next year or so, with a goal to eventually reach 10~\invfb.
Using the coherent decay techniques discussed
earlier, it is clear that BEPC-II and BES-III 
should be able to substantially improve our
knowledge of charm mixing 
in the very near future.

\subsection{LHCb}
\label{sec:LHCb}

At the time of this writing, LHCb has embarked on its charm physics
program using data taken in 2010 and 2011 and has reported results on
open charm production\cite{Gersabeck:2010bd,Hunt:2011zz} and other 
measurements.  Given the
detector design which is optimized for heavy-flavor physics, LHCb is 
expected to provide precision measurements of charm mixing and
\CP violation parameters in the next few years.  First results 
showing evidence for direct \CP violation by measuring the difference 
between the two time-integrated \CP asymmetries 
${\cal A}(\Dz\to\Kp\Km)$ and ${\cal A}(\Dz\to\pip\pim)$
have already been reported.\cite{Charles:2011nx,Aaij:2011in} Results
on $y_{CP}$ and $A_{\Gamma}$ are expected 
soon,\cite{Hunt:2011zz}
based on analyses of the higher-statistics 2011 and 2012 data samples.
Additional competitive
charm mixing and \CPV measurements are expected to be forthcoming.

\subsection{SuperKEKB and Belle II}
\label{sec:SuperKEKB}

After a decade-long successful program, the Belle detector and the KEKB
accelerator stopped operations in June 2010.\cite{Palka:2010zz}  
Construction of SuperKEKB
has started, and work has begun on the Belle~II detector.  An initial
data sample of 5~\invab is planned to be recorded starting in 2014 with
the eventual goal to reach 50~\invab by 2021--2022.\cite{Aushev:2012ed}
With these integrated 
luminosities, Belle~II will have an excellent opportunity to improve 
on current \DzdashDzb mixing and \CPV measurements.  With 5~\invab,
Belle~II is expected to improve the existing statistics-limited measurements
of $x$ and $y$ by approximately a factor of two; 
with 50~\invab, an additional factor of two.\cite{Abe:2010sj}  

\subsection{The Super Flavor Factory SuperB}
\label{sec:SuperB}

The recently approved SuperB project\cite{O'Leary:2010af}
in Italy will be able to contribute
substantially to our knowledge of charm mixing and \CP violation.  
Plans call for the SuperB facility to be able to run
at the $\psiprpr$, where a sample of
$2\times10^9$ \DzdashDzb pairs is expected to be accumulated.
Both avenues are likely to lead to greatly increased understanding of
the details of mixing and \CPV.  
Also, like
its predecessor \babar, SuperB will be able to make use of the large
charm production cross section near the $\Upsilon(4S)$.
Estimates of statistical 
uncertainties using both $K\pi$ and lifetime ratio methods range from
$6\times$ to $12\times$ improvements over existing measurements, and
possibly even better, depending on how much SuperB's improved
decay-time resolution contributes.

\section{Summary}
\label{sec:Summary}

Evidence for charm mixing at the level of 1\% in the mixing
parameters, first reported in 2007 by the \babar\ and Belle 
experiments, along with the recent evidence for direct \CPV
obtained by the LHCb Collaboration,
has created renewed interest in the
charm sector as a window to new physics.
In the near future, BES-III and LHCb should be 
reporting new charm results, along with final contributions from
\babar, Belle, CDF, and CLEO.  In the next several
years, SuperKEKB and SuperB should 
improve the precision of mixing and \CPV measurements by a factor of ten or
more.  This will be an exciting time for anyone interested in charm 
physics or precision flavor physics in general.

\section*{Acknowledgments}

The authors would like to thank their colleagues for helpful conversations
and feedback while preparing this article, including I. Bigi, P. Fisher,
K. Flood, J. Hewett, A. Kagan, B. Meadows, M. Peskin, A. Schwartz, 
M. Sokoloff and W. Sun.
The authors also gratefully acknowledge support by the U.S. Department of Energy, and would like to
thank CERN and the SLAC National Accelerator Laboratory for their kind
hospitality.

\newpage
\appendix

\bibliographystyle{unsrt} 
\bibliography{D_mixing_review}

\end{document}